\documentclass[fleqn,usenatbib]{mnras}
\hypersetup{linkcolor=blue,citecolor=blue,filecolor=cyan,urlcolor=magenta}

\usepackage{newtxtext,newtxmath}
\usepackage[T1]{fontenc}

\DeclareRobustCommand{\VAN}[3]{#2}
\let\VANthebibliography\thebibliography
\def\thebibliography{\DeclareRobustCommand{\VAN}[3]{##3}\VANthebibliography}

\usepackage{graphicx}	% Including figure files
\usepackage{amsmath}	% Advanced maths commands
\usepackage{bm}

\defcitealias{Harding2021}{AH21}
\defcitealias{Harding2015}{AH15}
\defcitealias{Cerutti2016}{CS16}
\defcitealias{Vigano2015}{VT15}
\defcitealias{Kelner2015}{KP15}
\defcitealias{Barnard2022}{BH22}
\defcitealias{DuPlessis2024}{DP24}
\defcitealias{DuPlessis2025}{DP25}

\title[AR Sco Calibration]{Towards Modelling AR Sco: Calibration -- Reproducing High-Energy Pulsar Emission and Testing Convergence to Aristotelian Electrodynamics}

\author[L. Du Plessis et al.]{
L. Du Plessis,$^{1}$\thanks{E-mail: louisdp95@gmail.com}
C. Venter,$^{1,2}$
A.K. Harding$^{3}$
Z. Wadiasingh$^{4,1}$
and C. Kalapotharakos$^{4}$
\\
$^{1}$Centre for Space Research, North-West University, Private Bag X6001, Potchefstroom 2520, South Africa\\
$^{2}$National Institute for Theoretical and Computational Sciences, South Africa\\
$^{3}$Theoretical Division, Los Alamos National Laboratory, Los Alamos, NM 58545, USA\\
$^{4}$Astrophysics Science Division, NASA Goddard Space Flight Center, Greenbelt, MD 20771, USA
}

\date{Accepted 2025 June 4. Received 2025 June 4; in original form 2025 February 25}

\pubyear{2024}

\begin{document}
\label{firstpage}
\pagerange{\pageref{firstpage}--\pageref{lastpage}}
\maketitle

\begin{abstract}
In recent years, kinetic simulations have been crucial to further our understanding of pulsar electrodynamics. Yet, due to the large-scale separation between the gyro-period vs the stellar rotation period, resolving the particle gyration has been computationally unfeasible for realistic pulsar parameters. The main aim of this work is comparing our gyro-phase-resolved model with a gyro-centric pulsar model, where our model solves the general equations of motion with included radiation reaction using a higher-order numerical solver with adaptive time steps. Specifically, we aim to (i) reproduce a pulsar's high-energy emission maps, namely one with $10\%$ the surface $B$-field strength of Vela, and spectra produced by an independent gyro-centric pulsar emission model; (ii) test convergence of these results to the radiation-reaction limit of Aristotelian Electrodynamics. (iii) Additionally, we identify the effect that a large $E_{\parallel}$-field has on the trajectories and radiation calculations. We find that we can reproduce the curvature radiation emission maps and spectra well, using $10\%$ field strengths of the Vela pulsar and injecting our particles at a higher altitude in the magnetosphere. Using sufficiently large $E_{\parallel}$-fields, our numeric results converge to the analytic radiation-reaction limit trajectories. Additionally, we illustrate the importance of accounting for the $\mathbf{E}\times \mathbf{B}$-drift in the particle trajectories and radiation calculations, validating the Harding and collaborators' model approach.
% We also found that the radiation reaction does not damp the pitch angle in the super-relativistic regime, and the latter increases due to the $\mathbf{E}\times \mathbf{B}$-drift. This illustrates the effect of using synchrotron and synchro-curvature radiation expressions derived assuming that no $E$-field is present, where a high $E_{\perp}$-field is actually present. 
Lastly, we found that our model deals very well with the high-radiation-reaction and high-field regimes present in pulsars.      
\end{abstract}


%%%%%%%%%%%%%%%%%%%%%%%%%%%%%%%%%%%%%%%%%%%%%%%%%%

%%%%%%%%%%%%%%%%% BODY OF PAPER %%%%%%%%%%%%%%%%%%

\section{Introduction} \label{sec:1}
Over the years, various pulsar emission models have been proposed to explain the multi-wavelength signatures observed from millisecond pulsars. Local acceleration gap models such as the slot gap \citep{Arons83, Muslimov03} and the outer gap \citep{Chen1984, Romani95} could produce reasonable light curves and spectra, but fell short of explaining pair production, where particles are accelerated, global current patterns, and the multi-wavelength emission. These topics are still somewhat uncertain, even with the significant advancements of modern pulsar emission models, but they have highlighted the importance of solving the Maxwell equations self-consistently, including the particle dynamics and the pulsed emission. One of the important advancements for these models is the force-free electrodynamics (FFE) framework \citep{Contopoulos1999, Spitkovsky2006, Timokhin2006, Kalapotharakos2009} where the plasma is solely governed by the Lorentz force, meaning inertia and gas pressure are ignored. Also, no gaps would form in this plasma-filled magnetosphere, so in principle there can be no particle acceleration and subsequent radiation. This regime therefore represents an idealised limiting case. An important addition to the FFE framework were the dissipative force-free formulations \citep{Gruzinov2008, Kalapotharakos2012, Li2012} that allowed for gaps\footnote{Regions of charge density lower than the local Goldreich-Julian density} or dissipative regions to form, giving hints to the distribution of the accelerating electric fields, but these models still struggled to address the exact emission regions, particle acceleration, and pair formation, being dependent on the assumption of the spatial distribution of the (macroscopic) conductivity. From these works and modern pulsar magnetosphere simulation solutions, the common belief is that most pulsar magnetospheres in which sufficient pair production occurs are nearly force-free and the dominant dissipation takes place in the current sheet, beyond the light cylinder.

Kinetic models were therefore developed to address the required kinetic-scale plasma physics, particle dynamics, and radiation reaction from first principles. Particle-in-cell (PIC) models therefore effectively integrate the micro-physics to model relativistic pulsar magnetospheres \citep{Chen2014, Philippov2014, Cerutti2015, Belyaev2015, Philippov2015}. PIC models have also been used to explain the high-energy emission in pulsar observations \citep{Cerutti2016, Kalapotharakos2018, Brambilla2018, Philippov2018, Kalapotharakos2023}. But PIC codes are computationally very demanding, since they solve the full particle gyration with included classical radiation-reaction forces (RRFs). Moreover, a major limiting factor in these simulations is the large-scale separation between the gyro-period compared to the stellar rotation period. PIC models approach this limitation by scaling up the gyro-period to computationally realistic scales by lowering the electromagnetic field values, particle Lorentz factors $\gamma$, and RRF by orders of magnitude. These parameters are many orders of magnitude different to what is realised in actual pulsars, and one cannot recreate the true pulsar environment by such simulations. Simply re-scaling the parameters and forces to higher values after the simulation is questionable, since different considerations come into play when one is in the RRF limit at high $\gamma$ and field strengths \citep{Petri2023}. 

An alternative approach to the pulsar problem is that of \citet[hereafter AH15]{Harding2015} and \citet[hereafter AH21]{Harding2021}, which we will mainly use in this work against which to compare our results. In their approach, they use the force-free (FF) $B$-field and $E$-field grids produced in the FF inside and dissipative outside (FIDO) model of \citet{Kalapotharakos2014} and calculate the gyro-centric\footnote{The centre of the particle's orbit as it gyrates around the $B$-field, assuming circular motion and slow drifting of this point \citep{Burger1985}.} particle trajectory using the following equations of motion presented (compare with Eq.~[\ref{AE}]) in the same work:
\begin{equation} \label{Alice_traj}
    \frac{\mathbf{v}}{c} = \frac{\mathbf{E}\times \mathbf{B}}{B^{2} + E_{0}^{2}} + f\frac{B}{\mathbf{B}}, 
\end{equation} 
where $\mathbf{v}$ is the particle velocity, $\mathbf{B}$ is the $B$-field, $\mathbf{E}$ is the $E$-field, and $f$ is a scaling factor obtained by assuming $v=c$, with $c$ the speed of light in a vacuum. Here $E_{0}$ is one of the field invariants calculated using $E_{0} = \sqrt{\sqrt{\left(P/2\right)^{2} + Q^2} - P/2}$\footnote{This is the $E$-field strength in the frame where $\mathbf{E}$ is parallel to $\mathbf{B}$.}, where $P=B^{2} - E^{2}$, $Q=\mathbf{E}\cdot\mathbf{B}$, and $E_{0}>0$. Since Equation~(\ref{Alice_traj}) assumes $v=c$ to solve for $f$, this equation is only used to obtain the trajectory and direction of the particle; thus, separate transport equations are necessary to calculate the $\gamma$ and momentum of the particle needed for the emission calculations. The transport equations in the observer frame are given by \citet{Harding2015} as, 
\begin{equation} \label{Alice_transport}
\begin{aligned} 
\frac{d\gamma}{dt} &= \frac{eE_{\parallel}}{mc} - \frac{2e^{4}B^{2}p^{2}_{\perp}}{3m^{3}c^{5}} - \frac{2e^{2}\gamma^{4}}{3\rho^{2}_{c}}, \\
\frac{dp_{\perp}}{dt} &= -\frac{3cp_{\perp}}{2r} - \frac{2e^{4}B^{2}p^{3}_{\perp}}{3m^{3}c^{5}\gamma},
\end{aligned}
\end{equation} 
where $e$ is the particle charge, $m$ is the mass of the particle, $r$ is the radial distance, $\rho_{\rm c}$ is the particle trajectory radius of curvature, and $p$ the momentum. In the original \citet{Harding2005} work, these equations were derived in the observer frame and with respect to the local $B$-field. In the modern models of \citetalias{Harding2015, Harding2021} there is a perspective shift where these calculations are done with respect to the particle trajectory. This thus accounts for the $\mathbf{E}\times\mathbf{B}$-drift effect, assuming equivalence between the calculations in the observer frame vs the frame following the $\mathbf{E}\times\mathbf{B}$-drift curve (particle trajectory). Hence, these equations yield the components of the momentum with respect to the particle trajectory. Therefore the particles have a deviation angle (generalised pitch angle) between the particle's velocity and the gyro-centric $\mathbf{E}\times\mathbf{B}$-drift curve particle velocity\footnote{This particle velocity direction is proposed to be described by the AE equations.} as it gyrates around this curve, vs the traditional definition of pitch angle $\theta_{\rm p}$ being the angle between the particle velocity and the local $B$-field. Equations~(\ref{Alice_transport}) were not rigorously re-derived with respect to the particle trajectory, thus our work will additionally validate their approach. Additional assumptions of these equations are super-relativistic particles, small pitch angles, and phase-averaged pitch angles.
% \LD{The parallel and perpendicular momentum components are specified with respect to the local $B$-field. These equations assume super-relativistic particles, small pitch angles, and phase-averaged pitch angles.} In this manuscript, the pitch angle $(\theta_{\rm p})$ is defined as the angle between the particle velocity and the local $B$-field direction. 
In Equation~(\ref{Alice_transport}) we have excluded the resonant photon absorption due to the difficulty of including this effect into the equations of motion. This modelling approach has a massive computational advantage compared with the full particle dynamics calculation, since it does not resolve the particle gyrations. Further details on these models will be discussed in Section~\ref{sec:2} to explain how we compared various aspects of our new model vs this model.

A third approach uses the concept of Aristotelian Electrodynamics (AE), proposing that the particle is quickly accelerated by the parallel $E$-field\footnote{Unless specified, all parallel and perpendicular components are with respect to the local $B$-field.} to a critical $\gamma$ where the Lorentz force and RRF are in equilibrium as the particle follows the principal null direction (i.e., moving at the speed of light $c$). The advantage is that one can make approximations for the super-relativistic particle trajectories and avoid integrating the full equations of motion. \citet{Petri2023} blends the FFE and AE approaches, which allows for avoiding the integration of the equations of motion and instead uses a particle pusher, similar to the gyro-centric approach discussed in the previous paragraph. In this model, one balances the Lorentz force with a radiative force that is linear in velocity, reducing to the AE result in the limit of $v=c$. A similar approach is that of \citet{Yangyang2022}, using the ideas of AE and letting the particles follow the principal null direction and equating the spatial component of the light-like moving particle to the radiation-reaction-limited velocity. The AE trajectories are given by \citet{Gruzinov2012}:
\begin{equation} \label{AE}
\mathbf{v}_{\pm} = \frac{\mathbf{E}\times \mathbf{B} \pm \left( B_{0}\mathbf{B} + E_{0}\mathbf{E} \right)}{B^{2} + E_{0}^{2}},
\end{equation}
where $E_{0}$ is defined as above and $B_{0} = \left(Q/|Q|\right) \sqrt{\sqrt{\left(P/2\right)^{2} + Q^2} + P/2}$, namely the other field invariant. One can then balance the power gained due to the accelerating $E$-field and the power lost due to curvature radiation (CR) to obtain the critical $\gamma$ as:
\begin{equation} \label{gam_CRR}
\gamma_{\rm c} = \left( \frac{3E_{0}\rho_{\rm c}^{2}}{2\vert e\vert }\right)^{1/4}.    
\end{equation}
This assumes that CR dominates the particle losses, but one can also balance the gained power with the synchro-curvature power radiated as given in \citet[hereafter VT15]{Vigano2015} for a more general expression for $\gamma_{\rm c}$. AE is an equilibrium solution following the principal null direction, thus the particle has to be quickly accelerated to $\gamma_{\rm c}$ for this limit to be applicable. Notably, the AE trajectory describes the particle's gyro-centric trajectory, which needs to be taken into consideration when comparing to gyro-phase-resolved trajectories. Additionally, the RRF can exceed the Lorentz force in the observer frame, as discussed by other authors \citep{landau1975, Cerutti2012, Vranic2016, Yangyang2022}, meaning that these equations only apply once the particle enters equilibrium. An advantage of AE is that it traces out the particle trajectory as it heads out to infinity due to the influence of the fields, incorporating the $\mathbf{E}\times \mathbf{B}$-drift, which is important for synchro-curvature radiation (SCR), synchrotron radiation (SR), and CR calculations. The $\mathbf{E}\times \mathbf{B}$-drift effect on the trajectory and radiation needs to be accounted for in pulsar-like sources, since the standard SR calculations are derived in the absence of an $E$-field \citep{Blumenthal1970}, so it is technically not applicable in this case. This drift is especially relevant in pulsar magnetospheres, where $E_{\perp}$ can be a significant fraction of the $B$-field. The assumption is that the deviation between the velocity and the local AE velocity yields the general pitch angle, since the AE trajectory is assumed to be equivalent to the gyro-centric $\mathbf{E}\times\mathbf{B}$ curve of the particle as it flows outward from the stellar surface. Comparing Equation~(\ref{Alice_traj}) to the AE trajectory in Equation~(\ref{AE}), one sees that they are equivalent, where the motion parallel to the $B$-field is encapsulated in the $f$ factor of the second term of Equation~(\ref{Alice_traj}). This second term is equivalent to the second and third term in Equation~(\ref{AE}), namely the motion parallel to the local $B$-field, assuming $E_{0}$ is small.  

In our previous work \citep[hereafter DP24]{DuPlessis2024}, we detailed the implementation of an efficient and accurate solver of the general equations of motion of charged particles with included classical RRF. This was done using a higher-order numerical solver with adaptive time steps to save computational time and obtain higher accuracy. In this work, we will use this to solve the general particle gyro-motion (gyro-resolved). We will compare our results to those of the \citetalias{Harding2015, Harding2021} models for a pulsar scenario using FF $B$-fields with an additional accelerating $E$-field. We will also compare our results with the AE results of \citet{Gruzinov2012} and \citet[hereafter KP15]{Kelner2015}, assessing if and when the particle reaches equilibrium where the Lorentz force and RRF are equal. Our goal in comparing our results with the AE results is to assess under which conditions the particle enters or approaches equilibrium in a short enough timescale, justifying the use of the AE equations. This is also important for models using the AE trajectories to include the $E_{\perp}$-drift effect when calculating the correct particle SCR. Additionally, we will be assessing two SCR calculation methods, namely that of \citet[hereafter CS16]{Cerutti2016} / \citetalias{Kelner2015} vs that of \citetalias{Vigano2015}, to determine if they yield similar and reliable results, and which method to use for our future modelling. The goal in comparing our results to those of the \citetalias{Harding2015, Harding2021} models is to replicate their results for a Vela-like pulsar (with $10\%$ the surface $B$-field strength of Vela), in order to assess the reliability of our emission and spectral calculations. This will also assess the validity of their approach using Equations~(\ref{Alice_transport}) with respect to the particle trajectory to model the parameters required for the SCR. %Thus, the light cylinder scale is of main interest in these simulations. 
We also want to compare the differences in using our gyro-phase-resolved modelling vs their gyro-centric modelling, as well as considering each model's shortcomings. This will furthermore give us insight into where each type of modelling is applicable. All of the mentioned comparisons and assessments will allow us to confidently produce emission maps and spectra for AR Sco (the first white dwarf pulsar)\footnote{The general equations of motion are required for this source due to the magnetic mirroring of the particles suggested for the source \citep{Takata2017}.}, pulsar-like sources, and pulsars in future work. AR Sco will be the first source to be modelled using our code (Du Plessis et al., in prep.), since this source motivated the development of the code. Limitations of our approach will be discussed in the text.         

In Section~\ref{sec:2}, we will discuss the methods used to compare the respective trajectories, emission phase corrections, radiation calculations, emission maps, and spectra. For a full description of how we solve our particle trajectories, see \citetalias{DuPlessis2024}. In Section~\ref{sec:3}, we will present and discuss the various comparisons with the \citetalias{Harding2015, Harding2021} models and AE results. Finally, we will make closing remarks in Section~\ref{sec:4}.     

\section{Method}\label{sec:2}
In this section, we will discuss some details and assumptions that allowed us to compare our results with those produced by the models of \citetalias{Harding2015} and \citetalias{Harding2021}. We will also discuss how we compared our results with the AE results, and also describe the SR, CR, and SCR calculations used to produce our results. For context on how our particle trajectories and the radiation reaction are calculated, see \citetalias{DuPlessis2024}.

\subsection{Retarded dipole (RD) and FF Fields} \label{sec:2.1}
Similar to \citetalias{DuPlessis2024} for the RD fields, we use Equations~$3.11 - 3.13$ from \citet{Dyks2004} for the $B$-field, and the $E_{\perp}$-field from \citet{GJ1969}
\begin{equation}\label{E-Dipole}
\mathbf{E}_{\perp} = -\frac{\mathbf{\Omega} \times \mathbf{R}}{c} \times \mathbf{B}.   
\end{equation}
Above, $\mathbf{\Omega}$ is the angular velocity of the star and $R$ the radial position.

For the FF fields, we reproduce the fields and methods used by \citetalias{Harding2015, Harding2021} for a better comparison of our respective results. The FF fields used are the $B$-field and $E_{\perp}$- field grids generated by the FIDO model of \citet{Kalapotharakos2014}. The field grid spacing is $0.05R_{\rm LC}$, where tri-linear\footnote{One linear interpolation for each spatial dimension.} interpolation is necessary, since the grids are so coarse. Here $R_{\rm LC}$ is the light cylinder radius, where the corotation speed equals the speed of light. Due to the FF grid not starting at the surface of the star, the RD fields specified above are used from $R=(0.0 - 0.2) R_{\rm LC}$. 
To transition from the RD fields to the FF fields, a transition region is implemented using a ramp function. Therefore, from $R=(0.2 - 0.4)R_{\rm LC}$, a linear joining (ramp) function is used to transition from RD to FF. The linear joining function consists of adding the two function values and weighing each component by the normalised distance to its respective boundary. Finally, above $R=0.4R_{\rm LC}$, the FF solutions are used. See \citetalias{Harding2015} for the full description.

An important requirement for kinetic simulations is that the $B$-fields are divergence-free. When assessing $\nabla\cdot\mathbf{B}$, we found the $B$-fields used by \citetalias{Harding2015, Harding2021, Barnard2022} to be significantly divergent at the ramp function region. This is because the joining function is not sufficiently smooth on the scale of resolving the full particle gyration and does not conserve the divergence when bridging the two fields. Therefore, we initialise our equations of motion at $0.4R_{\rm LC}$, after the joining function at the FF field region to avoid the ramp function region, which is discussed in Section~\ref{sec:2.3}. We initialise the particle parameters using the FF-field at $0.4R_{\rm LC}$ instead of initialising at a lower altitude to ensure that our fields are as similar as possible to the \citetalias{Harding2015, Harding2021, Barnard2022} models, since changing the field structure would change the trajectories and emission maps. 

It is not feasible to generate field grids on the particle gyro-radius scale, therefore we defer self-consistent field calculations to a future work. This will additionally prevent the fields from becoming divergent due to oversampling the $B$-field grids, since linear interpolation does not necessarily ensure the fields stay divergence-free \citep{Schlegel2020}. The divergence, $\nabla\cdot\mathbf{B}$, was evaluated in our simulations to ensure the fields are divergence-free over the whole particle trajectory. We would advise the \citetalias{Harding2015, Harding2021, Barnard2022} models to monitor for divergence at the ramp function region. However, due to their models being gyro-centric and using much larger step lengths, their particle trajectories are less susceptible to being scattered by these divergences, although they could still be slightly altered from the true trajectories. We additionally found that making changes to the radial range of the ramp function region caused significant changes in the particle trajectories of our results and those of the \citet[hereafter BH22]{Barnard2022} model results. For a detailed discussion of the divergence and the accompanying results see \citet[hereafter DP25]{DuPlessis2025}.

\subsection{Classical Radiation Reaction Force} \label{sec:2.2}
We have discussed the implementation and results for the classical RRF in our previous methods work (\citetalias{DuPlessis2024}). We include the equations here for easier discussion of the AE comparison results in this work. The classical RRF is given by \citet{landau1975} as:
\begin{equation} \label{RRF}
\begin{aligned}
\mathbf{f} =& ~\frac{2e^{3}\gamma}{3mc^{3}}\left\lbrace\left(\frac{\partial}{\partial t} + \frac{\mathbf{p}}{\gamma m}\cdot\nabla \right)\mathbf{E} + \frac{\mathbf{p}}{\gamma mc}\times\left(\frac{\partial}{\partial t} + \frac{\mathbf{p}}{\gamma m}\cdot\nabla \right)\mathbf{B} \right\rbrace \\
+&\frac{2e^{4}}{3m^{2}c^{4}}\left\lbrace \mathbf{E}\times\mathbf{B} +\frac{1}{\gamma mc}\mathbf{B}\times\left(\mathbf{B}\times\mathbf{p} \right) +\frac{1}{\gamma mc}\mathbf{E}\left(\mathbf{p}\cdot\mathbf{E} \right) \right\rbrace \\
-&\frac{2e^{4}\gamma}{3m^{3}c^{5}}\mathbf{p}\left\lbrace \left(\mathbf{E} + \frac{\mathbf{p}}{\gamma mc}\times\mathbf{B} \right)^{2} - \frac{1}{\gamma^{2}m^{2}c^2}\left(\mathbf{E}\cdot\mathbf{p} \right)^2 \right\rbrace. 
\end{aligned}  
\end{equation}
We have neglected the first term, since the temporal and spatial changes in the fields have a negligible contribution, similar to what has been done by other authors \citep{Cerutti2013, Vranic2016, Tamburini2010}. Even though the third term (the radiative losses, e.g. SCR) dominates over the second term by a factor of $\gamma^{2}$, the second term is included since it is important for obtaining the correct curvature cooling time \citep[hereafter CS16]{Cerutti2016}. As discussed in our methods work (\citetalias{DuPlessis2024}), we monitor that the $E$-field experienced by the particle $E_{\rm f}$ is below the Schwinger limit $E_{\rm S}$ to ensure that the RRF remains in the classical regime. This is given by \citet{Sokolov2010} as:
\begin{equation} \label{field_exp}
   E_{\rm f} = \frac{\vert \mathbf{p}\times \mathbf{B} \vert}{mc},  
\end{equation}
and $E_S = m_{\rm e} c^2 /(\vert e\vert \lambdabar_c)$, with $\lambdabar_c$ the reduced Compton wavelength and $m_{\rm e}$ the electron mass. A similar approach is followed by \citet{Cerutti2012}. Several PIC models in the literature revert to setting a limit on $\gamma$ using the $\gamma_{\rm max}$ derived from the polar cap (PC) potential drop \citep{Cerutti2012, Brambilla2018, Cruz2023}. 

In our investigation we found that when using the \citetalias{Barnard2022} model parameters and initialising our equations of motion after the ramp function at $0.4R_{\rm LC}$, $E_{\rm f}$ exceeded $E_{\rm S}$ when using $B_{\rm S} = 8\times10^{11} \, \rm{G}$ and $B_{\rm S} = 8\times10^{12} \, \rm{G}$. This is caused by initialising our equations close to or beyond the Schwinger limit, since if one initialises close to the Schwinger limit the particle will very quickly be accelerated to beyond $E_{\rm S}$ by the large $E_{\perp}$ and $E_{\parallel}$ fields. When $E_{\rm f} > E_{\rm S}$ the RRF enters the non-classical regime and the simulations encounter a numerical runaway solution. However, if the particle is initialised with a lower $E_{\rm f}$ the particle can settle into equilibrium of the Lorentz force and RRF without exceeding $E_{\rm S}$. Ideally one would like to initialise the particle with realistic parameters at the surface and let it naturally relax into equilibrium. This runaway solution is due to the RRF significantly dominating in the Quantum regime, leading to the particles rapidly radiating more energy than they have, leading to a loss of energy conservation. This is discussed and assessed for our particle dynamics implementations in \citetalias{DuPlessis2024} and more generally discussed in \citet{landau1975, Vranic2016}. In future work, we will use Quantum Electrodynamics (QED) considerations to calculate the RRF in non-classical regimes similar to \citet{Grismayer2016, Vranic2017}, since this would address this issue elegantly. When decreasing the Lorentz factor to $\gamma = 4$ and initialising at $0.65R_{\rm LC}$ we found we could simulate the trajectories with $E_{\rm f} < E_{\rm S}$ for the $B_{\rm S} = 8\times10^{11} \, \rm{G}$ case. Interestingly, when limiting the Lorentz factor with $\gamma<\gamma_{\rm c}$ we found that we could initialise at $0.4R_{\rm LC}$ using the \citetalias{Barnard2022} model $\gamma$, while $E_{\rm f} < E_{\rm S}$. Importantly, in Section~\ref{sec:3.3} we show that limiting $\gamma$ during the simulations leads to numerical problems in the results. Unfortunately, we were not able to keep $E_{\rm f} < E_{\rm S}$ for the $B_{\rm S} = 8\times10^{12} \, \rm{G}$ case, even when reducing $\gamma$ and increasing the starting altitude\footnote{For most of the field lines in this case, the particles could be initialised at $~0.9R{\rm LC}$, but then we lose the whole internal magnetosphere from our simulations as well as the time required for the particle to settle into equilibrium.}. In this work, we will thus investigate the $B_{\rm S} = 8\times10^{11} \, \rm{G}$ case for the comparison of the results to avoid the particles entering the non-classical RRF regime. For a detailed discussion of these results evaluating $E_{\rm f}$, see \citetalias{DuPlessis2025}.

\subsection{Particle Trajectory Calibration} \label{sec:2.3}
For the FF-fields scenario, \citetalias{Harding2015, Harding2021} start by calculating the position of the PC rim and subsequently calculate the active region (where the production of pairs near the neutron star occurs and primaries are injected) for the slot gap.
%\NT{This is due to the conditions $J/J_{\rm GJ} >1$ or $J/J_{\rm GJ}<0$, where $J$ is the current density and $J_{\rm GJ}$ is the Gouldreich-Julian current density.}
These calculations are discussed in detail in \citet{Dyks2004b}, who defined the open volume coordinates $r_{\rm ovc}$ and $l_{\rm ovc}$ (polar and azimuthal coordinates relative to the polar cap angle and ring length) to calculate the non-circular active regions. They thus define a region interior (closer to the magnetic pole) to the separatrix (between open and closed field lines) between $r_{\rm ovc}^{\rm min}$ and $r_{\rm ovc}^{\rm max}$ in which the particles are injected (these coordinates are defined on the stellar surface). This then gives the starting surface position (footpoints) of the $B$-fields to be probed in the calculations. We take these starting positions as output from their model and use them as initial positions for solving our equations of motion to avoid any discrepancy. 

In Sections~\ref{sec:2.1} and~\ref{sec:2.2} we explained why we have to initialise our model's equations of motion at a higher altitude in the magnetosphere ($\sim 0.4R_{\rm LC}$). To solve the problem of having to start implementing our equations of motion at a higher altitude, we used Equations~(\ref{Alice_traj}) to trace out the particle trajectories up to a specified altitude ($\sim 0.4R_{\rm LC}$, after the ramp function where the fields are FF), which gives the particle position and direction needed for initialising our equations. Using the particle $\gamma$ from Equation~(\ref{Alice_transport}) at the specified altitude from the output results of \citetalias{Barnard2022}, we then had the missing momentum needed for solving our equations of motion, which we used to simulate the rest of the particle trajectory.
%We thus ensured that our initial position and particle direction for the same field line aligned with the parameter values of \citetalias{Barnard2022}. 
Starting the particle at a higher altitude also saves an enormous amount of computational time, since close to the surface for large $B_{\rm S}$ values the gyro radius is very small, requiring unfeasibly small time steps to resolve the particles gyrations. This is especially problematic for the $B_{\rm S} = 8\times10^{11} \, \rm{G}$ and $B_{\rm S} = 8\times10^{12} \, \rm{G}$ cases, but using our numerical solver we are able to inject particles at the surface for $B_{\rm S} = 8\times10^{8}  \, \rm{G}$ and $\gamma = 1$ with a reasonable computational time\footnote{Using $B_{\rm S} = 8\times10^{8}  \, \rm{G}$ and higher fields were numerically feasible if we initialised particles with a higher $\gamma$ value or invoked sufficient $E_{\parallel}$-fields to accelerate the particles.}. This is an improvement on current pulsar PIC models using $B_{\rm S} = (10^{6} -10^{8})  \, \rm{G}$ and starting at higher altitudes.

To calculate the particle trajectories \citetalias{Harding2015, Harding2021} use Equations~(\ref{Alice_traj}) and~(\ref{Alice_transport}) to calculate $\gamma$ and $p_{\perp}$, as discussed in Section~\ref{sec:1}. It is important to note for comparison that these equations are decoupled, meaning any radiation self-force or feedback does not affect the particle trajectories. Equations~(\ref{Alice_traj}) and~(\ref{Alice_transport}) are also gyro-centric and assume super-relativistic particles, while Equation~(\ref{Alice_transport}) assumes small, gyro-averaged particle $\theta_{\rm p}$. 
After calculating the trajectories with our code and the \citetalias{Harding2015} code, we can then compare both the position and particle-direction components to see if there are any deviations. We compare our results to theirs for a Vela pulsar scenario, assuming the CR scenario, since this was thoroughly explored by \citetalias{Barnard2022} using the \citetalias{Harding2015} code. Thus, using the \citetalias{Barnard2022} code we use a spin period of $P_{\rm S} = 8.933\times 10^{-2} \, \rm{s}$, magnetic inclination angle $\alpha = 75^{\circ}$, the distance to the source $d = 0.29 \, \rm{kpc}$, and surface $B$-field of $B_{\rm S} = 8\times10^{12} \, \rm{G}$, where $B_{\rm S}$ is used to scale the FF-fields. In their simulations \citetalias{Harding2015, Harding2021} add constant $E_{\parallel}$-fields (parallel to the local $B$-field) to accelerate the particles to high enough $\gamma$ values to produce CR. We emulate their setup by defining $E_{\parallel}$ in terms of the acceleration per length scale $(d\gamma/dl)$ $R_{\rm acc}=eE_{\parallel}/mc^{2}$,  using $R_{\rm acc}^{\rm min}=4.0\times 10^{-2} \,\rm{cm}^{-1}$ inside the light cylinder and $R_{\rm acc}^{\rm max}=2.5\times 10^{-1} \, \rm{cm}^{-1}$ outside the light cylinder. The constant $E_{\parallel}$-field is used to approximate the accelerating fields present in the slot gap regions where the local conditions deviate from FF for an accelerating $E_{\parallel}$-field to exist. For a more detailed discussion on the origin of these fields in the context of a dissipative PIC model, see \citet{Kalapotharakos2018}. It is important to note that \citetalias{Harding2015, Harding2021} use these $E_{\parallel}$-fields in Equation~(\ref{Alice_transport}) to accelerate the particles to high enough $\gamma$ values for the radiation calculations, but these $E_{\parallel}$-fields are not taken into account for their trajectory calculations in Equation~(\ref{Alice_traj}). To get similar particle acceleration (high enough $\gamma$) we thus need to include the $E_{\parallel}$-fields into our equations of motion, thus this means that if $E_{\parallel}$ is large, it affects our particle trajectories but does not affect those of the \citetalias{Harding2015, Harding2021} models.

Equations~(\ref{Alice_transport}) account for the energy losses and gains in the first equation and similarly for the perpendicular momentum in the second. As the model approaches equilibrium, where the acceleration and the losses balance, which term dominates at this point starts to oscillate. To approximate the $\gamma$ value in this state, the \citetalias{Harding2015, Barnard2022, Harding2021} models use the equilibrium $\gamma_{\rm c}$ value. The SR scenario is more complicated due to the $p_{\perp}$ that is influenced by the absorption of radio photons, see \citetalias{Harding2015} for more details. We are mainly interested in Equation~(\ref{gam_CRR}) to compare that $\gamma$-value to our model's $\gamma$-value due to the AE assumption of a CR-dominated scenario for pulsars. To calculate the SR-reaction-limited $\gamma$ as an additional value to compare our $\gamma$ to, we use
\begin{equation} \label{gam_SRR}
\gamma_{\rm SRR} = \left(\frac{6\pi eE_{0}}{\sigma_{T}B^{2}}\right)^{1/2},   
\end{equation}
where $\sigma_{\rm T}$ is the Thomson cross-section and $E_{0}$ is the same quantity as defined in Equation~(\ref{Alice_traj}). 
% The approximation in Equation~(\ref{gam_CRR}) is mainly investigated in the comparison scenarios below, since CR dominates in the chosen scenarios.

In our previous work (\citetalias{DuPlessis2024}), we show and discuss how we compare our results to the AE results given by Equation~(\ref{AE}) for the uniform-field cases. In this work, we will compare our results to the AE results but for non-uniform RD and FF-fields. We thus define a deviation angle between our model velocity and the local AE velocity given by Equation~(\ref{AE}) as $\theta_{\rm VA}$ to assess if our results converge to the AE results, as well as to directly compare the velocity components. 
% Additionally, we also compare our results to the AE results from \citetalias{Kelner2015}, where the respective velocity directions are given by:
% \begin{equation} \label{AE_Kelner}
%     v_{\rm k} =\frac{1}{Q}\mathbf{E}\times\mathbf{B} \pm \left(\frac{\sqrt{Q-B^{2}}}{Q}\mathbf{E} + \frac{\sqrt{Q-E^{2}}}{Q}\mathbf{B}  \right),
% \end{equation}
% where $Q = (E^{2} + B^{2})/2 + \sqrt{ \left( E^{2} - B^{2}\right)^{2}/4 +(\mathbf{E}\cdot\mathbf{B})^{2}}$. Similar to what was done before, we also calculate a deviation angle using Equation~(\ref{AE_Kelner}). 
Additionally, \citetalias{Kelner2015} gives a theoretical equation for the particle pitch angle while following this AE trajectory 
\begin{equation} \label{Kel_pitch}
    \sin\theta_{\rm p} = \frac{\sigma\sin\theta_{\rm EB}}{\sqrt{0.5\left(\sqrt{(1+\sigma^{2})^{2} -4\sigma^{2}\sin^{2}\theta_{\rm EB}} + (1+\sigma^{2})\right)}},
\end{equation}
where $\sigma = E/B$ is the ratio of the $E$-field to the $B$-field, and $\theta_{\rm EB}$ is the angle between these fields.
Additionally, we also calculate the different $E$-field and $B$-field drift effects to assess what impact they have on the particle $\theta_{\rm p}$ alongside the RRF. We are interested in what phenomena affect the $\theta_{\rm p}$, since in \citetalias{DuPlessis2024} we found that for the uniform field cases, the RRF did not significantly affect the particle $\theta_{\rm p}$ until the $\gamma$ was quite low. This is because the dominant term in the RRF that scales with $\gamma^{2}$ is directly opposite to the particle velocity, only scaling the velocity vector and not changing the velocity direction. Thus in the super-relativistic case, the particle $\theta_{\rm p}$ is not decreased by the RRF. This is also discussed by \citet{landau1975,Harding2006}. Close to the neutron star surface $\theta_{\rm p}$ is decreased by RRFs via quantum SR due to the very high field values \citep{Harding2006}. The gradient, curvature, and $\mathbf{E}\times \mathbf{B}$-drift velocities can be found in \citet{Chen1984}, where we used the drift velocity due to a general force $v_{\rm f} = \mathbf{F}\times\mathbf{B}/(qB^{2})$ to obtain the relevant forces for each of these drift components (we will assess these forces as a function of distance in Figure~\ref{FF_AE_rho_c_8e11}). 
%These drift velocities are given in \citet{Chen1984} where 
The $\mathbf{E}\times \mathbf{B}$-drift velocity is calculated by
\begin{equation}
\mathbf{v}_{\rm EB} = \frac{c\mathbf{E}\times \mathbf{B}}{B^{2}}, 
\end{equation}
the $\nabla \mathbf{B}$-drift velocity is calculated by 
\begin{equation}
\mathbf{v}_{\rm grad} = \frac{r_{\rm L}v_{\perp}}{2}\frac{\mathbf{B}\times\nabla\mathbf{B}}{B^{2}},    
\end{equation}
where $r_{\rm L} = \gamma mc v_{\perp}/eB$ is the Larmor radius, and the curvature drift velocity is calculated by
\begin{equation}
\mathbf{v}_{\rm curv} =  \frac{mcv^{2}_{\parallel}}{eB^{2}}\frac{\bm{\rho}_{\rm c}\times\mathbf{B}}{\rho^{2}_{\rm c}}.  
\end{equation}
All the drift velocities above are given in cgs units.

\subsection{CR and SR Calculations} \label{sec:2.5}
To calculate the radiation emitted along the particle's trajectory, we initially look at the two limit cases, namely CR and SR. For CR, the particle is assumed to have zero $\theta_{\rm p}$ as the particle follows the curved $B$-field line. To calculate the emitted CR spectrum and reproduce the CR results of \citetalias{Barnard2022}, we implement the standard calculations for CR and SR as discussed in \citetalias{Harding2015}. Thus, the approximate CR spectrum emitted by a single particle is given by
\begin{equation} \label{CR_spec}
    \frac{\dot{N}_{\rm CR}(\epsilon)}{d\epsilon} = \frac{\alpha_{\rm f}}{\left(\lambdabar mc\right)^{1/3}}\left( \frac{c}{\rho_{\rm c}}\right)^{2/3}\epsilon^{-2/3}\exp{\left( \frac{\epsilon}{\epsilon_{\rm CR}}\right)},
\end{equation}
where $\epsilon$ is the emitted photon energy normalised to $m_{\rm e}c^{2}$, $\epsilon_{\rm CR}= 3c\gamma^{3}/(2\rho_{\rm c})$ is the CR cut-off energy, $\rho_{\rm c}$ is the particle's radius of curvature, and $\alpha_{\rm f}$ is the fine structure constant. To determine the radius of curvature of the particle trajectory in the inertial observer frame, one uses the inverse of the second derivative of the position along the path $\mathbf{l}$
\begin{equation} \label{AL_rho_c}
    \rho_{\rm c} = \Big| \frac{d^{2}\mathbf{x}}{d\mathbf{l}^{2}}\Big|^{-1},
\end{equation}
where $\mathbf{x}$ is the position vector. The total power radiated via CR by a single particle is given as:
\begin{equation} \label{pow_CR}
    P_{\rm CR} = \frac{2e^{2}c\gamma^{4}}{3\rho_{\rm c}^{2}}.
\end{equation}

For SR, the particle is assumed to have a non-zero $\theta_{\rm p}$ with respect to the local $B$-field. To calculate the approximate single-particle SR spectrum, we use the equation given in \citetalias{Harding2015},
\begin{equation} \label{SR_spec}
    \dot{N}_{\rm SR}(\epsilon) = \frac{2^{2/3}}{\Gamma\left(\frac{1}{3}\right)}\alpha_{\rm f} B'\sin{\theta_{\rm p}}\epsilon^{-2/3}\epsilon_{\rm SR}^{-1/3}\exp{\left(-\frac{\epsilon}{\epsilon_{\rm SR}}\right)},
\end{equation}
where $B'= B/B_{\rm cr}$, $B_{\rm cr} = 4.4\times10^{13} \, \rm{G}$ is the critical $B$-field strength, $\Gamma$ is the Gamma function, and $\epsilon_{\rm SR}= (3/2)\gamma^{2}B'\sin\theta_{\rm p}$ the critical synchrotron frequency. Both Equations~(\ref{CR_spec}) and~(\ref{SR_spec}) use the asymptotic limits for the integral of the modified Bessel function of the second kind $K_{n}$
\begin{equation} \label{Besel}
\kappa(x) = x\int^{\infty}_{x} K_{5/3}(x')dx'    
\end{equation}
to approximate the spectra. \citetalias{Harding2015, Harding2021} use the coupled transport equations in Equation~(\ref{Alice_transport}) for their radiation calculations. Thus, to obtain $\theta_{\rm p}$ for their radiation calculations, they use $\sin\theta_{\rm p} = p_{\perp}/p$ and $p^{2} = \gamma^{2} - 1$. \footnote{The resonant radio photon absorption to boost the particle $\theta_{\rm p}$ for the SR was not included in our modelling, in contrast to what was done in \citetalias{Harding2015, Harding2021}. We would have to calculate the equivalent force to add to the Lorentz force to achieve this effect.} The total power radiated via SR is given by
\begin{equation} \label{pow_SR}
    P_{\rm SR} = \frac{2e^{4}(\gamma^{2}-1)B^{2}\sin^{2}\theta_{\rm p}}{3m^{2}c^{3}}.
\end{equation}

We are mainly interested in comparing the primary electrons' CR and SR, contrary to the \citetalias{Harding2021} model that includes inverse-Compton radiation, synchrotron-self-Compton radiation, and also models radiation from secondary pairs that form in the particle cascades. In this updated model they also use SCR, which we will discuss in the next subsection.  

\subsection{SCR Calculations} \label{sec:2.6}
The need for calculating SCR is firstly to have a spectrum calculation that transitions between the limiting cases of CR and SR. Secondly, the standard SR expressions are derived in the absence of an $E$-field \citep{Blumenthal1970}, which is problematic for pulsar environments due to the large $E_{\perp}$-field component that is present. We have thus decided to investigate two approaches to model the SCR spectrum and will assess their applicability and convergence to the CR and SR cases throughout this work.

The first approach is that of \citetalias{Vigano2015}, which is a more compact version of the calculations done by \citet{Cheng1996} and is implemented in the \citetalias{Harding2021} model. Importantly, the \citet{Cheng1996} model assumes an artificial, static, circular $B$-field structure for their derivations to approximate dipole-like fields for astrophysical applications. \citetalias{Vigano2015} give the critical SRC energy as $\epsilon_{\rm SCR} = 3\hbar cQ_{2}\gamma^{3}/2$, with $\hbar$ the reduced Planck constant, and
\begin{equation}
    Q_{2}^{2} = \frac{\cos^{4}\theta_{\rm p}}{\rho_{\rm c}^{2}}\left( 1 + 3.0\zeta + \zeta^{2} \frac{R_{\rm g}}{\rho_{\rm c}}\right), 
\end{equation}
where $R_{\rm g}$ is the relativistic particle gyro radius and the synchro-curvature parameter is defined as
\begin{equation}
    \zeta = \frac{R_{\rm g}\sin^{2}\theta_{\rm p}}{\rho_{\rm c}\cos^{2}\theta_{\rm p}}.
\end{equation}
To calculate the SCR spectrum, one implements
\begin{equation} \label{Vig_SCR_spec}
    \frac{dP_{\rm SCR}(\epsilon)}{d\epsilon} = \frac{\sqrt{3}e^{2}\gamma y}{4\pi\hbar\rho_{\rm eff}}\left[ (1+z)\kappa(y) - (1-z)K_{2/3}(y) \right],
\end{equation}
where $y=\epsilon/\epsilon_{\rm SCR}$, $z = \left(Q_{2}\rho_{\rm eff}\right)$, and
\begin{equation} \label{Vig_rho_eff}
 \rho_{\rm eff} = \frac{\rho_{\rm c}}{\cos^{2}\theta_{\rm p}}\left( 1 + \zeta + \frac{R_{\rm g}}{\rho_{\rm c}}\right)^{-1}.   
\end{equation}
To calculate the total power radiated by the particle via SCR, one uses
\begin{equation} \label{SCR_pow_Vig}
    P_{\rm SCR} = \frac{2e^{2}\gamma^{4}c}{3\rho_{\rm c}^{2}}g_{\rm r}, 
\end{equation}
where the synchro-curvature correction factor is given as
\begin{equation}
    g_{\rm r} = \frac{\rho_{\rm c}^{2}}{\rho_{\rm eff}^{2}}\frac{\left( 1+7z\right)}{8\left( Q_{2}\rho_{\rm eff}\right)^{-1}}.
\end{equation}
Equation~(\ref{SCR_pow_Vig}) thus scales between Equations~(\ref{pow_CR}) and (\ref{pow_SR}).
This approach does not include an $E_{\perp}$-component in its derivations and assumes that $R_{\rm g}\ll \rho_{\rm c}$, where the latter holds in the outer magnetosphere. To apply these equations where an $E_{\perp}$-field is present, one would have to follow the particle as it gyrates around the AE trajectory, namely the considerations taken by \citet{Kelner2015, Cerutti2016, Kalapotharakos2019}. This can theoretically be done for the \citetalias{Vigano2015} model by using the parameters in the AE frame ($\mathbf{E}\parallel\mathbf{B}$-frame). This does however assume equivalence between the particle gyrating around this circular B-field and the particle gyrating around the curved trajectory due to the $\mathbf{E}\times\mathbf{B}$-drift. We will show in the results section that this is effectively what is happening in the \citetalias{Harding2021} model, namely using $\rho_{\rm c}$ and $\theta_{\rm VA}$, which we will study in this work (though using a general pitch angle and the AE parameters is not discussed by \citetalias{Vigano2015}). 

Importantly, the \citetalias{Harding2015, Harding2021} model's $\rho_{\rm c}$ is that of the particle gyro-centric trajectory, since it is obtained from the FFE trajectory in Equation~(\ref{Alice_traj})\footnote{Due to calculating the $\rho_{\rm c}$ with the AE directions they are calculating the radius of curvature of the AE trajectory curve.}; conversely, our model takes the particle gyrations into account, thus our $\rho_{\rm c}$ will oscillate, but the trend will follow that of the gyro-centric trajectory. %Another factor to take into consideration for the result comparison is that in the
Moreover, the \citetalias{Barnard2022} and \citetalias{Harding2021} $\rho_{\rm c}$ is artificially smoothed to remove any oscillation in $\rho_{\rm c}$. Thus in the \citetalias{Harding2021} model the assumptions are that they are following the AE trajectory and modelling the AE parameters that are used in their \citetalias{Vigano2015} SCR calculations. 
Notably, in \citetalias{Harding2021} the CR and SR losses are replaced by the SCR losses from \citetalias{Vigano2015} but only for $d\gamma/dt$, not $dp_{\perp}/dt$. Our model comparison will assess the validity of this approach, since our modelling is from first principles. 

The second approach is that of \citetalias{Cerutti2016} implementing a simplified SCR spectral formula from \citetalias{Kelner2015}. This modelling takes the approach of AE, namely calculating the equations of motion by balancing the Lorentz force and the leading term in the RRF. This thus calculates the SCR along the particle trajectory curve governed both by the $B$-field and $E$-field. The model was derived by \citetalias{Kelner2015} in terms of a $\rho_{\rm c}$ calculated from this trajectory that scales between the SR and CR scenarios, but was implemented and rewritten in terms of $\tilde{B}_{\perp}$ by \citetalias{Cerutti2016} for convenience and simplicity. This model and the AE approach are also implemented by \citet{Kalapotharakos2019}. To calculate the particle spectrum, \citetalias{Cerutti2016} uses
\begin{equation} \label{Cer_SCR_spec}
    \frac{dP_{\rm SCR}}{d\nu} = \frac{\sqrt{3}e^{3}\tilde{B}_{\perp}}{mc^{2}}\kappa(y),  
\end{equation}
with $y = \nu/\nu_{\rm SCR}$ and the critical frequency $\nu_{\rm SCR} = 3e\gamma^{2}\tilde{B}_{\perp}/4\pi mc$. Upon comparing Equations~(\ref{Vig_SCR_spec}) and~(\ref{Cer_SCR_spec}), one sees that the latter only contains one $\kappa$-term vs two in the prior. This second $\kappa$-term also appears in the quantum electrodynamics (QED) asymptotic formula in \citet{Harding1987}, suggesting that Equation~(\ref{Vig_SCR_spec}) could be more accurate in the very-high energy and high $B$-field regimes. It is not possible to test this while solving the general equations of motion as we do due to the impractically small time steps required to resolve the particle gyrations and the RRF from \citet{landau1975} that was derived in the classical limit. One would therefore need to incorporate QED RRF solutions into the equations of motion to probe this regime. Equation~(\ref{Cer_SCR_spec}) is the same as the standard SR spectrum given in \citet{Blumenthal1970}, but using $\tilde{B}_{\perp}$ and thus being more general, since it is valid for SCR. This augmented $B$-field parameter is given by \citetalias{Cerutti2016} as
\begin{equation} \label{B_tilde}
    \tilde{B}_{\perp} = \sqrt{\left( \mathbf{E+\beta\times\mathbf{B}}\right)^{2} - \left( \beta\cdot\mathbf{E}\right)^{2}}.
\end{equation}
The parameter $\tilde{B}_{\perp}$ in the SR regime is the relativistically invariant $B$-field perpendicular to the particle motion and in the CR case scales with the local $\rho_{\rm c}$ of the particle \citep{Cerutti2016}. Equation~(\ref{B_tilde}) is obtained from the leading term in Equation~(\ref{RRF}). To calculate the effective $\rho_{\rm c}$ of the trajectory, one can use the equation from \citetalias{Kelner2015}
\begin{equation} \label{Cer_rho_eff}
    \rho_{\rm eff} = \frac{\gamma mc^{2}}{e\tilde{B}_{\perp}}.
\end{equation}
We also calculate a $\theta_{\rm eff}$ assuming $\tilde{B}_{\perp} = B\sin\theta_{\rm eff}$ to see if this value is close to $\theta_{\rm VA}$. We expect these values to be closest in the SR-dominated cases.

\subsection{Emission Maps and Spectra} \label{sec:2.7}
\begin{figure}
\centering
\includegraphics[width=.5\textwidth]{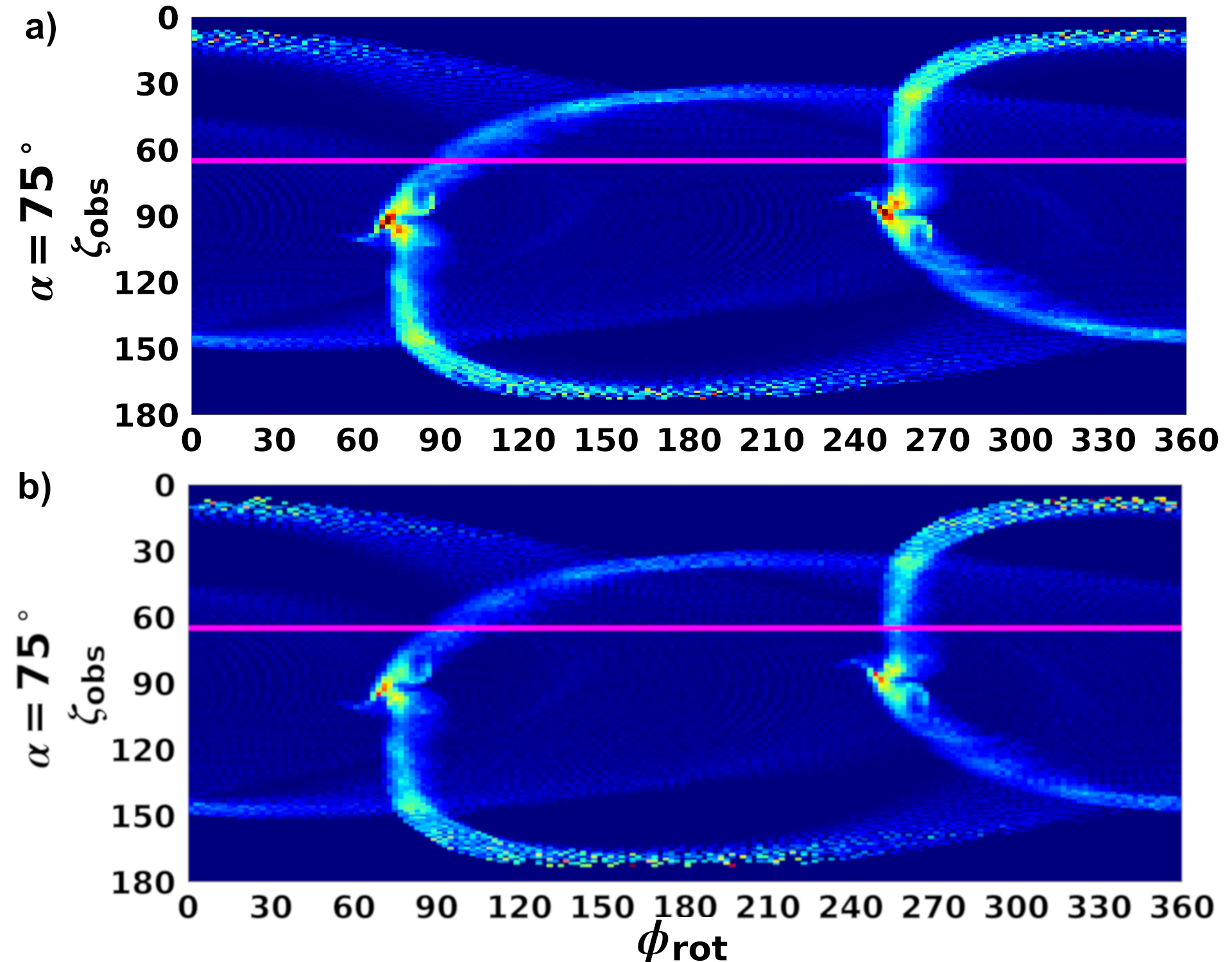}
\caption{Emission maps generated using the code from \citetalias{Harding2015, Barnard2022}, and using the parameters discussed in Section~\ref{sec:2.3}, a photon energy range of $100~\rm{MeV} - 50~\rm{GeV}$, and using the unscreened $r_{\rm ovc}$ region between $r_{\rm ovc} = 0.9$ and $r_{\rm ovc} = 0.96$. Panel a) shows the CR emission map using $B_{\rm S} = 8\times10^{12} \, \rm{G}$ and panel b) shows the CR emission map using $B_{\rm S} = 8\times10^{11} \, \rm{G}$.}
\label{Al_B12_skymap}
\end{figure}

To produce our emission maps and spectra, we use the same techniques as \citetalias{Harding2015}, where our specific implementation can be found in \citetalias{DuPlessis2025}. 
In the \citetalias{Harding2015, Harding2021, Barnard2022} models, only the northern hemisphere is calculated to save computational time. Thus the emission from the northern hemisphere is flipped and shifted by $180^{\circ}$ for the southern hemisphere's emission, assuming that the emission is mirrored but equal from the two magnetic poles. They found that calculating the southern hemisphere's emission yielded identical results to flipping and shifting the northern hemisphere's emission, thereby justifying this approach.   
In Figure~\ref{Al_B12_skymap}, we show the emission maps produced by the same code that was used to produce the results of \citetalias{Harding2015} and \citetalias{Barnard2022}, using the parameters for Vela as discussed in Section~\ref{sec:2.3}. We used $B_{\rm S} = 8\times10^{12} \, \rm{G}$, $R_{\rm acc}^{\rm min}=4.0\times 10^{-2} \,\rm{cm}^{-1}$, and $R_{\rm acc}^{\rm max}=2.5\times 10^{-1} \, \rm{cm}^{-1}$ to produce panel a) and $B_{\rm S} = 8\times10^{11} \, \rm{G}$ with the same $R_{\rm acc}^{\rm min}$ and $R_{\rm acc}^{\rm max}$ to produce panel b).  
%The resolution of $\zeta_{\rm obs}$ and $\phi_{\rm rot}$ are lower than in \citetalias{Barnard2022} as well as using rougher divisions of $r_{\rm ovc}$. This is done to save computational time, since we are interested in comparing these results with our own, not reproducing high-resolution results. 
Additionally, to produce Figure~\ref{Al_B12_skymap}, we  used $0.5$ divisions per degree in $\zeta_{\rm obs}$ and $\phi_{\rm rot}$, $\delta r_{\rm ovc}=0.01$, $r_{\rm ovc}$ ranging from $0.9 - 0.96$, and four divisions per decade in our photon energies. The photon energy range 
%for Figures~\ref{Al_B12_skymap} and ~\ref{Al_B11_skymap} 
was $100~\rm{MeV} - 50~\rm{GeV}$, as was used by \citetalias{Barnard2022}. 
%\citetalias{Barnard2022} did not produce SR emission maps thus for clarity we use the same code they used to produce the SR emission maps in panels~c) for Figures~\ref{Al_B12_skymap} and~ \ref{Al_B11_skymap}. The reason the SR emission map looks so different to the CR is because the \citetalias{Harding2015, Harding2021, Barnard2022} models have a small initial $p_{\perp}$ and is decreased drastically by the SR meaning the SR emission is only located close to the stellar surface. Decreasing the photon energy range for the SR emission map had no effect beyond the emission region around the surface becoming slightly broader.    

\section{Results}\label{sec:3}
In this Section, we show our results when attempting to reproduce the CR scenario presented for Vela in \citetalias{Barnard2022}. We will start by (i)~comparing our particle trajectories (position and direction) and photon emission phase corrections using FF-fields.
%, additionally showing how these results converge to the AE trajectories. 
%This will show us 
We will next (ii)~assess whether these results indeed converge to those of AE, which will allow us to investigate the applicability and assumptions of using the AE equations. We will then move on to the SCR results (iii)~to compare the \citetalias{Vigano2015} calculation method to those of \citetalias{Cerutti2016, Kelner2015} for both the CR and SR cases. This will allow us to investigate if these two methods yield similar results and which method is more applicable for our use case. Lastly, we will (iv)~show our emission maps for a constant emissivity, CR, SR, and SCR emissivity, and end the Section by comparing our spectra to those produced by the \citetalias{Harding2015, Barnard2022} models. 

In all of these results, our model results are labelled in the legend as DPM and the \citetalias{Barnard2022} model results as BHM. In the appendix all our model results have RRF included. We have also normalised the radial positions to $R_{\rm LC}$. As clarification, we mainly compare our results to the model results of \citetalias{Barnard2022}, which is essentially the same as those of the \citetalias{Harding2015} model but includes a smoothed $\rho_{\rm c}$ and CR, where the \citetalias{Harding2021} model includes the smoothed $\rho_{\rm c}$ and
%is the same as the \citetalias{Barnard2022} model but 
the SCR calculations from \citetalias{Vigano2015}. The SR results generated with the \citetalias{Barnard2022} model are the same calculations discussed in Section~\ref{sec:2.5} for the \citetalias{Harding2015} model without the resonant photon absorption. The figure labels are emphasised below. \\ 

\begin{center}
\begin{tabular}{|c|c|} 
\hline 
    BHM & Results using \citetalias{Barnard2022} model.  \\  
    \hline
    DPM & Results using our model.  \\ 
    \hline
    DPM-RRF & Results using our model with RRF included. \\ 
    \hline 
    AE & AE results with RRF excluded. \\
    \hline
    AE-RRF & AE results with RRF included. \\
\hline
\end{tabular}
\end{center}

\subsection{FF Trajectories} \label{sec:3.2}
\begin{figure}
\centering
\includegraphics[width=.5\textwidth]{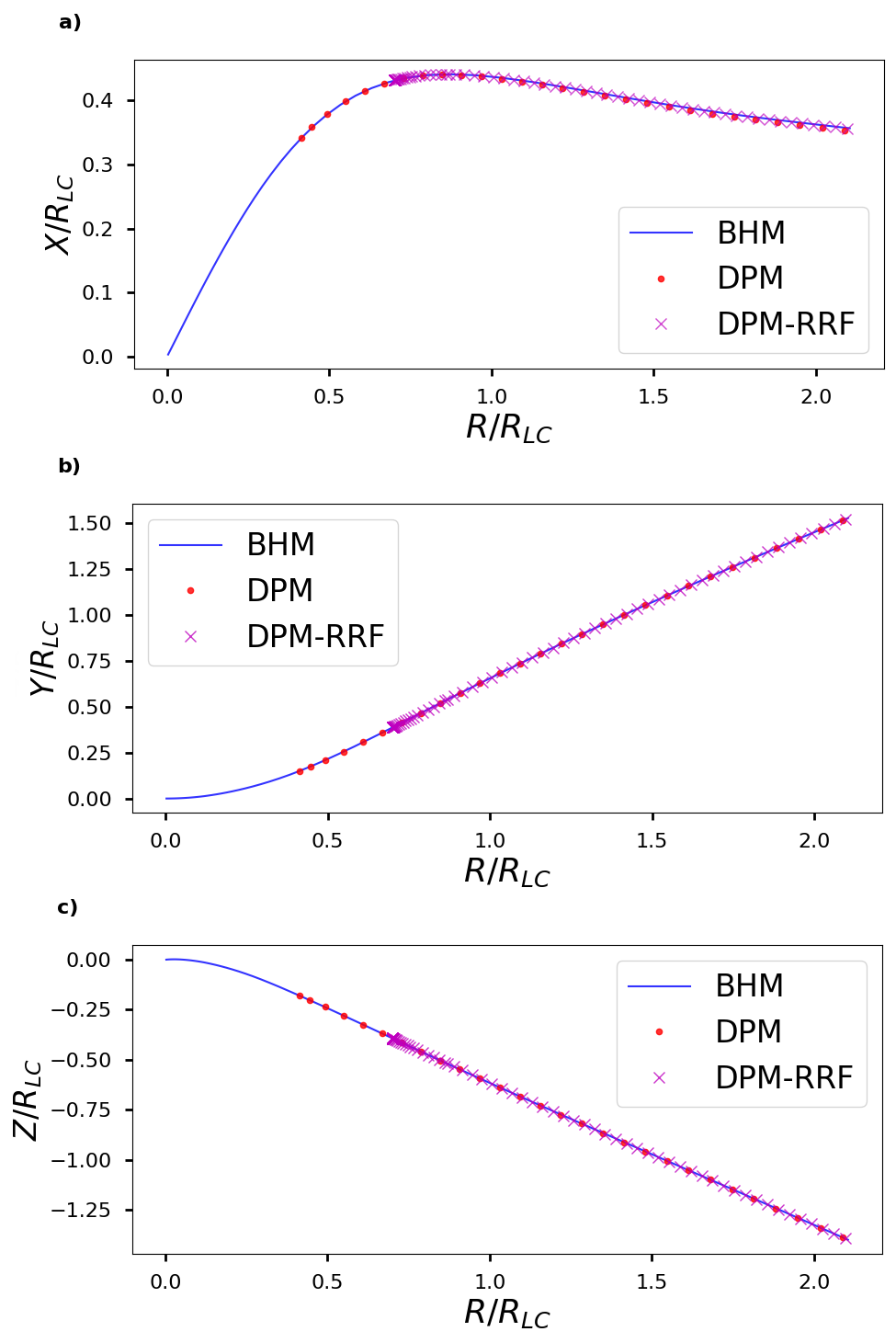}
\caption{Particle position plot for the Vela-like calibration case using $B_{\rm S} =8\times 10^{11} \, \rm{G}$, where the red dots represent our model results without RRF, the magenta crosses the case with RRF, and the blue line the position components of the \citetalias{Barnard2022} model. In panel a) we show the $x$-component of the particle position, in panel b) the $y$-component, and in panel c) the $z$-component.}
\label{FF_Pos_8e11}
\end{figure}

Here we show our particle trajectory comparison results for the Vela-like $B_{\rm S} = 8\times 10^{11} \, \rm{G}$ case. The justification for why we focus on this case and not the $B_{\rm S} = 8\times 10^{12} \, \rm{G}$ case is mentioned in Sections~\ref{sec:2.1} and~\ref{sec:2.2}. In these figures, we labelled with `RRF' the plots that show results with RRF included in the calculation. All other results in the article have RRF included, unless specified that it was excluded. For these results, we have followed the field line starting at PC phase $0^{\circ}$, using $R_{\rm acc}^{\rm min} = 4.0\times 10^{-3} \rm{cm}^{-1}$ and $R_{\rm acc}^{\rm max} = 2.5\times 10^{-2} \rm{cm}^{-1}$. When neglecting the RRF, we used the initial $\gamma_{0} = 6\times 10^{6}$ and when including the RRF, we used $\gamma_{0} = 4$. We thus start our particle without the RRF at $0.41R_{\rm LC}$ and the particle with RRF at $0.65R_{\rm LC}$, 
%due to the classical RRF limits 
in order to remain in the classical RRF regime, as explained previously. The results when limiting $\gamma$ with $\gamma_{\rm c}$ and using higher $R_{\rm acc}$ are included in the Appendix.
%and we will refer to them for applicable comparison.
Figure~\ref{FF_Pos_8e11} indicates that %for each position component vs normalised radial position, 
our model's particle position agrees well with that of the \citetalias{Barnard2022} model, both when including and excluding the RRF. 

\begin{figure}
\centering
\includegraphics[width=.5\textwidth]{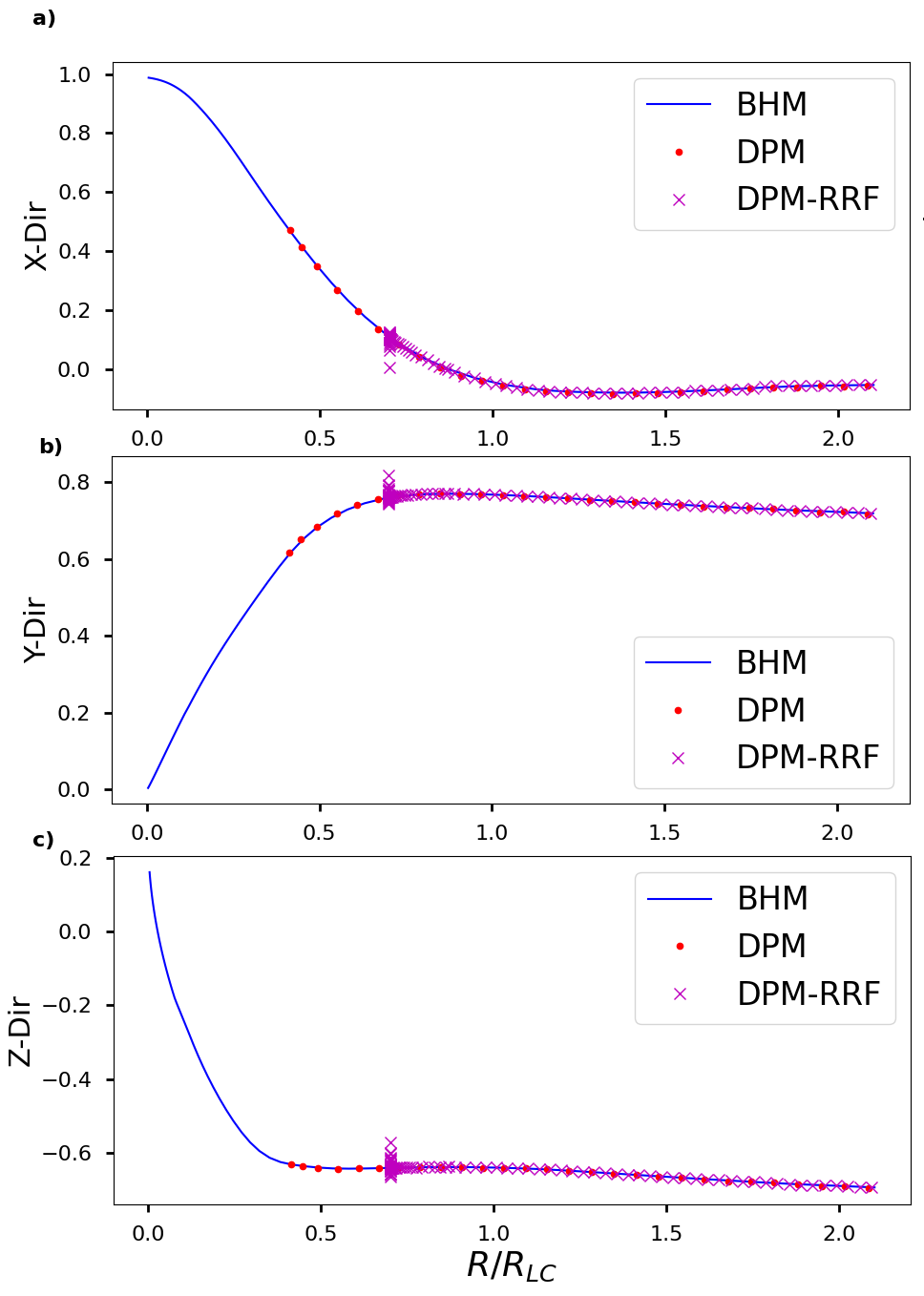}
\caption{Components of particle direction for the same case as in Figure~\ref{FF_Pos_8e11}, and the symbols having the same meaning.
%where the red dots represents our model results without RRF, the magenta crosses the case with RRF, and the blue line that of the \citetalias{Barnard2022} model. In panel a) we show the $x$-direction component, in panel b) the $y$-component, and in panel c) the $z$-component. 
The large spike at $0.6R_{\rm LC}$ is due to initialising with a large $\theta_{\rm p}$ and small $\gamma$ as explained in Section~\ref{sec:3.2}.}
\label{FF_Dir_8e11}
\end{figure}

In Figure~\ref{FF_Dir_8e11}, we plotted the three particle direction components. 
%We see that in all three panels, o
Our model results agree well with those of the \citetalias{Barnard2022} model, both when including and excluding the RRF. There does seem to be some difference between including and excluding the RRF 
%in the particle direction components 
as one approaches $2.0R_{\rm LC}$. If one were to zoom in on each of our resulting curves, one would also notice the curves oscillating around the \citetalias{Barnard2022} model results due to the particle gyrations. There is furthermore a large initial oscillation observed in all the particle direction components when including the RRF. The most prominent cause is due to the particle being initialised with a larger $\theta_{\rm p}$ as it starts further out in the magnetosphere, since we initialise our particle direction with the trajectory from Equation~(\ref{Alice_traj}) used by \citetalias{Harding2015, Harding2021, Barnard2022}.
%, giving the particle a large initial $\theta_{\rm p}$. 
The particle is also initialised to the gyro-centric values of Equation~(\ref{Alice_traj}), meaning it experiences an initial acceleration due to the $E_{\perp}$ field before it starts oscillating around the gyro-centric trajectory. Thus, this is a problem due to not initialising with trajectory results that take particle gyro-phase, namely the direction and position, into consideration. 

\begin{figure}
\centering
\includegraphics[width=.5\textwidth]{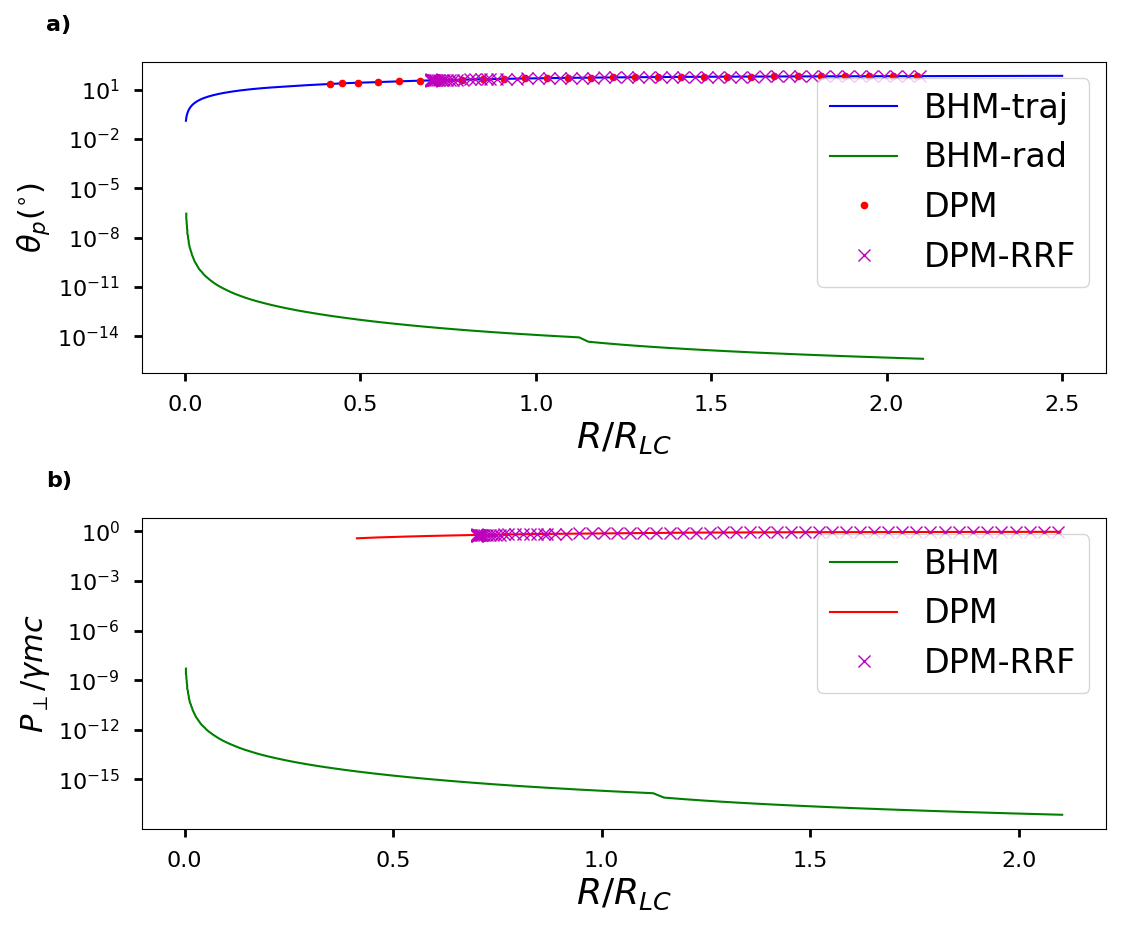}
\caption{Results for the Vela-like case as in Figure~\ref{FF_Pos_8e11}, with panel a) showing $\theta_{\rm p}$ for our model without RRF in red (dots), with RRF in magenta (crosses), the calculated $\theta_{\rm p}$ from the trajectories of Equation~(\ref{Alice_traj}) in blue, and the $\theta_{\rm p}$ (their modelled $\theta_{\rm VA}$) calculated from Equation~(\ref{Alice_transport}) in green. Panel b) shows the normalised $p_{\perp}$ from our model without RRF in red, with RRF in magenta, and the $p_{\perp}$ calculated from Equation~(\ref{Alice_transport}) in green.}
\label{Discrep}
\end{figure}

To investigate this phenomenon, we have included Figure~\ref{Discrep}, showing the $\theta_{\rm p}$ from our results in red and magenta, the $\theta_{\rm p}$ calculated from Equation~(\ref{Alice_traj}) in blue, and the $\theta_{\rm p}$ from Equation~(\ref{Alice_transport}) in green. 
%In Figure~\ref{Discrep}, one sees that 
Our $\theta_{\rm p}$ agrees well with the one calculated from the trajectory in the \citetalias{Harding2015, Harding2021, Barnard2022} models, but there is a discrepancy of many orders of magnitude between this one and the $\theta$ used in their radiation calculations. 
% As mentioned in the methods section, their trajectory and radiation calculations are decoupled, meaning they have no feedback on one another. 
%For their trajectories, t
They are following the FFE trajectories where the perpendicular $\mathbf{E}\times \mathbf{B}$-motion becomes dominant as one approaches $R_{\rm LC}$, causing $\theta_{\rm p}$ to increase, which is what we find with our general equations of motion as well. 
% As is discussed by \citet{landau1975, Harding2006} and showed in \citetalias{DuPlessis2024}, the RRF does not decrease the particle $\theta_{\rm p}$ in the super-relativistic limit, since the $\gamma^{2}$ dominating term is opposite to the particle velocity in Equation~(\ref{RRF}). 
On the other hand, the \citetalias{Harding2015, Harding2021, Barnard2022} models are assumed to model the AE parameters, thus $\theta_{\rm VA}$ (the general pitch angle), which is assumed to be quickly depleted by SR losses and then to remain small, only increasing once the particle absorbs radio photons. Therefore, rather than being interested in the usual pitch angle $\theta_{\rm p}$, if there is a large $E_{\perp}$ relative to $B$, one is interested in the angle between the AE trajectory and the particle velocity $(\theta_{\rm VA})$ for the SCR calculations. %(encapsulating both SR and CR limits, and valid for any general pitch angle).
In panel b) where we plot the $p_{\perp}$ normalised to $\gamma mc$, we see the same vast difference\footnote{The large difference in $p_{\perp}$ with respect to the local $B$-field and the particle trajectory limits our ability to compare our results with the standard SR calculations in \citetalias{Harding2015}, but we will focus on the CR and SCR for most of the comparison.} between our results and those of \citetalias{Harding2015, Harding2021, Barnard2022}. Similar to panel a), this is due to the \citetalias{Harding2015, Harding2021} models calculating the $p_{\perp}$ with respect to the particle trajectory, not the local $B$-field. 
We will discuss using the AE parameters for the spectrum calculations more in Section~\ref{sec:3.4} for the application of the \citetalias{Vigano2015} SCR calculations to compare if these SCR results converge to the \citetalias{Cerutti2016} SCR results. This will allow us to validate the \citetalias{Harding2021} SCR model approach.     

\begin{figure*}
\centering
\includegraphics[width=.9\textwidth]{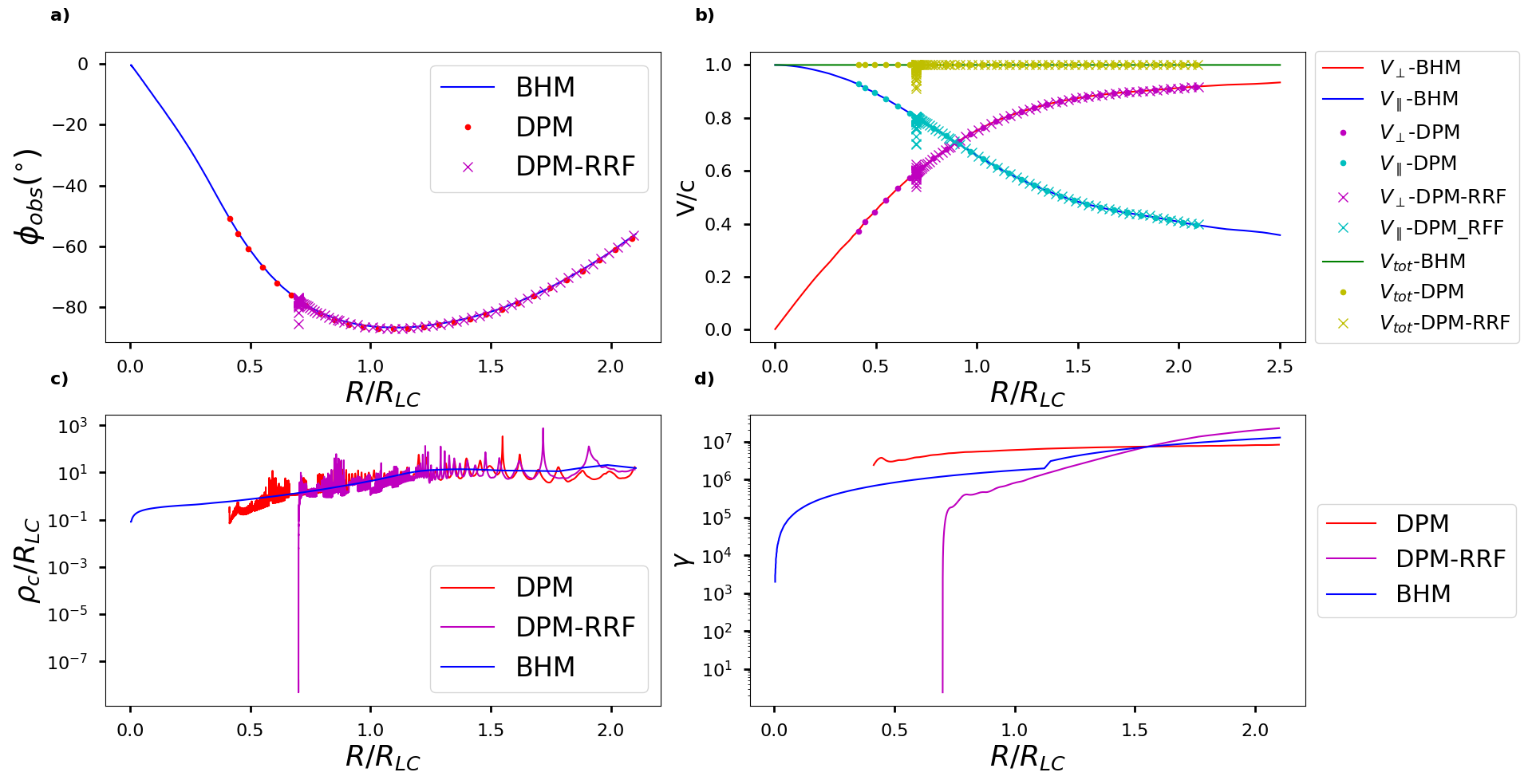}
\caption{Calibration case of Figure~\ref{FF_Pos_8e11}, showing the observer-corrected emission phase in panel a), normalised particle velocity components in panel b), $\rho_{\rm c}$ in panel c), and $\gamma$ in panel d). Here DPM represents our results without RRF, DPM-RRF our results with RRF, and BHM the \citetalias{Barnard2022} model results. In panel b), we show the perpendicular velocity and parallel velocity to the local $B$-field as well as the total velocity normalised to $c$. We use $B_{\rm S}=8\times10^{11}$~G, without limiting $\gamma$ by $\gamma_{\rm c}$.} 
\label{FF_Emission_phase_8e11}
\end{figure*}

In Figure~\ref{FF_Emission_phase_8e11} panel a) we show the corrected observer phase of the photon emission
%using Equation~(\ref{obs_phase}) 
for our model and the \citetalias{Barnard2022} model. Our results agree very well with their results, both with RRF included and excluded. In panel b) we plot the perpendicular velocity $V_{\perp}$ and parallel velocity $V_{\parallel}$ with respect to the local $B$-field, as well as the total velocity magnitude, where all of the velocities are normalised to $c$. Our velocity components agree well with those of the \citetalias{Barnard2022} model, and the total velocity components add up to $c$ everywhere, as one would expect. One can also see that the particle starts with a small $V_{\perp}$, which increases quite rapidly due to the $\mathbf{E}\times \mathbf{B}$ drift and becomes dominant over $V_{\parallel}$ as the particle approaches $R_{\rm LC}$ and beyond. This is where the effect of limiting $\gamma$ with $\gamma_{\rm c}$ is seen, as is evident in Figure~\ref{FF_Emission_phase_8e11_CRR} panel~b) for the $B_{\rm S} =8\times 10^{11} \, \rm{G}$ case. In Figure~\ref{FF_Emission_phase_8e11_CRR}, as the particle starts at $0.65R_{LC}$ (red line), one can see that the total velocity initially fluctuates around $c$ as the $\gamma$ is limited by $\gamma_{\rm c}$, until it relaxes closer to an equilibrium state where it converges to $c$. These are unphysical results, since we use relativistic equations of motion. This does not happen in the cases where $\gamma$ is not limited, as shown in Figure~\ref{FF_Emission_phase_8e11}, or where this limitation is not needed, as we found when using $B_{\rm S} =8\times 10^{10} \, \rm{G}$. This effect is even worse when limiting $\gamma$ and modelling the higher field case $B_{\rm S} =8\times 10^{12} \, \rm{G}$\footnote{These additional $B$-field cases can be found in \citetalias{DuPlessis2025}}. The re-normalisation (limiting) of $\gamma$ and implementation of the new velocity in the equations of motion seems to be what is affecting the results, since it happens with any limiting of $\gamma$, but is more noticeable in the extreme field and $\gamma$ cases. This is the main reason why we focus on calibrating our model results in the $B_{\rm S} =8\times 10^{11} \, \rm{G}$ case without limiting $\gamma$, and thus starting at a higher altitude and using a lower initial $\gamma$. 

In Figure~\ref{FF_Emission_phase_8e11} one sees that the $\rho_{\rm c}$ from our model oscillates around the results of \citetalias{Barnard2022}, but follows the trend of their results. This is due to their results being gyro-centric, where $E_{\perp}$ in our modelling oscillates the gyro-radius, changing it as the particle is moving parallel to $E_{\perp}$ in one half of the gyro-orbit and anti-parallel in the other half of the gyro-orbit. We discuss and show this in more detail for the uniform field cases in \citetalias{DuPlessis2024}. This is also shown by \citetalias{Kelner2015} in their appendix where they calculate the $\rho_{\rm c}$ in the ultra-relativistic limit using the drift approximation, which also reduces to taking the derivative of the AE trajectory. When excluding the RRF, our model $\rho_{\rm c}$ is initially lower than that of \citetalias{Barnard2022}, since the gyro-radius of the particle is much smaller in the lower magnetosphere.\footnote{Due to the high $B$-field and lower $\gamma$.} This effect 
%of the small gyro-radius 
is more evident from the magenta curve in panel c) where we initialised the particle with a much lower $\gamma=4$ for the RRF inclusion, but we see as the particle is quickly accelerated to high $\gamma$, therefore $\rho_{\rm c}$ increases and converges to our red curve and the \citetalias{Barnard2022} result. 
%This is also seen in Figure~\ref{FF_Emission_phase_8e10} panel c) for the lower-fields case, where we started at $0.41R_{\rm LC}$. 
Thus the \citetalias{Harding2015, Harding2021, Barnard2022} models yield the smoothed gyro-centric $\rho_{\rm c}$, whereas our model includes the particle oscillations and $\mathbf{E}\times \mathbf{B}$ drift effects. This discrepancy in $\rho_{\rm c}$ is a factor to take into account for the CR skymaps and spectral calibration due to the $\rho_{\rm c}$ dependence in Equation~(\ref{CR_spec}). This effect is accounted for in the $\rho_{\rm eff}$ calculations of SCR in Equations~(\ref{Vig_rho_eff}) and~(\ref{Cer_rho_eff}), scaling between the gyro-radius and $\rho_{\rm c}$. In panel d) of Figure~\ref{FF_Emission_phase_8e11}, we plot our model $\gamma$ with and without RRF and the $\gamma$ of \citetalias{Barnard2022} from their radiation calculations using Equation~(\ref{Alice_transport}). Since this $\gamma$ is decoupled from their trajectories, we use this to investigate how close our modelling results are to the $\gamma$ used for their radiation calculations. In panel d) we see that the magenta RRF curve starts with a low $\gamma$ and is accelerated quickly to similar values as the \citetalias{Barnard2022} model. Thus, even though our $\gamma$ values differ from \citetalias{Barnard2022}, we are in the same order of magnitude, which is encouraging, since we use completely different equations of motion and methods to model the RRF. When limiting $\gamma$ with $\gamma_{\rm c}$, we find in Figure~\ref{FF_Emission_phase_8e11_CRR} that when the particle enters equilibrium, our model $\gamma$ converges reasonably close to the \citetalias{Barnard2022} model results. Looking at the blue curve in panel d) one sees a jump in $\gamma$ at $1.0R_{\rm LC}$ where the $E_{\parallel}$ field is increased. This jump is much more prominent in our results where we use larger $E_{\parallel}$-fields, namely panel d) of Figure~\ref{FF_Emission_phase_8e11_CRR}. All sporadic jumps in the parameters at $1.0R_{\rm LC}$ are most likely due to the sudden increase in $E_{\parallel}$ there.    

\subsection{AE Results} \label{sec:3.3}
In this Section, we compare our model results to those of the \citet{Gruzinov2012} AE model labelled with `AE', and the AE results of \citetalias{Kelner2015} labelled with `-k'. These are the same trajectories and parameters from our previous results for the $B_{\rm S} =8\times 10^{11} \, \rm{G}$ case, but we are now investigating to see if our numerical model converges to the analytic AE limit / radiation-reaction limit results. The direction components for both AE models are the local direction of the AE trajectory, meaning that we use our model position and the local field values to calculate the AE trajectories. %Thus, we 
We calculate the AE results for our model with and without RRF. The AE trajectories are approximated to be solely governed by the electromagnetic fields, therefore, the AE results only consider the energy gained from acceleration, whereas the AE-RRF results account for the acceleration and RRF. The $\gamma_{\rm c}$ and $\gamma_{\rm SRR}$ curves are the values calculated using Equations~(\ref{gam_CRR}) and ~(\ref{gam_SRR}), respectively, at each point along the particle trajectory, and not our particle $\gamma$ limited by these values. One expects the particle $\gamma$ to be reasonably close to $\gamma_{\rm c}$ at equilibrium in a CR-dominated scenario, since the power gained by the particle via acceleration is balanced with the CR power radiated to obtain Equation~(\ref{gam_CRR}).

\begin{figure*}
\centering
\includegraphics[width=.9\textwidth]{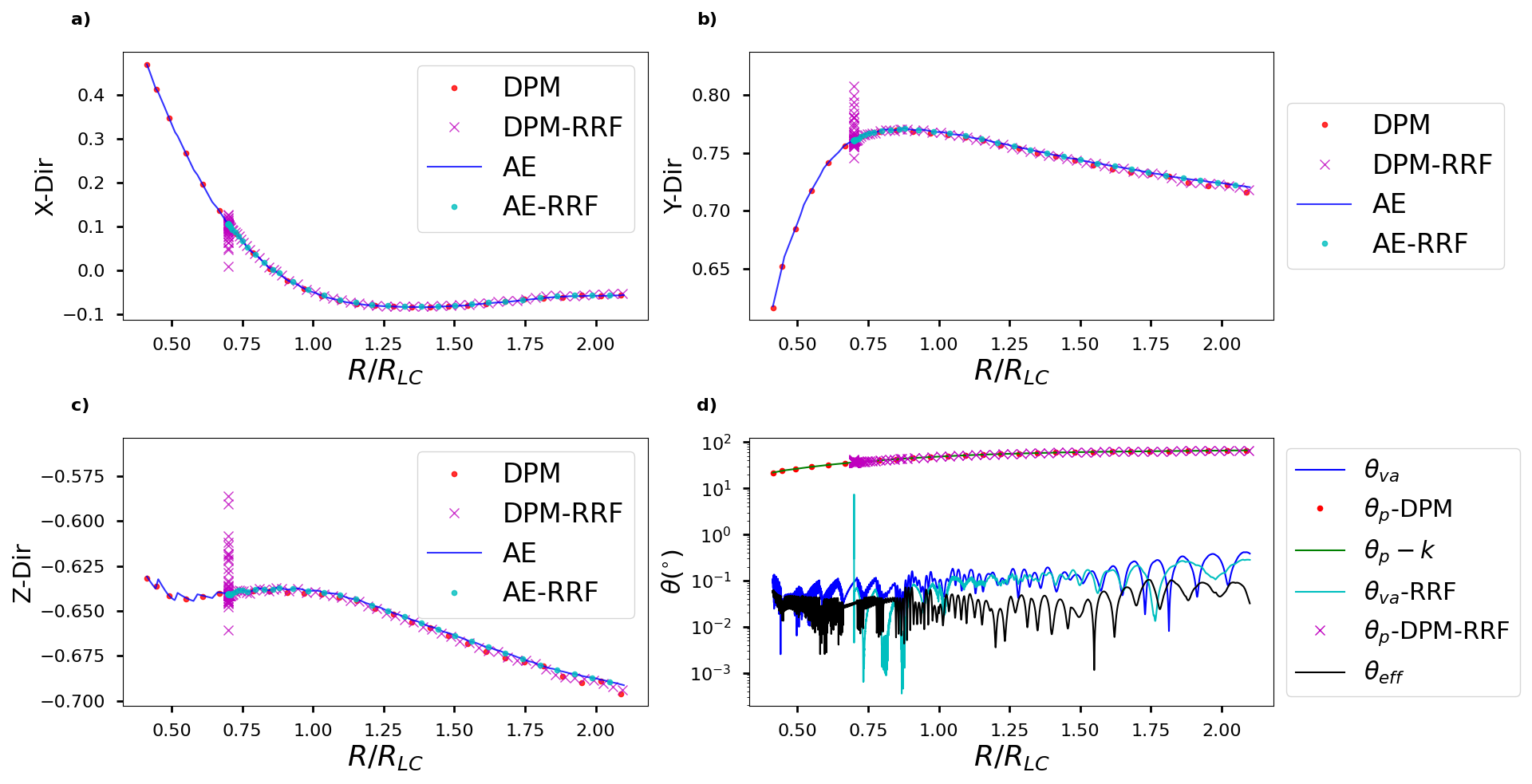}
\caption{The AE convergence results for the case as in Figure~\ref{FF_Pos_8e11}. In this plot, DPM labels our model results, and AE  those of \citet{Gruzinov2012}. Panel a) shows the particle $x$-direction, panel b) the $y$-direction, and panel c) the $z$-direction of the particle's motion. In panel d), we show the various angles discussed in Section~\ref{sec:2}. Here $\theta_{\rm p}$-DPM (red dots) overlaps with $\theta_{\rm p}$-DPM-RRF (magenta crosses) and $\theta_{\rm p}$-k (green line); this is also the case for  $\theta_{\rm VA}$ (blue line) which overlaps with $\theta_{\rm VA}$-RRF (cyan line). The RRF at the end of a plot legend indicates that RRF was included. The large initial spikes (magenta crosses) are due to initialising with a large $\theta_{\rm p}$ as discussed in the text.}
\label{FF_AE_vel_8e11}
\end{figure*}

In panels a), b) and c) of Figure~\ref{FF_AE_vel_8e11}, we plot the three particle direction components, finding that our model results to converge very well to the AE results of both \citet{Gruzinov2012} and \citetalias{Kelner2015}. We see a large initial oscillation in our RRF results, as explained in the previous Section, due to the large initialised $\theta_{\rm p}$, as well as how our particle oscillates around the AE trajectory due to initialising with gyro-centric values. In panel d), we plot the various angles discussed in the methods Section. In this panel, $\theta_{\rm VA}$ is underneath $\theta_{\rm VA-k}$, where we see these values are $\sim0.1^{\circ}$, meaning that our trajectory is very close to the AE trajectory. As discussed in \citetalias{DuPlessis2024}, $\theta_{\rm VA}$ does not decrease to zero as the results converge, since the AE trajectory is gyro-centric. This means that at equilibrium, our results will converge to a constant $\theta_{\rm VA}$, where the particle gyrates with a constant radius around the AE trajectory curve as the gyro-centre. In panel~d), we also plot our model $\theta_{\rm p}$ in red/magenta vs that obtained from Equation~(\ref{Kel_pitch}), showing that they agree very well. Lastly, we plot the $\theta_{\rm eff}$ discussed in Section~\ref{sec:2.6}, showing that it is very close to $\theta_{\rm VA}$, but we do not expect an exact match, since we believe they would be much closer in the pure SR regime. Investigating another case in Figure~\ref{FF_AE_vel_8e11_CRR} where there is a higher $E_{\parallel}$ compared to the local $B$-field, we see that at $1.0R_{\rm LC}$, there is a jump in the AE trajectories due to the change in $E_{\parallel}$, causing a divergence from our model results. The results show that our particle is kicked out of equilibrium by this jump in $E_{\parallel}$ and has to settle into a new equilibrium state. Similar to the $\gamma$ results at $1.0R_{\rm LC}$, we see a noticeable increase in $\theta_{\rm VA}$ due to the sudden increase in $E_{\parallel}$, which means that the particle has to converge to a new equilibrium, and thus deviates from the AE results. This jump is much more apparent in Figure~\ref{FF_AE_vel_8e11_CRR}, where there is a larger $E_{\parallel}$ relative to the $B$-field. 

\begin{figure*}
\centering
\includegraphics[width=.9\textwidth]{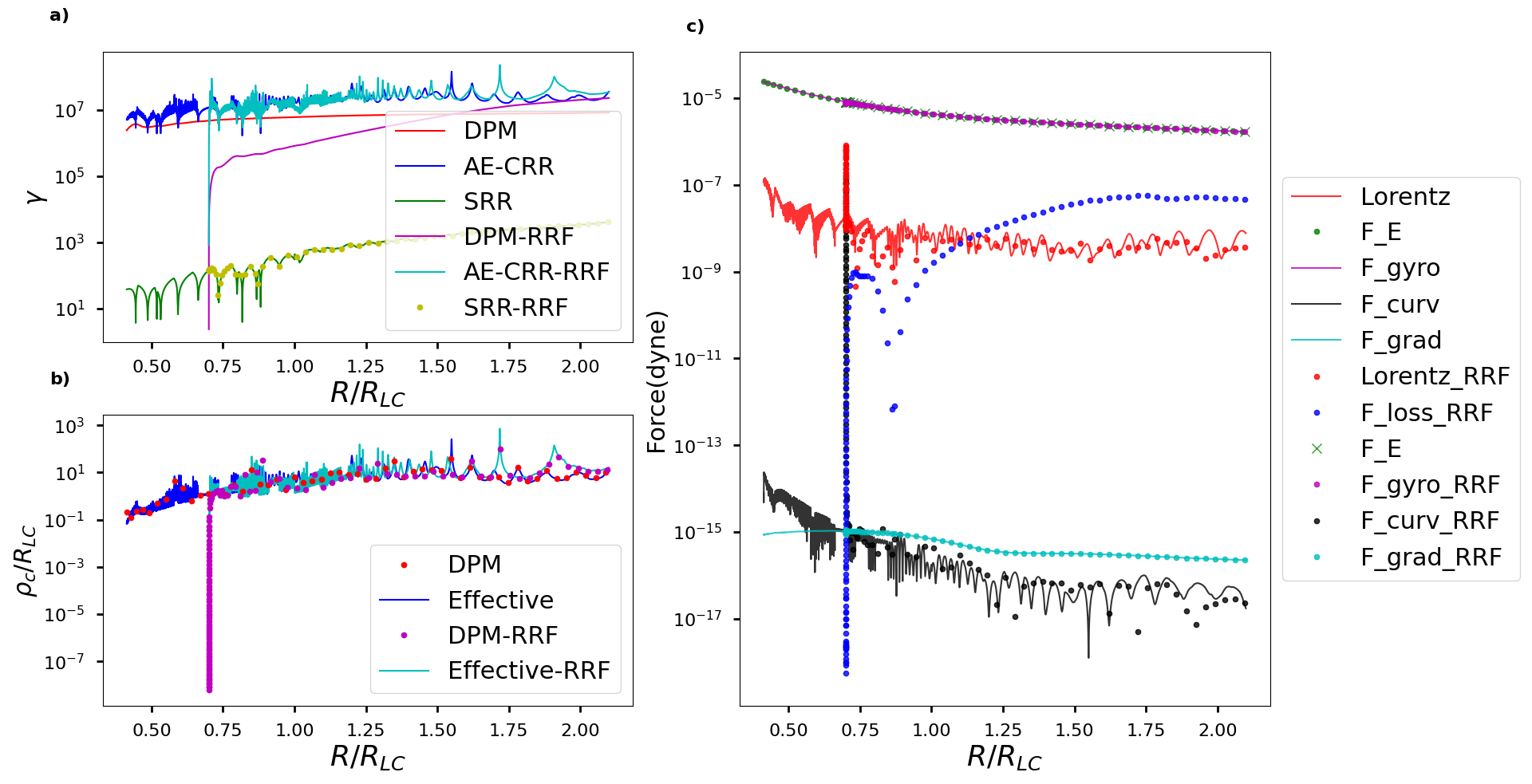}
\caption{Results for the Vela-like case as in Figure~\ref{FF_Pos_8e11}, with panel a) showing our $\gamma$ results in red, $\gamma_{\rm c}$ in blue, and $\gamma_{\rm SRR}$ in green. Panel b) shows our particle $\rho_{\rm c}$ in red and the effective $\rho_{\rm c}$ from \citetalias{Kelner2015} in blue. Panel c) shows the different force components, namely the Lorentz force in red, the RRF in blue, the gyro-component of the Lorentz force in magenta, the $E$-field component of the Lorentz force in green, the curvature drift in black, and the gradient drift in cyan. The RRF at the end of a plot legend indicates that RRF was included.}
\label{FF_AE_rho_c_8e11}
\end{figure*}

In Figure~\ref{FF_AE_rho_c_8e11} panel a), we plot our model $\gamma$ values, now including the theoretical $\gamma_{\rm c}$ from Equation~(\ref{gam_CRR}), and theoretical $\gamma_{\rm SRR}$ from Equation~(\ref{gam_SRR}). We see that our model $\gamma$ values with and without the RFF are below $\gamma_{\rm c}$, where the RRF case in magenta approaches this critical value at $2.0R_{\rm LC}$. We thus see that the particle $\gamma$ value lies between $\gamma_{\rm c}$ and $\gamma_{\rm SRR}$ and converges to $\gamma_{\rm c}$ as the particle converges to equilibrium in this CR-dominated scenario. In the other Vela-like case shown in the Appendix, namely panel a) of Figure~\ref{FF_AE_rho_c_8e11_CRR}, one can see where the $\gamma$ is limited by $\gamma_{\rm c}$. We can also see that the $\gamma$ values converge to $\gamma_{\rm c}$ as the particles get closer to equilibrium. This thus conforms to the approximation of $\gamma_{\rm c}$ as the $\gamma$ value at equilibrium.
% In Figure~\ref{FF_AE_rho_c_8e12}, we can see that the model $\gamma$ is almost entirely limited by $\gamma_{\rm c}$ along its trajectory corresponding to the regions with unphysical velocity in panel b) of Figure~\ref{FF_Emission_phase_8e12}. %Interestingly, we see the low-field case in Figure~\ref{FF_AE_rho_c_8e10} is still not close to equilibrium, even at $2.0R_{LC}$, thus one would need higher $E_{\parallel}$-fields to reach equilibrium faster as is required by AE, but this would also affect the particle trajectory. 

The question arises whether adding the large jump in $E_{\parallel}$ at $1.0R_{\rm LC}$ kicks the particle out of the AE equilibrium and requires time to reach a new equilibrium, since the AE trajectories are only valid in equilibrium. It would have been better to include a smooth $E_{\parallel}$ or a smaller jump in $E_{\parallel}$. In panel~b) of Figure~\ref{FF_AE_rho_c_8e11}, we plot our model $\rho_{\rm c}$ in red and magenta and the $\rho_{\rm eff}$ from Equation~(\ref{Cer_rho_eff}) in blue and cyan. One sees that our model values are almost exactly equal to $\rho_{\rm eff}$. Thus, our model yields $\rho_{\rm eff}$ when using Equation~(\ref{AL_rho_c}), namely the instantaneous radius of curvature of the particle, not the radius of curvature of the gyro-centric curved trajectory as obtained by the \citetalias{Harding2015, Harding2021, Barnard2022} models using the FFE trajectory. In panel b) of Figure~\ref{FF_AE_rho_c_8e11_CRR}, we see that our $\rho_{\rm c}$ conforms very well to $\rho_{\rm eff}$, even in the $\gamma_{\rm c}$-limited cases and interestingly where the particles are not in equilibrium. 

In panel~c) of Figure~\ref{FF_AE_rho_c_8e11}, we have plotted the various magnitudes of the forces and drift forces acting on the particle as discussed in Section~\ref{sec:2.3}. 
Here we plot the total Lorentz force in red, the $E$-field component of the Lorentz force $(F=qE)$ in green, the gyro-component $(\propto\mathbf{p}\times \mathbf{B})$ of the Lorentz force in magenta, the force due to the curvature drift in black, the force due to the gradient drift in cyan, and the RRF in blue. Dots are used to show the results when including RRF (labelled with `RRF'), but these results mostly overlap with our model results when excluding the RRF. This is also why there is only a dotted blue curve. Comparing the forces of the drift components (i.e, $\mathbf{E}\times\mathbf{B}$, curvature, and gradient drift) and the RRF tells us which component is dominant in changing $\theta_{\rm p}$ since the drifts are perpendicular the local $B$-field\footnote{The RRF's dominant component is anti-parallel to the particle velocity and will thus not affect the particle$\theta_{\rm p}$ significantly.}. Since our model uses FF-fields, one expects the $E$-field component of the Lorentz force to be equal or very close in magnitude to the gyro-component of the Lorentz force, since the fields were defined to be FF, i.e., $F=0$, meaning these two components should cancel one another. Thus, the FF $E_{\perp}$-field is defined so that $q\mathbf{E}$ is opposite to the Lorentz force gyro-component. This means the inclusion of an $E_{\parallel}$-field deviates from the FF solution, but using a small $E_{\parallel}$ causes the fields to be close to FF, which is sufficient for our modelling purposes.
% The dotted lines are used to show the results when including RRF, but these results mostly overlap with our model results when excluding the RRF. In panel~c), we have plotted the $E$-field component of the Lorentz force in green,
% the gyro-component $(\mathbf{p}\times \mathbf{B})$ of the Lorentz force in magenta, the total Lorentz force in red, the RRF in blue, curvature drift in black, and the gradient drift in cyan. 
%In panel c) we see that the $E$-field and gyro-component of the Lorentz force are almost equal, namely the green and magenta curves overlap, meaning that the total Lorentz force is much lower than these two components as seen by the much lower red curve and expected in FF-fields. 
The curvature and gradient drifts are found to be much lower than the $E$-field component, thus the $\mathbf{E}\times \mathbf{B}$-drift is the dominating contribution to the change in $\theta_{\rm p}$ in this case. Looking at the Lorentz force and RRF, we can now directly investigate if or how close the particle is to equilibrium, since these forces are expected to balance at equilibrium. In panel~c), we see that the particle is starting to close in on equilibrium at $1.0R_{\rm LC}$ (since the red Lorentz force and blue RRF curves are close), but it has to achieve a new equilibrium after encountering the larger $E_{\parallel}$ at $1.0R_{\rm LC}$. Even at $2.0R_{\rm LC}$ the particle has yet to achieve the new equilibrium state. A larger $E_{\parallel}$ would decrease the time for the particle to achieve equilibrium, as argued by the AE models. The problem is calibrating with the \citetalias{Harding2015, Harding2021} models using a large $E_{\parallel}$, since their particle trajectory is unaffected by the $E_{\parallel}$-field. The second problem is accelerating the particle too quickly into the non-classical regime of the RRF. In panel~c) at $\sim 0.88R_{\rm LC}$ there is a large decrease and increase in the RRF blue curve. This is also observed in panel~c) of Figure~\ref{FF_AE_rho_c_8e11_CRR}. These features correspond to divergence spikes at the same radius in each case's corresponding $\nabla\cdot\mathbf{B}$ plot shown in \citetalias{DuPlessis2025}. Thus, it would appear this correlates to the $y$-point in the $B$-field structure as a possible explanation for the divergence spike.   
%Analysing the other Vela cases and starting with the lower-field case in Figure~\ref{FF_AE_rho_c_8e10}, we see in panel~c) that the particle is still far away from equilibrium as was mentioned looking at the $\gamma$ plot in panel a) as well. This could be due to the particle needing to be accelerated to higher $\gamma$ values to enter equilibrium more quickly. 
In the $\gamma_{\rm c}$-limited case for $B_{\rm S} = 8\times 10^{11} \, \rm{G}$ of Figure~\ref{FF_AE_rho_c_8e11_CRR}, we see in panel~c) that the particle is quite close to equilibrium between $1.7R_{\rm LC} - 2.0R_{\rm LC}$. This is most likely due to the higher $E_{\parallel}$ used in this region. 
% In the higher-field case of Figure~\ref{FF_AE_rho_c_8e12}, we see that the particle is in equilibrium and interestingly this is the point where the problems in the velocity subside in Figure~\ref{FF_Emission_phase_8e12}. 
It is thus clear that one needs a large enough $E_{\parallel}$ to accelerate the particle to high enough $\gamma$ values to achieve equilibrium quickly for the AE trajectories and assumptions to hold. Looking at the force plots of Figure~\ref{FF_AE_rho_c_8e11_CRR}, we also see that limiting the particle's $\gamma$ with $\gamma_{\rm c}$ does not force the particle into equilibrium, but the assumption that $\gamma = \gamma_{\rm c}$ is valid once the particle is in equilibrium. As shown in Figure~\ref{FF_AE_rho_c_8e11} and in \citetalias{DuPlessis2024} for uniform fields, this is due to the RRF exceeding the Lorentz force and an oscillatory effect of which term dominates until equilibrium is reached. %One should also use a smooth $E_{\parallel}$ since the sharp change causes a sharp change in equilibrium, meaning that the particle has to settle into the new equilibrium state.    

We have lastly added the AE convergence results in the Appendix using RD fields for a pulsar scenario, and using RD fields for a magnetic mirror scenario. We have initialised the particle on the PC of the stellar surface for the first case and at $0.5R_{\rm LC}$ for the second case while using $B_{\rm S} = 8 \times 10^{8} \, \rm{G}$ for both cases. For the full parameter and scenario setups, see the Appendix. In Figure~\ref{RD_AE_vel}, one sees that our model results converge very well to the AE trajectories. This shows how well our results converge to the AE limit using these analytic fields, since it is quite difficult to accurately calculate the RRF in these extreme radiation reaction regimes. On the other hand, in the mirror scenario in Figure~\ref{AE_vel_mirror}, one sees that as the particle travels inward from $0.5R_{\rm LC}$, our model results are completely different from the AE trajectory, but after turning around and travelling outward, our model results converge with the AE trajectory again. This is because Equation~(\ref{AE}) is defined for particles travelling outward from the stellar surface, with the sign defined for opposite charges\footnote{Upon deeper investigation, changing the sign of Equation~(\ref{AE}) when the particle mirrors, mimicking a turnaround with the AE equations, the AE equations were unable to model the correct particle trajectory for this scenario.}. This would suggest that these trajectory descriptions do not apply to a magnetic mirror scenario. Remarkably, looking at Figure~ \ref{AE_rho_c_mirror}, we see that our model $\rho_{\rm c}$ agrees very well with $\rho_{\rm eff}$ from Equation~(\ref{Cer_rho_eff}), even though the Lorentz force and RRF are not in equilibrium yet. Thus, using $\rho_{\rm c}$ from the \citetalias{Kelner2015} SCR model description or $\tilde{B}_{\perp}$ from the \citet{Cerutti2016} seems to work well.   

\subsection{SCR Comparison} \label{sec:3.4}
In this Section, we investigate a CR-dominant scenario and an SR-dominant scenario to compare the single-particle spectral results to the SCR spectra calculations of \citetalias{Vigano2015} and \citetalias{Cerutti2016, Kelner2015}. For the CR-dominant case, 
%we follow the field line at PC phase $0^{\circ}$ and 
we use $B_{\rm S} = 8\times 10^{10} \, \rm{G}$, $R_{\rm acc}^{\rm min} = 4.0\times 10^{-4} \rm{cm}^{-1}$, $R_{\rm acc}^{\rm max} = 2.5\times 10^{-3} \rm{cm}^{-1}$, and initialising our equations of motion at $0.4R_{\rm LC}$. For the SR-dominant case, we use RD fields with $B_{\rm S} =8\times 10^{6} \, \rm{G}$, $R_{\rm acc} = 4.0\times 10^{-4} \rm{cm}^{-1}$, $\gamma_{0} = 1\times 10^{6}$, and initial $\theta_{\rm p} = 60^{\circ}$. For both of the cases, we use an infinitesimally small slot gap at $r_{\rm ovc} = 0.96$. The particle position was initialised in both cases to start at the stellar surface following the PC phase $0^{\circ}$ $B$-field line. Additionally, we also investigate this same SR-dominant case, but excluding the $E_{\perp}$-field. In all of these comparison cases, we have set the spectral normalisation to one. The figure labels from Section~\ref{sec:3.4} to~\ref{sec:3.6} are emphasised below.\\ 

\begin{center}
\begin{tabular}{|c|c|} 
\hline 
    SR/CR/SCR & SR/CR/SCR calculated using our own model.  \\  
    \hline
    CR-BHM$\_$rho & CR calculated using our model but also \\ &using the \citetalias{Barnard2022} $\rho_{\rm c}$. \\
    \hline
    Vig$\_$p & Calculations using the \citetalias{Vigano2015} equations, \\ &while using $\theta_{\rm p}$.  \\ 
    \hline
    Vig$\_$VA & Calculations using the \citetalias{Vigano2015} equations, \\ &while using $\theta_{\rm VA}$. \\
    \hline
    Vig$\_$AE & Calculations using the \citetalias{Vigano2015} equations, while \\ &using $\theta_{\rm VA}$ and the AE $\rho_{\rm c}$. \\
    \hline
    SCR-Vig* & SCR calculated using \citetalias{Vigano2015} equations \\ &and indicated deviation angle. \\ 
    \hline
    BHM$\_$CR/SR & CR/SR calculated using the \citetalias{Barnard2022} model. \\
    
\hline
\end{tabular}
\end{center}

\begin{figure}
\centering
\includegraphics[width=.5\textwidth]{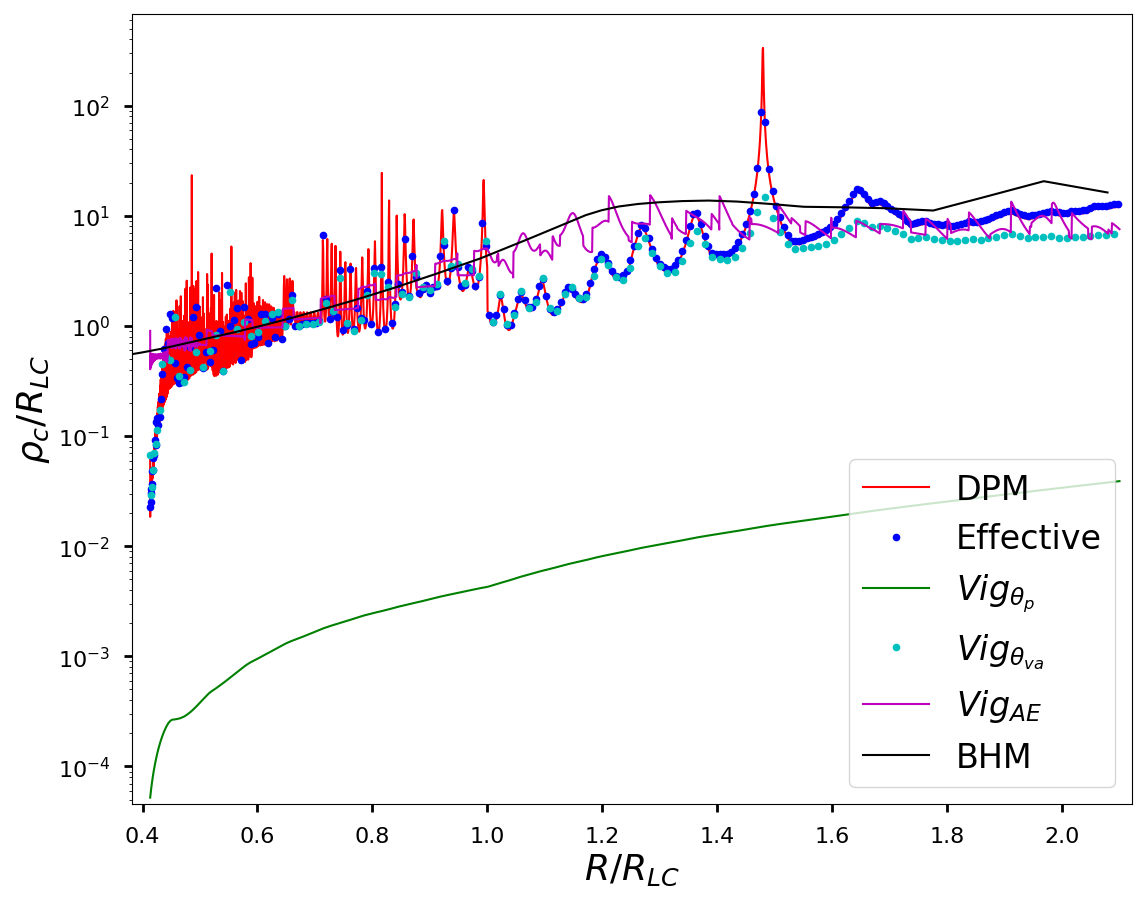}
\caption{Results for the $\rho_{\rm c}$ values for the $B_{\rm S} =8\times 10^{10} \, \rm{G}$ case. All the results are calculated using our model using the different equations from Section~\ref{sec:2}, except for the black curve. Our model results are shown in red, the $\rho_{\rm c}$ from the \citetalias{Barnard2022} model in black, $\rho_{\rm eff}$ from \citetalias{Kelner2015} using blue dots, $\rho_{\rm eff}$ from \citetalias{Vigano2015} using $\theta_{\rm p}$ and our model $\rho_{\rm c}$ in green, their $\rho_{\rm eff}$ using $\theta_{\rm VA}$ and our $\rho_{\rm c}$ using cyan dots, and their $\rho_{\rm eff}$ using $\theta_{\rm  VA}$ and $\rho_{\rm c}$ from the AE trajectories in magenta.}
\label{SCR_rho_c}
\end{figure}

In Figure~\ref{SCR_rho_c}, we plotted the different model $\rho_{\rm eff}$ values normalised to $R_{\rm LC}$. In red, we show our model $\rho_{\rm c}$, in black the \citetalias{Barnard2022} model $\rho_{\rm c}$, and in blue $\rho_{\rm eff}$ from Equation~(\ref{Cer_rho_eff}). For the \citetalias{Vigano2015} $\rho_{\rm eff}$ in Equation~(\ref{Vig_rho_eff}), we show results from three different calculation methods, namely using $\theta_{\rm p}$ and our model $\rho_{\rm c}$ in green, using $\theta_{\rm VA}$ and our model $\rho_{\rm c}$ in cyan, and using $\theta_{\rm VA}$ and the $\rho_{\rm c}$ calculated from the AE trajectory in magenta. One can see that our model $\rho_{\rm c}$ agrees very well with the \citetalias{Kelner2015} $\rho_{\rm eff}$ as was mentioned for Figure~\ref{FF_AE_rho_c_8e11}, but surprisingly the cyan $\rho_{\rm eff}$ agrees well with our results and only deviates slightly in the extended magnetosphere. What is visible is the large deviation between the cyan and green $\rho_{\rm eff}$, where the green curve is far too low. This indicates that one can not use the $\theta_{\rm p}$ as implemented in the \citetalias{Vigano2015} model when including large $E_{\perp}$-fields. Thus, one has to rather do this calculation following the AE curve where we see that using $\theta_{\rm  VA}$ instead of $\theta_{\rm p}$ yields a massive improvement, correlating with the \citetalias{Kelner2015} $\rho_{\rm eff}$. We also implemented the AE trajectory $\rho_{\rm c}$ for the magenta curve, where this magenta $\rho_{\rm eff}$ corresponds well to the $\rho_{\rm c}$ from the \citetalias{Barnard2022} model in black. Thus, the \citetalias{Harding2015, Harding2021} models indeed use the AE trajectory $\rho_{\rm c}$. The variation in the magenta $\rho_{\rm eff}$ is due to the varying gyro-radius due to $E_{\perp}$ in the observer frame. Thus we find that using $\theta_{\rm VA}$ and the AE $\rho_{\rm c}$ already has a big impact on the \citetalias{Vigano2015} SCR calculations. 

\begin{figure}
\centering
\includegraphics[width=.5\textwidth]{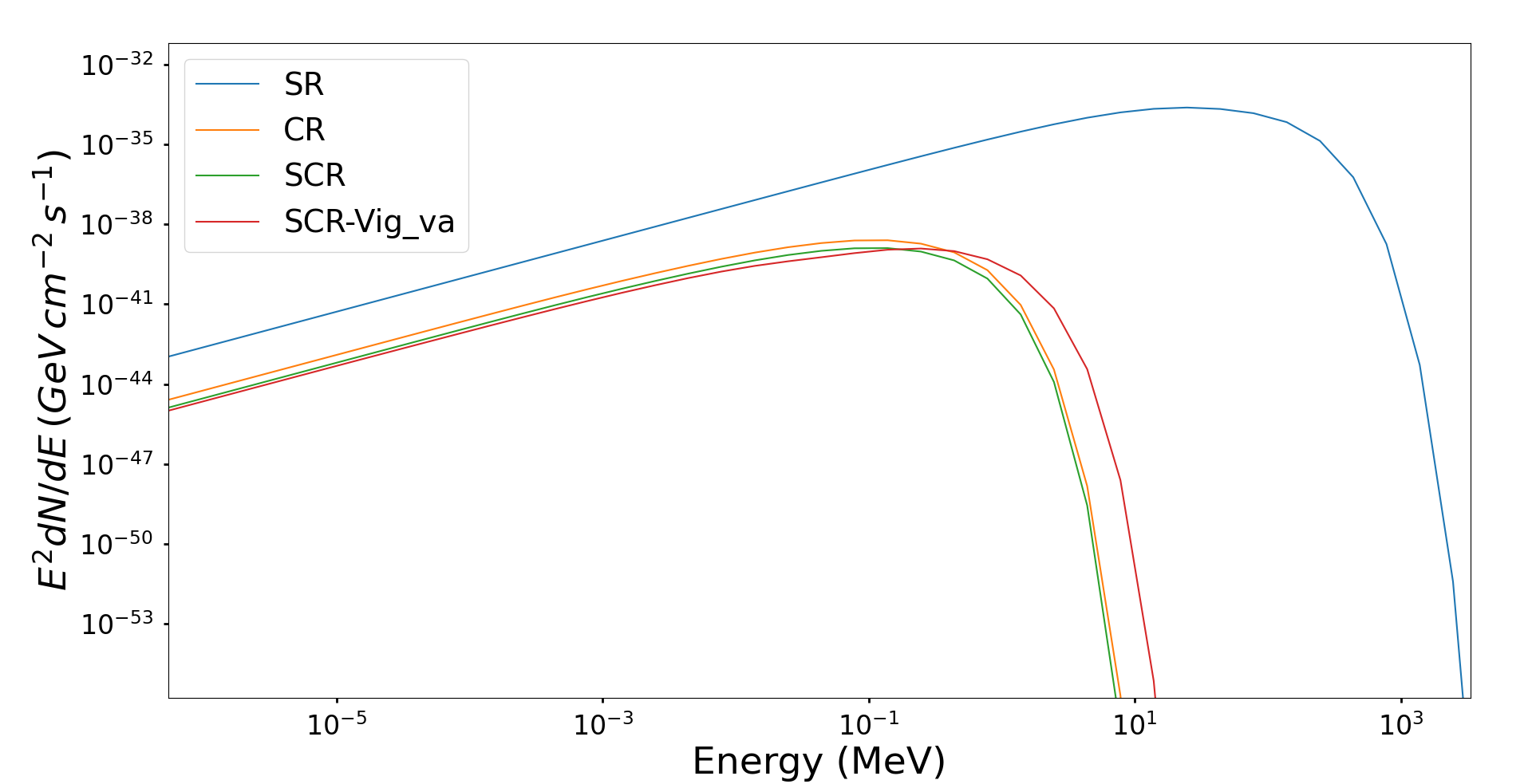}
\caption{The single-particle spectra for the same CR-dominant case as Figure~\ref{SCR_rho_c}. We plot the standard CR spectrum in orange and the standard SR spectrum in blue as discussed in Section~\ref{sec:2.5}. The SCR spectrum for \citetalias{Cerutti2016} is plotted in green and the \citet{Vigano2015} SCR spectrum is plotted in red using $\theta_{\rm  VA}$ and the $\rho_{\rm c}$ calculated from the AE trajectory. Here the SR spectrum cutoff is so high due to the large $\theta_{\rm p}$ and using the inapplicable standard SR formulae, where one has a large $E_{\perp}$-field as well as large $B$-fields and particle $\gamma$.}
\label{SCR_CR_case}
\end{figure}

In Figure~\ref{SCR_CR_case}, we plot the single-particle spectra for the CR-dominant case. We plot the SR spectrum in blue (using $\theta_{\rm p}$), the CR spectrum in orange, the \citetalias{Cerutti2016} SCR spectrum in green, and the \citetalias{Vigano2015} SCR spectrum using $\theta_{\rm VA}$ and the AE $\rho_{\rm c}$ in red. Both SCR spectra are reasonably close to the CR spectrum, with the \citetalias{Cerutti2016} SCR spectrum being much closer. We do not expect an exact match, but the SCR should be quite close to the CR spectrum in this case. Using the standard SR formulae causes the spectrum to be much higher and peak at much larger energies than the CR spectra, due to the large increasing $\theta_{\rm p}$ shown in Figure~\ref{Discrep}. This means that $p_{\perp}$ is much larger than expected, causing a larger energy cutoff and higher spectral flux for the SR spectrum. It is therefore problematic to use the standard SR equations that neglect an $E_{\perp}$-field in their derivations. Here we have a scenario with high $B$-fields and $\gamma$ where the large $E_{\perp}$-field significantly alters the particle trajectory as well as oscillates all of the particle parameters, meaning the standard SR calculations would lead to unphysical results. 
%We therefore illustrate the importance of using valid SR calculations when a large $E_{\perp}$-field is present. 

\begin{figure}
\centering
\includegraphics[width=.5\textwidth]{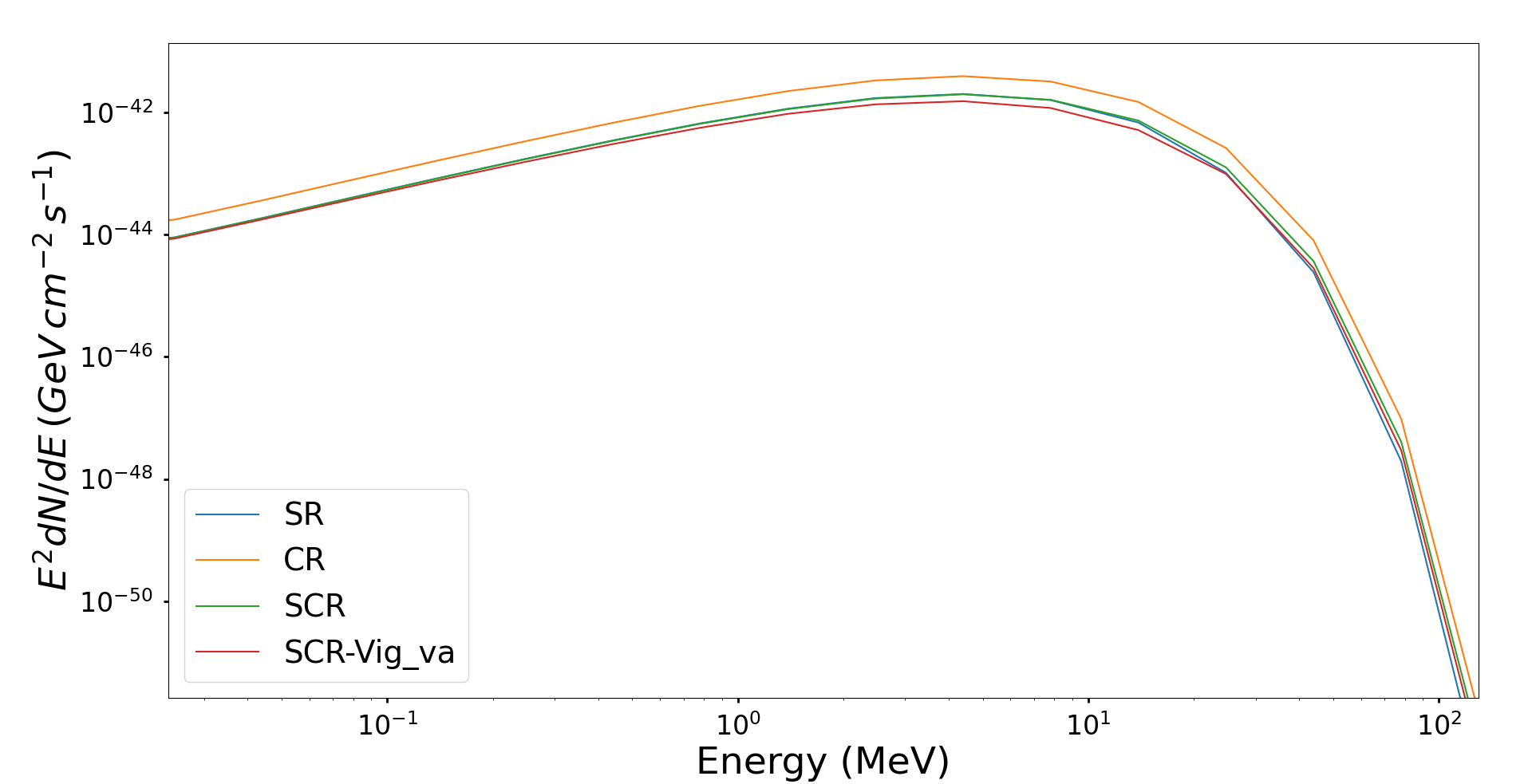}
\caption{The single-particle spectra for an SR-dominant case using RD fields with $B_{\rm S} =8\times 10^{6} \, \rm{G}$, $R_{\rm acc}^{\rm min} = 4.0\times 10^{-4} \rm{cm}^{-1}$, $R_{\rm acc}^{\rm max} = 2.5\times 10^{-3} \rm{cm}^{-1}$, $\gamma_{0} = 1\times 10^{6}$, and initial $\theta_{\rm p} = 60^{\circ}$. We have plotted the standard CR spectrum in orange and the standard SR spectrum in blue. The SCR spectrum for \citetalias{Cerutti2016} is plotted in green and the \citetalias{Vigano2015} SCR spectrum is plotted in red using $\theta_{\rm VA}$ and the $\rho_{\rm c}$ calculated from the AE trajectory.}
\label{SCR_SR_EB_case}
\end{figure}

In Figure~\ref{SCR_SR_EB_case}, we plot the single-particle spectra for the SR-dominant case when including an $E_{\perp}$-field. 
%In this figure, we see 
The standard SR spectrum in blue and CR spectrum in orange are reasonably close, with the SR spectrum being slightly lower. We plotted the \citetalias{Cerutti2016} SCR spectrum in green and the \citetalias{Vigano2015} SRC spectrum in red, where we used $\theta_{\rm VA}$ and the AE $\rho_{\rm c}$.
%for the calculation. 
Both SCR spectra were found to be quite close to the SR spectrum.
%, with minor deviations in some parts of the spectra, namely at the exponential cut-off. 
The CR spectrum is found to be close to the SR spectrum, since in Equation~(\ref{AL_rho_c}) our calculated $\rho_{\rm c}$ is closer to the gyro-radius due to the large $\theta_{\rm p}$. 

\begin{figure}
\centering
\includegraphics[width=.5\textwidth]{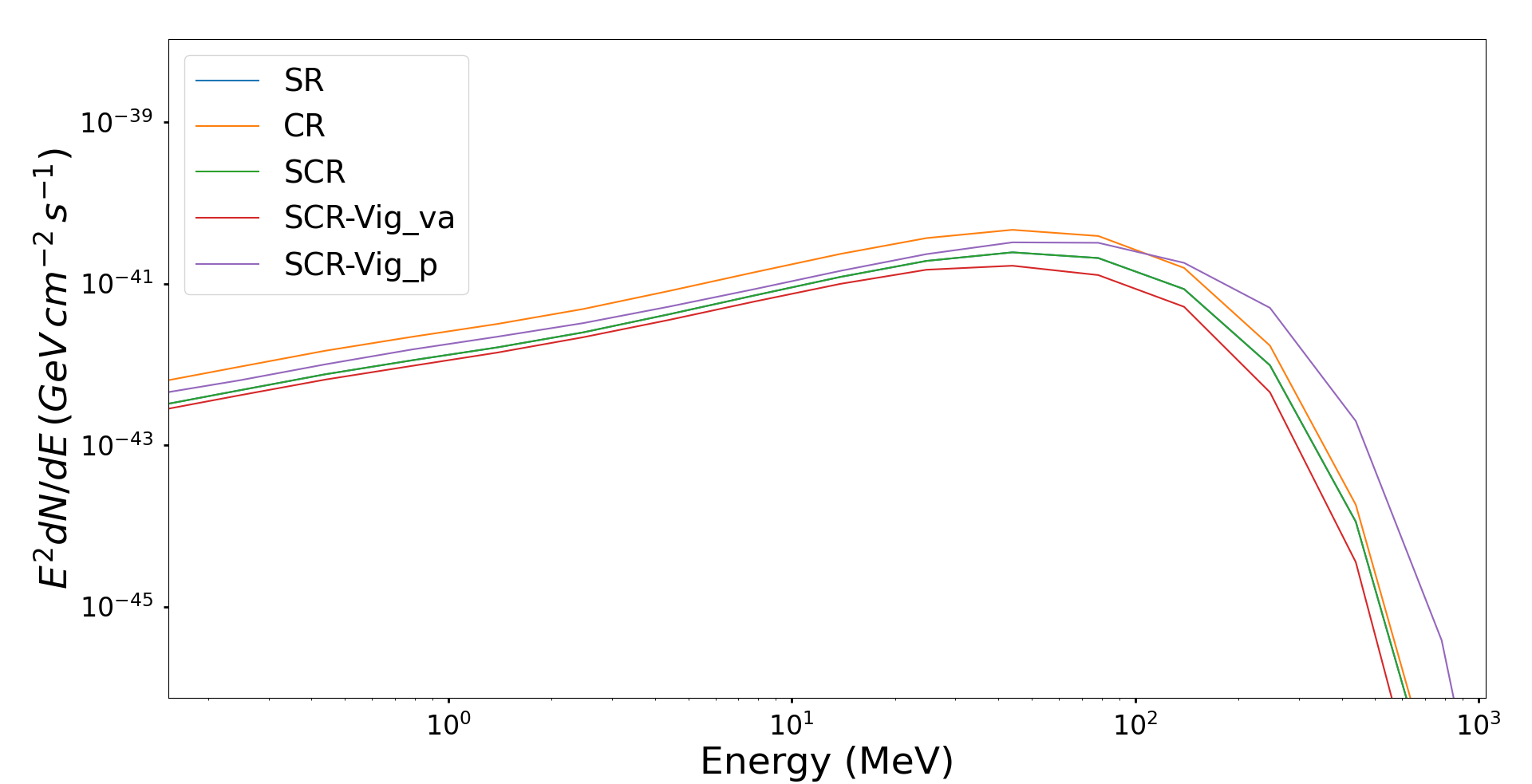}
\caption{The single-particle spectra for the same case as in Figure~\ref{SCR_SR_EB_case}, but excluding the $E_{\perp}$-field. We have plotted the standard CR spectrum in orange and the standard SR spectrum in blue. The SCR spectrum for \citetalias{Cerutti2016} is plotted in green (overlapping the blue) and the \citetalias{Vigano2015} SCR spectrum is plotted in red using $\theta_{\rm  VA}$ and the $\rho_{\rm c}$ calculated from the AE trajectory.}
\label{SCR_SR_no_E_case}
\end{figure}

Since the \citet{Cheng1996} and by extension the \citetalias{Vigano2015} SCR calculations are done in the absence of an $E$-field, we also plotted the spectra for the same case as previously, but excluding $E_{\perp}$. In Figure~\ref{SCR_SR_no_E_case}, we plot the SR spectrum in blue, CR spectrum in orange, and the \citetalias{Cerutti2016} SCR spectrum in green. For the \citetalias{Vigano2015} SCR spectrum, we have used $\theta_{\rm VA}$ and the AE $\rho_{\rm c}$ plotted in red, and then using $\theta_{\rm p}$ and our model $\rho_{\rm c}$ plotted in purple. Here we see that the SR and \citetalias{Cerutti2016} SCR spectra overlap completely, but both \citetalias{Vigano2015} spectra deviate from the SR spectra. Using $\theta_{\rm  VA}$ for the calculations, the spectrum does seem to be closer to the SR spectrum. In all three cases, if we used $\theta_{\rm VA}$ and our $\rho_{\rm c}$ instead of the AE $\rho_{\rm c}$ to calculate the \citetalias{Vigano2015} SCR spectrum, we obtained very similar results. The biggest factor was using $\theta_{\rm VA}$ instead of $\theta_{\rm p}$.

To probe these two SCR methods further, we calculated the total power radiated via SCR for each method and compared it to the total radiated power as calculated with the RRF, namely $\mathbf{v}\cdot\mathbf{F}_{\rm RRF}$, where $\mathbf{F}_{\rm RRF}$ is the RRF\footnote{For more details, see \citetalias{DuPlessis2024}.}. To calculate the total power radiated for the \citetalias{Vigano2015} model, we use Equation~(\ref{SCR_pow_Vig}) and in the \citetalias{Cerutti2016} model, we use Equation~(\ref{pow_SR}) and replace $B\sin\theta_{\rm p}$ with $\tilde{B}_{\perp}$. The \citetalias{Vigano2015} model using $\theta_{\rm VA}$ and the AE $\rho_{\rm c}$ is labelled as `Vigano-AE', and using $\theta_{\rm p}$ and our model $\rho_{\rm c}$ is labelled as `Vigano-p'. In Table~\ref{Loss_table} for the SR case with no $E_{\perp}$-field, we see that the \citetalias{Cerutti2016} SCR is much closer to the RRF power radiated than both methods of the \citetalias{Vigano2015} SCR. For the SR case with an $E_{\perp}$-field, we see that the \citetalias{Cerutti2016} SCR and \citetalias{Vigano2015} SCR using $\theta_{\rm VA}$ are almost equally close to the RRF radiated power. Finally, in the CR case, the \citetalias{Cerutti2016} SCR is closer to the RRF power than the \citetalias{Vigano2015} SCR using $\theta_{\rm VA}$, with the \citetalias{Vigano2015} SCR using $\theta_{\rm p}$ being very inaccurate. In the CR cases, we found the SCR relative errors to be a bit high, but found them to decrease when using lower fields. This could be due to the approximations used in both methods of the SCR calculations or it could be due to the fact that in the higher fields, the second term of the RRF starts to become more relevant. Notably, when adding larger $E_{\parallel}$-fields, the \citetalias{Vigano2015} SCR results, even when using $\theta_{\rm VA}$, became much more inaccurate\footnote{This could be due to the assumptions used to derive the equations, the frame equivalence discussed in Section~\ref{sec:2.5} or the larger stepwise $E_{\parallel}$-field, kicking the particle out of equilibrium, deviating from the AE velocity and therefore increasing $\theta_{\rm VA}$.}, as is seen in Figure~\ref{Spec_plot} that is discussed in Section~\ref{sec:3.6}. Through our investigation and implementation of the two SCR models, we found the \citetalias{Cerutti2016} method to be more accurate and reliable. It also has a much lower computational cost to calculate than the \citetalias{Vigano2015} method, thus for our SCR emission maps in the next section and for our future modelling, we will be using the \citetalias{Cerutti2016} method to calculate our SCR spectra.      

\begin{table}
\begin{center}
\begin{tabular}{||c|c|c|c||} 
 \hline
  & Relative error: & Relative error: & Relative error: \\ [0.5ex] 
 Case & Cerutti/Kelner & Vigano-AE & Vigano-p \\ [0.5ex] 
 \hline\hline
 SR no $E_{\perp}$ & 0.022 & 0.25 & 0.43 \\ 
 \hline
 SR & 0.019 & 0.023 & 0.77\\
 \hline
 CR  & 0.43 & 0.92 & 59.1\\
 \hline
\end{tabular}
\caption{\label{Loss_table} The relative error in total power radiated via SCR, vs the total power radiated calculated using the RRF. Each row shows each of the cases, namely SR with no $E_{\perp}$-field, SR with an $E_{\perp}$-field, and CR. The columns indicate the SCR model used, namely \citetalias{Cerutti2016}, \citetalias{Vigano2015} using $\theta_{\rm VA}$ and the AE $\rho_{\rm c}$, and \citetalias{Vigano2015} using $\theta_{\rm p}$ and our model $\rho_{\rm c}$.}
\end{center}
\end{table}

\subsection{Emission maps} \label{sec:3.5}
\begin{figure}
\centering
\includegraphics[width=.5\textwidth]{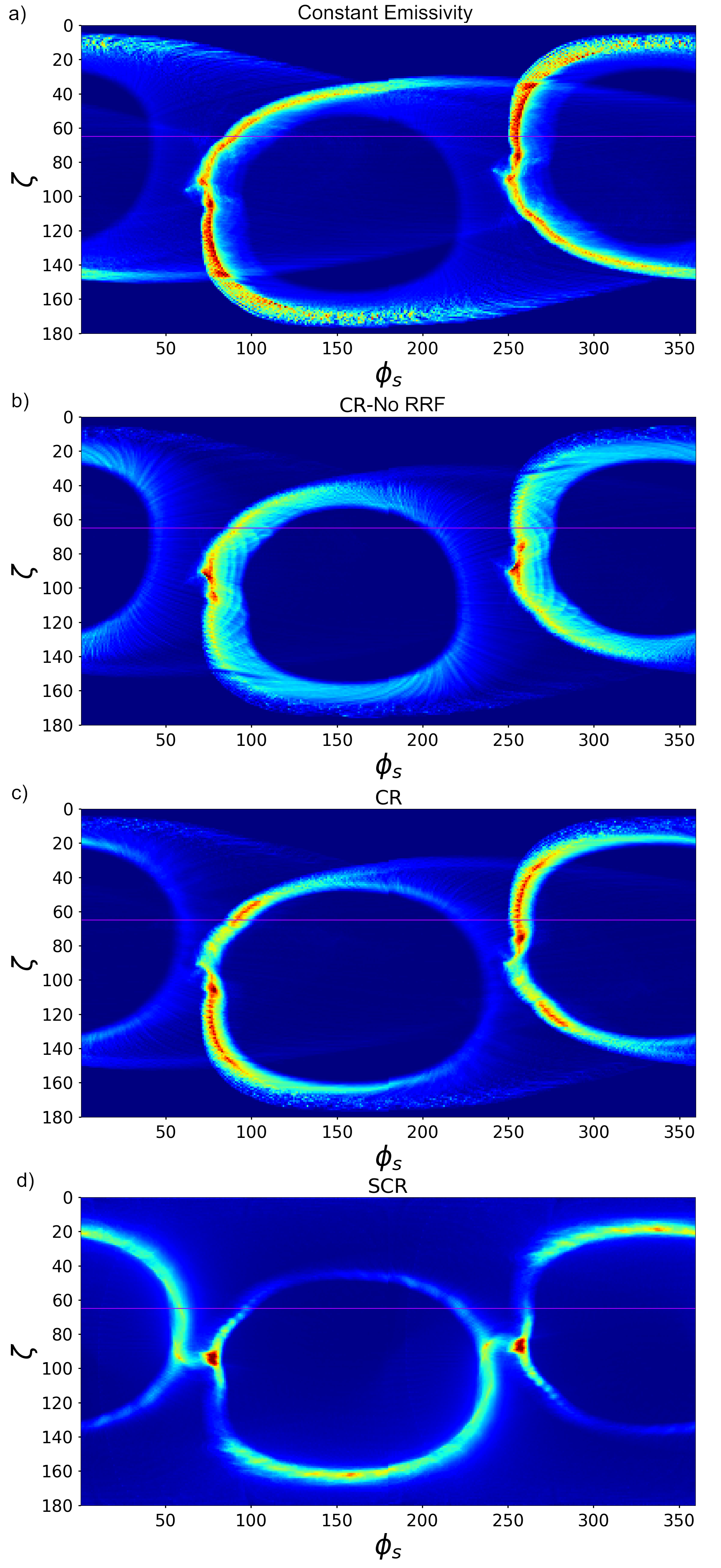}
\caption{CR emission maps produced by our model for the Vela-like $B_{\rm S} =8\times 10^{11} \, \rm{G}$ case discussed in Section~\ref{sec:3.2} using a slot gap between $r_{\rm ovc} = 0.9$ and $r_{\rm ovc} = 0.96$. In panel a) we use a constant emissivity, thus only scaling the emission with the particle step length. In panel b) we use the standard CR calculations, implementing the $\rho_{\rm c}$ from the \citetalias{Barnard2022} model and excluding the RRF, whereas in panel c) we have included the RRF. In panel d) we show the CR emission maps with included RRF and using our model $\rho_{\rm c}$.}
\label{CR_emission_maps}
\end{figure}

For the emission maps in Figure~\ref{CR_emission_maps}, we use the Vela-like $B_{\rm S} =8\times 10^{11} \, \rm{G}$ case discussed in Section~\ref{sec:3.2} with magnetic inclination angle $\alpha = 75^{\circ}$. We used the photon energy range $1~\rm{MeV} - 50~\rm{GeV}$, where the lower limit is lower than that used in Figure~\ref{Al_B12_skymap}, meaning we will see fainter, lower-energy emission. \footnote{Similar to the \citetalias{Harding2015, Harding2021, Barnard2022} models, we only calculate the northern hemisphere to save computational time, thus flipping and shifting the emission for the southern hemisphere's emission.}
We start by showing the CR emission maps, since these are the easiest to use for calibration with the \citetalias{Harding2015, Barnard2022} model results, because the CR is less affected by the $\theta_{\rm p}$ vs $\theta_{\rm  VA}$ effects mentioned in the previous sections. We use the same energy range and $\delta r_{\rm ovc}$ as Figure~\ref{Al_B12_skymap} panel~b), but instead of using $0.5$ divisions per degree in $\zeta$ and $\phi$, we use one division per degree to increase the resolution without the massive computational increase of extra active region rings on the PC. To produce the $B_{\rm S} =8\times 10^{11} \, \rm{G}$ emission map in Figure~\ref{Al_B12_skymap} panel b), we needed to use $R_{\rm acc}^{\rm min}=4.0\times 10^{-2} \,\rm{cm}^{-1}$ and $R_{\rm acc}^{\rm max}=2.5\times 10^{-1} \, \rm{cm}^{-1}$ to get the caustics, thus the same $E_{\parallel}$-field of the $B_{\rm S} =8\times 10^{12} \, \rm{G}$ case is used to achieve large enough $\gamma$ values for the CR in the \citetalias{Harding2015, Barnard2022} models. As mentioned in the previous sections, these $E_{\parallel}$-fields are excluded in the \citetalias{Harding2015, Harding2021, Barnard2022} models' trajectory calculations, but do affect our general equations of motion. Thus, we lowered the acceleration rates so that these $E_{\parallel}$-fields do not affect the trajectories and still have similar trajectories to those in the \citetalias{Barnard2022} model. Therefore, as discussed in Section~\ref{sec:3.2}, we use $R_{\rm acc}^{\rm min}=4.0\times 10^{-3} \,\rm{cm}^{-1}$ and $R_{\rm acc}^{\rm max}=2.5\times 10^{-2} \, \rm{cm}^{-1}$ when excluding the RRF, and $R_{\rm acc}^{\rm min}=4.0\times 10^{-4} \,\rm{cm}^{-1}$ and $R_{\rm acc}^{\rm max}=2.5\times 10^{-3} \, \rm{cm}^{-1}$ when including the RRF. The main goal of the emission map calibration is to assess reproduction of the same caustics as shown in Figure~\ref{Al_B12_skymap} panel~b). 

In Figure~\ref{CR_emission_maps} we use the active region between $r_{\rm ovc} = 0.9$ and $r_{\rm ovc} = 0.96$, with $\delta r_{\rm ovc} = 0.01$ (a slot gap of width $\sim$10\% of the PC angle). For the emission map in panel a) we plot a constant emissivity, scaling the radiation flux with the particle step length along the trajectory, while in panel b) we plot the CR map with excluded RRF and using the $\rho_{\rm c}$ from the \citetalias{Barnard2022} model. In panel c) we plot the CR map including RRF and using the $\rho_{\rm c}$ from the \citetalias{Barnard2022} model, while in panel d) we plot the CR maps including RRF and using our model $\rho_{\rm c}$. We investigated using the \citetalias{Barnard2022} model $\rho_{\rm c}$, since as mentioned in Section~\ref{sec:3.2} our $\rho_{\rm c}$ is initially lower than their $\rho_{\rm c}$ because our model yields the $\rho_{\rm eff}$ due to resolving the particle gyrations and their model yields the AE gyro-centric $\rho_{\rm c}$. 

Comparing the caustics in panel a) of Figure~\ref{CR_emission_maps} for the constant emissivity to the CR caustics in Figure~\ref{Al_B12_skymap}, we see that they look very similar and appear in the same place on the emission maps. One also sees a higher emission region in the caustic due to the PC notch at $\zeta \sim 90^{\circ}$ and $\phi = 250^{\circ}$ when compared to the other emission maps. The constant-emissivity caustics in Figure~\ref{CR_emission_maps} are seen to have higher overall intensities and extend further into the arms of the caustics than in Figure~\ref{Al_B12_skymap} panel~b). Looking at panels b) of Figure~\ref{CR_emission_maps}, we see that the caustics are very close to those in Figure~\ref{Al_B12_skymap}, where the caustics and the notch seem to be in the correct position on the emission map. In panel c) of Figure~\ref{CR_emission_maps}, we see that the caustics also appear very similar to those in Figure~\ref{Al_B12_skymap}, but there seems to be a little higher intensity extending into the arms of the caustic. In panel d) of Figure~\ref{CR_emission_maps}, we do see most of the expected caustics, but there seems to be some extended emission in the top arm of the caustic. Most of the intensity is also present at the position where the particle is initialised, thus we believe that these effects are possibly due to our $\rho_{\rm c}$ initially being smaller than that of the \citetalias{Harding2015} model. What is interesting is that we still see the correct positions of the caustics and the notch emission region. 

Looking at Figure~\ref{CR_map_own} in the Appendix where we use our own $\rho_{\rm c}$ and the same acceleration rates as Figure~\ref{Al_B12_skymap} panel~b), for the CR emission map, we see that the caustic looks much better. We believe this could be due to the larger $E_{\parallel}$-field accelerating the particles to higher $\gamma$ values much more quickly, producing the caustics at the correct altitude. The particle will also enter equilibrium much more quickly. The problem is the large jump in $E_{\parallel}$ and limiting $\gamma$ with $\gamma_{\rm c}$ that cause issues for our particle dynamics, as discussed in previous sections. One effect that is visible due to the large $E_{\parallel}$-fields is that looking at the bottom arm of the caustic, one sees that the emission region is extended much lower in $\zeta$ than any of the other cases. The extended emission is also dragged much more to the right of the emission map in this plot due to the $E_{\parallel}$-fields affecting the particle trajectories.  

\begin{figure}
\centering
\includegraphics[width=.5\textwidth]{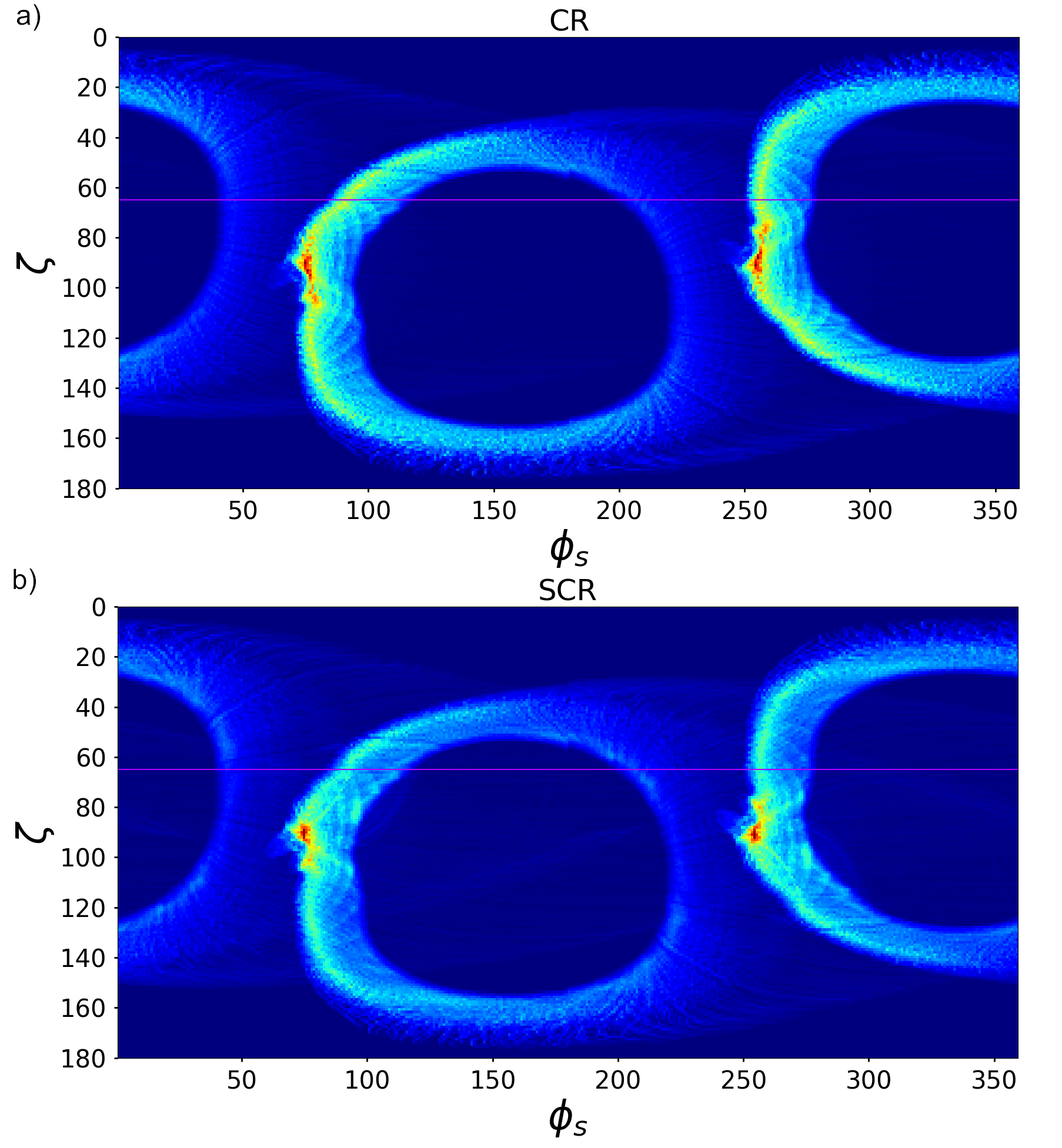}
\caption{Emission maps using our model parameters for the Vela-like $B_{\rm S} =8\times 10^{10} \, \rm{G}$ case using a slot gap between $r_{\rm ovc} = 0.9$ and $r_{\rm ovc} = 0.96$. In panel a) we plot the CR emission maps and in panel b) we plot the SCR emission maps.}  
\label{Own_emission_maps}
\end{figure}

For Figure~\ref{Own_emission_maps}, we have plotted our CR and SCR emission maps for the Vela-like $B_{\rm S} =8\times 10^{10} \, \rm{G}$ where $\alpha= 75^{\circ}$. We chose this case because we could start at lower altitudes in the magnetosphere with included RRF, without limiting $\gamma$ and using our own $\rho_{\rm c}$. Similar to the previous plot in the left column, we use the active region between $r_{\rm ovc} = 0.9$ and $r_{\rm ovc} = 0.96$, with $\delta r_{\rm ovc} = 0.01$. In panel a) of Figure~\ref{Own_emission_maps}, we show the CR emission map using our own $\rho_{\rm c}$ and in panel b), we show the SCR map using the method of the \citetalias{Cerutti2016} model. The emission maps look relatively similar, with minor deviations in the notch emission region and some of the extended emission. All caustics and notches appear in the correct positions and look very similar to the caustics in Figure~\ref{Al_B12_skymap}. 
%\LD{What is clearly visible is the difference in our SR emission maps vs the ones shown in Figures \ref{Al_B12_skymap} and \ref{Al_B11_skymap}. This is due to the discrepancy in $\theta_{\rm p}$ with the \citetalias{Harding2015, Harding2021, Barnard2022} models, as mentioned in the previous sections, since their $\theta_{\rm p}$ starts small and then decreases rapidly, meaning all the SR emission occurs very close to the stellar surface. This is in contrast to our $\theta_{\rm p}$ that increases with altitude, causing higher SR emission in the extended magnetosphere where $\theta_{\rm p}$ is larger. Thus this is why SCR should be implemented following the AE or the correct particle trajectory to calculate the correct mixture of CR and SR using the relevant $\theta_{\rm VA}$ or gyrating angle.}
Notably, the Vela SCR emission map in \citetalias{Harding2021} looks similar to the CR emission maps for Vela in \citetalias{Barnard2022}. Thus the fact that our SCR is similar to our CR is very encouraging since the SCR should be in the CR regime in this scenario. Due to the problems using $\theta_{\rm p}$ vs $\theta_{\rm VA}$ for the SR emission maps, we have omitted them from this work but these plots can be found in Du Plessis (2024b). We show the impact of using $\theta_{p}$ instead of $\theta_{\rm VA}$ for these high $E_{\perp}$ cases in the spectra of Figure~\ref{Spec_plot}.   

\subsection{Spectra} \label{sec:3.6}
\begin{figure*}
\centering
\includegraphics[width=.9\textwidth]{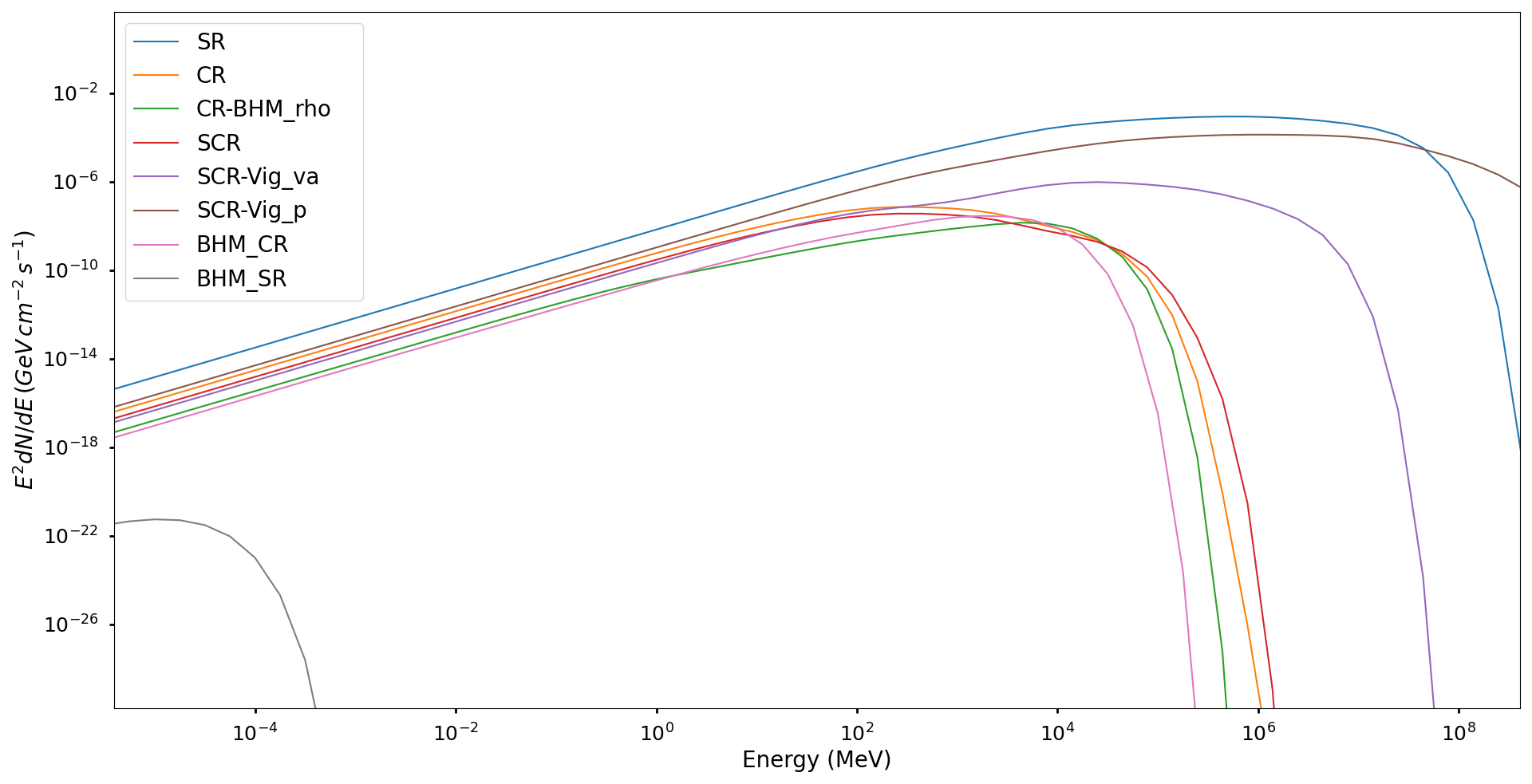}
\caption{Spectra for the Vela-like $B_{\rm S} =8\times 10^{11} \, \rm{G}$ case using $R_{\rm acc}^{\rm min}=4.0\times 10^{-2} \,\rm{cm}^{-1}$ and $R_{\rm acc}^{\rm max}=2.5\times 10^{-1} \, \rm{cm}^{-1}$. The spectra in this figure are produced with our model using the different radiation calculations from Section~\ref{sec:2}. The CR spectrum is plotted in orange, the SR spectrum plotted in blue, the CR spectrum using the \citetalias{Harding2015} model $\rho_{\rm c}$ is plotted in green, the \citetalias{Cerutti2016} SCR spectrum is plotted in red, the \citetalias{Vigano2015} SCR spectrum using $\theta_{\rm VA}$ and the AE $\rho_{\rm c}$ is plotted in purple, the \citetalias{Vigano2015} SCR spectrum using $\theta_{\rm p}$ and our model $\rho_{\rm c}$ is plotted in brown, the CR spectrum generated using the \citetalias{Barnard2022} model is plotted in pink, and the SR spectrum generated using the \citetalias{Barnard2022} model is plotted in grey.}  
\label{Spec_plot}
\end{figure*}

Here we present the best case we could compare our model's spectral results to those generated by the \citetalias{Harding2015, Barnard2022} model. For Figure~\ref{Spec_plot}, we show the various spectra for the Vela-like $B_{S} =8\times 10^{11} \, \rm{G}$ case discussed in Section~\ref{sec:3.2} using $R_{\rm acc}^{\rm min}=4.0\times 10^{-2} \,\rm{cm}^{-1}$, $R_{\rm acc}^{\rm max}=2.5\times 10^{-1} \, \rm{cm}^{-1}$, and using the slot gap between $r_{\rm ovc} = 0.9$ and $r_{\rm ovc} = 0.96$. As mentioned in the previous Section, we needed to use these higher acceleration rates for the \citetalias{Harding2015, Barnard2022} model to attain large enough $\gamma$ values, otherwise the caustics would not appear in the emission maps and the spectra would be very low. The only option we were thus left with was to compare spectra with the \citetalias{Harding2015, Barnard2022} model was the Vela-like $B_{\rm S} =8\times 10^{11} \, \rm{G}$ case with higher acceleration rates. 
%For the included spectra 
We plot SR in blue, CR in orange, CR using $\rho_{\rm c}$ from the \citetalias{Barnard2022} model in green, SCR using the \citetalias{Cerutti2016} model in red, SCR from the \citetalias{Vigano2015} model using $\theta_{\rm VA}$, and the AE $\rho_c$ in purple, SCR from the \citetalias{Vigano2015} model using $\theta_{\rm p}$ and our model $\rho_{\rm c}$ in brown, the CR spectrum produced by the \citetalias{Barnard2022} model in pink, and the SR spectrum produced by the \citetalias{Barnard2022} model in grey. In this figure, we see our model's SR spectrum is very high due to the large $\theta_{\rm p}$, illustrating why the
$\mathbf{E}\times\mathbf{B}$ considerations are important when large $E_{\perp}$-fields are present. As mentioned for Figure~\ref{SCR_CR_case}, due to the large effect the $E_{\perp}$-field has on the particle's trajectory, oscillating the particle parameters as seen in \citetalias{DuPlessis2024}, and causing a large $\theta_{\rm p}$ it leads to unphysical results when calculating the standard SR when encountering large fields and $\gamma$ values. This leads to photon energies in the spectrum cutoff that are beyond the particle energy. On the other hand, one sees that the \citetalias{Barnard2022} SR is very low due to the small $\theta_{\rm p}$ in their calculations. The Figure $y$-axis extends to reasonably low fluxes due to including the \citetalias{Barnard2022} model SR. The \citetalias{Vigano2015} SCR spectrum using $\theta_{\rm p}$ is also found to be very high due to the large $\theta_{\rm p}$ from the equations of motion and neglecting the $\mathbf{E}\times\mathbf{B}$ effects on the particle. Interestingly, our CR spectrum is found to be close to the \citetalias{Cerutti2016} SCR spectrum and both are found to be somewhat close to the \citetalias{Harding2015} model CR spectrum, but extending a bit further into the higher energies. The \citetalias{Vigano2015} SCR spectrum using $\theta_{\rm  VA}$ is found to be quite a bit higher than the \citetalias{Cerutti2016} SCR. We found the \citetalias{Vigano2015} SCR model to deviate more when including the higher $E_{\parallel}$-fields than the test cases shown in Section~\ref{sec:3.4}. %Looking at 
The green CR spectrum using the \citetalias{Barnard2022} $\rho_{\rm c}$
%, we see that it 
is reasonably close to the pink CR spectrum. The only difference is that it is slightly lower and extends to slightly higher photon energies. 
%Looking at 
All the spectra produced with our model
%, they 
exhibit a hump feature. This could be due to the two-step accelerating $E_{\parallel}$-fields as mentioned, since there are two distinct acceleration regions with largely different acceleration rates. The reason this feature is not present in the lower $E_{\parallel}$-field cases as shown in Figure~\ref{SCR_CR_case} is due to the acceleration rates being much lower. As seen from the lower $E_{\parallel}$-field cases' trajectories, there is much less of an abrupt jump in parameters namely $\gamma$ and $\rho_{\rm c}$ when the step increase in $E_{\parallel}$ is smaller compared to the local $E_{\perp}$-field and $B$-field. Another potential effect that can not be ruled out is the problems we encounter with our high-field cases when limiting $\gamma$ with $\gamma_{\rm c}$. 
%This is highlighted by the higher photon energy cutoff and higher energy spectral peak for the orange, red and green curves in Figure~\ref{Spec_plot}. This was also only noticed for the cases that we used higher $E_{\parallel}$-field values that also cause the mentioned double hump in the spectra. These higher energy peaks and cutoffs are also visible in \citetalias{Barnard2022} when using higher $E_{\parallel}$-fields. Thus ideally we should use a smooth or at least constant $E_{\parallel}$ and avoid cases that require limiting $\gamma$, preferably implementing QED RRF equations since part of the problem stems from exceeding the Schwinger limit. 

\section{Discussion and Conclusions}\label{sec:4}
In this work, we compare the results from our gyro-resolved emission code against those of the gyro-centric models of \citetalias{Harding2015, Harding2021, Barnard2022} using a Vela-like pulsar's high-energy emission maps and spectra. Furthermore, we tested the convergence of these results to the AE radiation-reaction limit. Additionally, we also investigated the effect of a large $E_{\perp}$-field on the trajectories and radiation calculations. 

We found that our particle trajectories, emission phase corrections, radiation calculations, emission maps, and spectra are similar to the results produced by the \citetalias{Barnard2022} model for a pulsar with $10\%$ of the surface $B$-field of Vela, and under certain conditions in extreme field cases. We show that even though there are differences in the phase-resolved vs gyro-centric approaches, we can reproduce the results of the gyro-centric model of \citetalias{Harding2015, Harding2021, Barnard2022} while also identifying a preferred SCR model for our use cases in future modelling. This work therefore validates the \citetalias{Harding2021} SCR approach following the AE trajectory in the limit of a small general pitch angle.

For the Vela-like $B_{\rm S} =8\times 10^{11} \, \rm{G}$ case with lower $E_{\parallel}$-fields, both our particle trajectory and direction agreed very well with those of the \citetalias{Barnard2022} model results, implying that our corrected observer phase also agreed very well. We highlighted the difference in the \citetalias{Harding2015, Harding2021, Barnard2022} model trajectory's $\theta_{\rm p}$ vs that used in their transport equation, which is the general pitch angle $\theta_{\rm VA}$. Our trajectory results aligned with the large $\theta_{\rm p}$ from their trajectory calculations. This is expected, since the $\mathbf{E}\times \mathbf{B}$ drift component dominates in the latter part of the extended magnetosphere as shown in Figure~\ref{FF_Emission_phase_8e11} panel b) for both our models. The RRF also does not decrease the $\theta_{\rm p}$ of the particle due to the $\gamma^{2}$ leading term being directly opposite to the particle velocity. 

If there is a large $E_{\perp}$-field relative to the local $B$-field, one is specifically interested in the angle of the particle gyrating around the $\mathbf{E}\times \mathbf{B}$ curve for the radiation calculations. This curve is postulated to be the AE trajectory, thus one is interested in the angle between the AE velocity and the particles' velocity, rather than the traditional pitch angle as discussed in detail in \citet{Kalapotharakos2019}. 
%Since Equation~(\ref{Alice_transport}) does not account for the $\mathbf{E}\times \mathbf{B}$ effect on the particle and assumes a small pitch angle in the derivation, the question arises why the SCR spectrum for Vela and other pulsars in \citetalias{Harding2021} converges to their CR spectrum when using their methods from \citetalias{Harding2015}.
%the question is if in \citetalias{Harding2021} they are using the AE trajectory parameters for the SCR calculations. 
%We believe this is due to them following the AE trajectory, emulating a small $\theta_{\rm VA}$ that converges to the AE trajectory and using the AE $\rho_{\rm c}$ for their SCR calculations as highlighted in Figure~\ref{SCR_rho_c}. Thus their $p_{\perp}$ is equivalent to the AE $p_{\perp}$. However, they are using an arbitrary small $\theta_{\rm VA}$ instead of calculating the angle between the particle velocity and the AE velocity.} 

We found our model $\rho_{\rm c}$ to converge but oscillate around the \citetalias{Barnard2022} model's $\rho_{\rm c}$ in the latter part of the extended magnetosphere, although it is lower in the initial low-altitude magnetosphere region. This is due to the \citetalias{Harding2015, Harding2021, Barnard2022} models using the AE $\rho_{\rm c}$ and our model yielding $\rho_{\rm eff}$, since we are resolving the particle gyrations. Thus, closer to the stellar surface where the gyro-radius is much smaller, our results yield a smaller $\rho_{\rm c}$. This was important to take into consideration for the CR calculations when comparing emission maps and spectra. This is why we also simulated the results using the \citetalias{Barnard2022} $\rho_{\rm c}$ in our calculations.

We found that limiting our $\gamma$ with $\gamma_{\rm c}$ does allow us to model the Vela-like $B_{\rm S} =8\times 10^{11} \, \rm{G}$ case with higher $E_{\parallel}$-fields to accelerate the particle to higher $\gamma$ values needed for CR. Unfortunately, upon further investigation, we found that limiting $\gamma$ causes problems in the velocity of the particle as seen in Figure~\ref{FF_Emission_phase_8e11_CRR} panel b). Our velocity was found to oscillate unphysically beyond $c$ due to limiting the $\gamma$ value. This is likely due to re-normalising the $\gamma$ value of our equation of motion between each step at this limit, causing numerical instabilities/inaccuracies when having to change the size of each velocity component during integration. We only found this to occur in the regions where $\gamma$ is limited. We thus caution other codes that limit their $\gamma$ values or other parameters, e.g. the RRF, during integration to assess if this causes numerical inaccuracies or features. This is mentioned since some PIC models are indeed limiting their $\gamma$ to a constant value during integration to avoid runaway solutions in extreme RRF cases \citep{Timokhin2024}, or they artificially scale the RRF down close to the stellar surface \citep{Soudais2024}, which may be problematic. 
Similar to our own results, limiting $\gamma$ does avoid numerical runaway of solutions due to entering the non-classical RRF regime but this causes other problems in one's results. The numerical impact of artificially scaling the RRF in PIC codes may also be problematic. These PIC simulations are typically run in low-field scenarios and then linearly scaled to emulate higher-field scenarios. %This is beyond the fact that 
There is indeed a difference in the particle dynamics in the high-field extreme RRF scenarios, where the particle is in the radiation-reaction limit regime, vs the low-field scenarios as seen in our AE comparison results and pointed out by \citet{Petri2023}. The reason that the \citetalias{Harding2015, Harding2021, Barnard2022} models' velocities are unaffected by the $\gamma_{\rm c}$ limit is that the particle trajectories are not influenced by the $\gamma$ value. %As mentioned, using the $\gamma_{\rm c}$ assumption or 
Moreover, limiting $\gamma$ by $\gamma_{\rm c}$ does not constrain the results to the classical RRF regime. The \citetalias{Harding2015, Harding2021, Barnard2022} models thus have no mechanism to test or constrain the results to the classical regime, causing us difficulty initialising our particle position and energy with the results from their $B_{\rm S} =8\times 10^{11} \, \rm{G}$ and $B_{\rm S} =8\times 10^{12} \, \rm{G}$ cases.

Upon comparing our model velocities to the local AE velocities of \citet{Gruzinov2012} and \citetalias{Kelner2015} in Figure~\ref{FF_AE_vel_8e11}, we found that our Vela-like $B_{\rm S} =8\times 10^{11} \, \rm{G}$ case with lower $E_{\parallel}$-fields converged very well. This thus yielded a small $\theta_{\rm  VA}$ angle between the two velocities (the actual particle velocity and the limiting one of AE). In the cases using larger $E_{\parallel}$-fields, we found that at $1.0R_{\rm LC}$ the particle deviated from the AE velocity and was kicked out of equilibrium due to the sharp increase in $E_{\parallel}$ there. We noticed this jump in other parameters at $1.0R_{\rm LC}$ as well. In future studies, we will use smooth $E_{\parallel}$-fields, but here we used the same $E_{\parallel}$-dependence to be able to reproduce the \citetalias{Barnard2022} model results. As mentioned in Section~\ref{sec:2}, the $E_{\parallel}$-fields are not included in their trajectory calculations. Thus we found that when using larger $E_{\parallel}$-fields, our particle trajectory deviated from their model trajectories. We therefore had to use lower $E_{\parallel}$-fields to obtain their particle trajectories and observer emission phases, leading to lower $\gamma$ values for the radiation calculations. 
% Excluding the $E_{\parallel}$-field from the particle trajectories could thus lead to incorrect trajectories, especially if large enough accelerating fields are required to quickly accelerate the particle to high $\gamma$ values.
% When investigating the Lorentz force and RRF equilibrium, we found that in the lower-field cases, the particles were not in equilibrium at $2.0R_{\rm LC}$. In the high-field cases, the particles however reached or were close to equilibrium at $1.6R_{\rm LC} - 2.0R_{\rm LC}$. As required by the AE assumptions, one needs a large enough $E_{\parallel}$-field to quickly reach equilibrium for the AE equations to hold. The particle should also not be kicked out of equilibrium, since AE describes an equilibrium state. 
What we thus found is that a large enough $E_{\parallel}$ is required to accelerate the particles before they reach the current sheet for AE to hold. But this $E_{\parallel}$ would affect the trajectories in Equation~(\ref{Alice_traj}) if $E_{\parallel}$ were included in the trajectory calculations. Close to the stellar surface, such a large $E_{\parallel}$ could also cause the particle to be accelerated into the non-classical RRF regime as well as cause a deviation from the FF-field solutions expected to hold inside the magnetosphere. Ideally, one would have to include a more physical modelling of these $E_{\parallel}$-fields similar to the dissipative models mentioned in Section \ref{sec:2.3}. 

When comparing our $\rho_{\rm c}$ to $\rho_{\rm eff}$ from the \citetalias{Kelner2015} model, we see that the results align almost completely, thus justifying the conclusion that our $\rho_{\rm c}$ yields $\rho_{\rm eff}$ due to resolving the particle gyrations. We see that our results align in all cases, even when the particle is not in equilibrium.
% , in agreement with the AE description used by \citetalias{Kelner2015}. 
This is an important result for our future modelling of AR Sco, since we found that when investigating a magnetic mirror scenario in Figure~\ref{AE_vel_mirror}, the AE velocities in Equation~(\ref{AE}) only agreed when the particle was moving outward, not inward as well. This is most likely due to the assumptions of outflowing ligth-like particles with $v=c$. Therefore, those equations are not applicable for a magnetic mirror scenario, but our $\rho_{\rm eff}$ agreed completely with that from \citetalias{Kelner2015} in this scenario, as seen in Figure~\ref{AE_rho_c_mirror}. Thus the \citetalias{Kelner2015} model seems to be reliable for a magnetic mirror scenario as well.

When investigating the $\rho_{\rm eff}$ from the SCR models of \citetalias{Cerutti2016} and \citetalias{Vigano2015}, we found the \citetalias{Cerutti2016} $\rho_{\rm eff}$ to agree with our model $\rho_{\rm c}$, where all these results oscillate around the AE $\rho_{\rm c}$ of \citetalias{Barnard2022}. The \citetalias{Vigano2015} value mostly agreed with our $\rho_{\rm c}$ when using $\theta_{\rm  VA}$ but when using the AE $\rho_{\rm c}$ as well for the $\rho_{\rm eff}$ calculation their value stopped oscillating as much and agreed more with the smooth \citetalias{Barnard2022} AE $\rho_{\rm c}$. However, the \citetalias{Vigano2015} $\rho_{\rm eff}$ deviated significantly when using $\theta_{\rm p}$ instead of the $\theta_{\rm VA}$. Upon testing CR and SR scenarios, we found the \citetalias{Cerutti2016} SCR to be overall more reliable concerning the spectral results and total SCR power when compared to the total power due to the RRF. This included the case with no $E_{\perp}$-field, where we expected the \citetalias{Vigano2015} SCR method to perform better. We found the \citetalias{Vigano2015} SCR spectrum to be inaccurate in the higher $E_{\parallel}$-field case, shown in Figure~\ref{Spec_plot}. Thus for future work, we will be using the \citetalias{Kelner2015, Cerutti2016} SCR method due to it being more reliable and much more computationally inexpensive to calculate, especially since one has to calculate the extra AE frame parameters, at a minimum $\theta_{\rm  VA}$, for the \citetalias{Vigano2015} model.

Comparing our CR spectra in Figure~\ref{CR_emission_maps} to those produced by the \citetalias{Barnard2022} model in Figure~\ref{Al_B12_skymap}, we find that when using a constant emissivity and the \citetalias{Barnard2022} model $\rho_{\rm c}$, our emission maps reproduce theirs very well. When using our model's $\rho_{\rm c}$, we found a similar shape for the caustic, but more emission at lower altitudes in the magnetosphere due to our smaller $\rho_{\rm c}$ values at the lower altitudes. The CR emission map in Figure~\ref{CR_map_own} with higher $E_{\parallel}$-fields does seem to produce a more similar caustic, possibly due to particles being accelerated to equilibrium faster. Thus when using their model $\rho_{\rm c}$, we can reproduce their CR emission map caustics for the Vela-like $B_{\rm S} =8\times 10^{11} \, \rm{G}$ case. We do not reproduce the Vela $B_{\rm S} =8\times 10^{12} \, \rm{G}$ case due to the non-classical RRF problems as well as the effects of limiting $\gamma$ to avoid runaway solutions. We are completely unable to compare the \citetalias{Barnard2022} model's standard SR spectrum with our standard SR spectrum, due to the mentioned problems in using $\theta_{\rm p}$ for the SR when the trajectories and radiation are highly affected by the $\mathbf{E}\times\mathbf{B}$-drift, since \citetalias{Harding2015,Barnard2022} have the small deviation angle with respect to the trajectory. It is thus better to compare and investigate the SCR results. Looking at the CR\footnote{Using our model $\rho_{\rm c}$.} and SCR emission maps we produced for the Vela-like $B_{\rm S} =8\times 10^{10} \, \rm{G}$ case, we see that the caustics look very similar to the CR caustics in Figure~\ref{Al_B12_skymap}. 
%Thus, our model produces the expected caustics for CR and SCR. The problem of using the standard SR was highlighted, since $\theta_{\rm p}$ is larger in the extended magnetosphere, yielding similar emission maps to CR and SCR.

In the spectra for the Vela-like $B_{\rm S} =8\times 10^{11} \, \rm{G}$ case with higher $E_{\parallel}$-fields shown in Figure~\ref{Spec_plot}, we found that our CR spectrum using the \citetalias{Barnard2022} model $\rho_{\rm c}$ was reasonably close to the CR spectrum of the \citetalias{Barnard2022} model. 
We do not expect them to align completely due to the differences in $\gamma$, thus this result gives us confidence in our spectral calculations. We found our CR spectrum and the \citetalias{Cerutti2016} SCR spectrum to be quite close to the \citetalias{Barnard2022} CR spectrum as well. The \citetalias{Vigano2015} SCR spectra were found to be quite different in this case. Looking at the \citetalias{Barnard2022} SR spectrum, one sees how low the spectrum is due to using the small general pitch angle vs our SR spectrum that is excessively high using the standard $\theta_{\rm p}$. This re-emphasises that one can not use the standard SR formulae when a large $E_{\perp}$-field compared to the $B$-field is present.

In this work, we have thus shown that we can reproduce the \citetalias{Harding2015, Harding2021, Barnard2022} models' trajectories, CR emission maps, and CR spectra under the discussed limitations. This represents a good calibration of our gyro-resolved code vs their gyro-centric one. Our trajectories tend to and follow the AE trajectories under the expected radiation-reaction-limit conditions, giving us confidence in our results in this regime. We have illustrated why the SCR calculations following the $\mathbf{E}\times \mathbf{B}$ drift are necessary in pulsar and pulsar-like sources in which there are large $E_{\perp}$-fields, identifying the appropriate SCR method for our model use. Thus even though we used the general equations of motion with the classical RRF, we were able to reproduce the results from a gyro-centric pulsar emission model reasonably well. This serves as a very good validation of the approach in \citetalias{Harding2021} in the limit of a small general pitch angle, following the AE trajectory and using the AE parameters for the SCR. However, our work does suggest using the \citetalias{Kelner2015, Cerutti2016} SCR instead of the \citetalias{Vigano2015} SCR. 

There are still outstanding questions that need to be assessed in the \citetalias{Harding2021} model approach. 1) Equations~(\ref{Alice_transport}) were not robustly re-derived to follow the particle $\mathbf{E}\times \mathbf{B}$ curve, with the $\mathbf{E}\times\mathbf{B}$-drift effect neglected in the original derivation (with respect to the local $B$-field). Yet, their results do hold very well in the limit of a small general pitch angle that we compared to in this work. 2) The equivalence of using $B\sin\theta$ ($\sin\theta = p/p_{\perp}$) vs $\tilde{B}_{\perp}$ in the SR regime. Actually, $\tilde{B}_{\perp}$ should be used instead of $B\sin\theta$ to accurately calculate the SCR radiation, as the particle follows the AE trajectory. 
%Due to the local $B$-field values used in the calculations of \citetalias{Harding2021} should be replaced with the frame-invariant $B_{0}$ to follow the definition of $\tilde{B}_{\perp}$ and to use the AE frame values. 
This was somewhat assessed in our calculations of $\theta_{\rm eff}$ and seems to be true, but it is difficult to test a pure SR regime. 3) The \citetalias{Harding2021} model only replaces the CR and SR losses with the SCR losses in $d\gamma/dt$ of Equations~(\ref{Alice_transport}) and not in the equation for $dp_{\perp}/dt$. This leads to problems in accurately calculating $p_{\perp}$ and the general pitch angle for the SCR equations. Thus, in the limit of a small general pitch angle, we have shown that the \citetalias{Harding2021} model approach works well and has a massive computational cost advantage due to being gyro-centric. Additional assessment is therefore required to determine the accuracy of the approach for larger general pitch angles due to resonant photon absorption or other mechanisms that may increase the general pitch angle, to determine if the SR contribution to SCR\footnote{Or if using their modelled general pitch angle for the standard SR is sufficiently accurate.} is sufficiently accurate for these regimes. 

We realise the limitations of our model compared to the gyro-centric models. For example, we can not inject the particles at the surface using realistic pulsar parameters, similar to other PIC models, due to how computationally heavy this method is because of the small time scales needed to resolve the gyro-radii in the high fields. We also need to address the RRF entering the non-classical regime for realistic pulsar fields without limiting $\gamma$. Our modelling method does have the advantage of having the particle dynamics coupled to the radiation calculations. We can therefore model the microphysics and relevant plasma physics much better. 

In future, we will incorporate charged particle feedback on the fields, similar to what is done in PIC models. From \citetalias{DuPlessis2024} and our current work, our numerical integration scheme is found to be more computationally efficient and accurate for larger electromagnetic fields and high RRF regimes, in our model implementation, than the Vay symplectic integrator which is commonly used in pulsar PIC models. This is due to our higher-order numerical scheme and adaptive time steps as discussed in \citetalias{DuPlessis2024}. We believe that this is due to the much higher accuracy of the numerical scheme that we have implemented. In \citetalias{DuPlessis2024} we discuss the significant accuracy problems of the Vay symplectic integrator in high $E_{\perp}$ cases relevant to pulsar electromagnetic fields. These problems are addressed in the Higuera Cary symplectic scheme \citep{Higuera2017} and thus would be of interest for implementation by pulsar PIC codes, as done by \citet{Torres2024}.
% Our model also seems to be computationally more efficient and accurate for larger electromagnetic fields than the PIC models, due to our higher-order numerical schemes and adaptive time steps as discussed in \citetalias{DuPlessis2024}. 
We have no problem dealing with the RRF in high-field cases of $B_{\rm S} =1\times 10^{7} \, \rm{G}$ and $B_{S} =1\times 10^{8} \, \rm{G}$ where the PIC models of \citet{Soudais2024, Timokhin2024, Mottez2024} have to scale down or limit their RRF during integration to avoid numerical runaway solutions. This is shown in Figure~\ref{RD_AE_vel}, where we injected particles at the stellar surface using no restrictions or scalings, and using $B_{\rm S} =8\times 10^{8} \, \rm{G}$ and $\gamma_{0} = 10^{4}$, a scenario which is reasonably close to the Schwinger limit. The main concern for these symplectic integrators used in pulsar PIC models is simply adding the RRF to the symplectic integrator without re-deriving the Hamiltonian, thus losing the symplectic nature of the integrator and leaving a second-order accuracy scheme\footnote{This is discussed in \citetalias{DuPlessis2024}.}. As discussed in Section~\ref{sec:2.2}, we do encounter problems at $B_{\rm S} =8\times 10^{11} \, \rm{G}$ and above, where we are entering the non-classical RRF regime. This can be addressed in future work using the QED considerations for the RRF as done by \citet{Vranic2017}. We thus believe it is worth investigating if implementing higher-order adaptive schemes in PIC codes would yield as significant computational and accuracy improvement as found in our modelling. Unfortunately, one would have to implement these schemes in a PIC code to concretely determine the improvements since including the adaptive step control is non-trivial and could affect how the grid calculations and charge deposition are computed. All of the above-mentioned calibrations and comparisons give us confidence in producing SCR emission maps and spectra in future work for the white dwarf `pulsar' AR Sco\footnote{Given its lower $B$-field and especially $\gamma$ values, the environment in the source is quite far away from the Schwinger limit.} and other sources requiring general equations of motion and RRF.     

\section*{Acknowledgements}
This work is based on research supported wholly / in part by the National Research Foundation of South Africa (NRF; Grant Number 99072). The Grant holder acknowledges that opinions, findings and conclusions or recommendations expressed in any publication generated by the NRF-supported research is that of the author(s), and that the NRF accepts no liability whatsoever in this regard.
L.~D. acknowledges partial support by the NRF under award numbers PMDS22060820024 and PMDS23042496534 as well as Anu Kundu and Paul Els for helpful discussions. C.~V. is funded by the research program "New Insights into Astrophysical and Cosmology with Theoretical Models Confronting Observational Data" of the National Institute for Theoretical and Computational Sciences (NITheCS) of South Africa. Z.~W. acknowledges support by NASA under award number 80GSFC21M0002.
%%%%%%%%%%%%%%%%%%%%%%%%%%%%%%%%%%%%%%%%%%%%%%%%%%

\section*{Data Availability}
The data underlying this article will be shared on reasonable request to the corresponding author.


\bibliographystyle{mnras}
\bibliography{ARsco_refs} 

\section*{Appendix}
%\CV{Ek stem saam met die inhoud van die hoofteks, en dat jy dit mooi ingekort het. Maar in alle liefde, my eerlike opinie is dat jy die resultate in hierdie bylaag op die aangewese plekke in die hoofteks kortliks kan opsom in woorde... dit sal die artikel meer verteerbaar maak, en jy kan na jou proefskrif verwys. Eerder as wat dit nog inligting byvoeg tot 'n reeds intense en lang artikel, maak dit die lees selfs moeiliker en benadeel dit eerder jou saak / hoofargument. Ek weet jy het hard gewerk en voel dat elke nuanse belangrik is, dus skop jy bietjie hierteen, maar doen my 'n guns en dink ten minste nog een keer hard en lank hieroor voordat jy hierdie kommentaar uitvee ;o) Jy kan dalk die `mirror results' na die hoofteks skuif en die res dan net in woorde beskryf... dink maar.}
In this Appendix, we include additional plots for extra clarity and comparison with our main results. As a reminder, all these results in the appendix include RRF.

\subsection*{Vela-like FF $B_{\rm S}= 8\times 10^{11}$-case, CRR-limited.}
\begin{figure*}
\centering
\includegraphics[width=.9\textwidth]{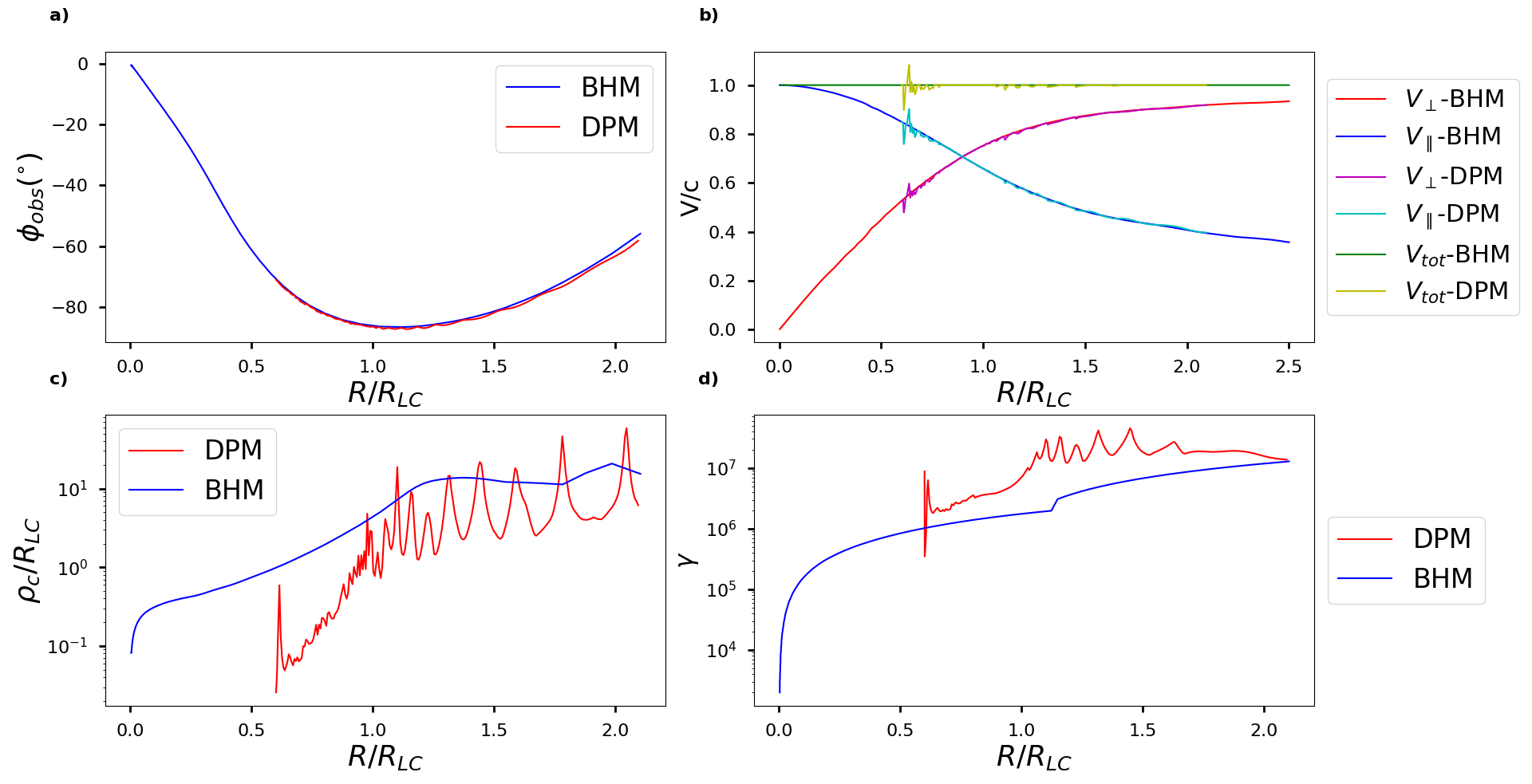}
\caption{Vela-like calibration case using $B_{\rm S} =8\times 10^{11} \, \rm{G}$, when limiting with $\gamma$ by $\gamma_{\rm c}$, showing the corrected observer emission phase in panel a), normalised particle velocity components in panel b), $\rho_{\rm c}$ in panel c), and $\gamma$ in panel d). Here DPM represents our results and BHM the \citetalias{Harding2015} model results. In panel b) we show the perpendicular and parallel velocity components with respect to the local $B$-field, as well as the total speed normalised to $c$.} 
\label{FF_Emission_phase_8e11_CRR}
\end{figure*}

Here we illustrate the effect limiting $\gamma$ during integration has on the numerical results, the effect of using a larger $E_{\parallel}$-field has on the particle trajectories\footnote{ Due to being neglected in the trajectories of the \citetalias{Harding2015,Harding2021,Barnard2022} models}, and the effect the sudden increase in $E_{\parallel}$ has on the convergence to equilibrium. In Figure~\ref{FF_Emission_phase_8e11_CRR}, we use $B_{\rm S} =8\times 10^{11} \, \rm{G}$ as discussed in Section~\ref{sec:3} and limit our model $\gamma$ by $\gamma_{\rm c}$. %These results are for comparison when limiting with $\gamma_{\rm c}$. 
In panel a), we have plotted the corrected observer phase where we see our model results in red agree with the \citetalias{Barnard2022} model results in blue. Our model's velocity component results in panel b) also agree well with those of the \citetalias{Barnard2022} model, except for the initial part of the particle trajectory. Here we see the total velocity in yellow oscillates slightly around $c$, going beyond the particle velocity limit of $c$. See Section~\ref{sec:3.2} for the discussion of this unphysical feature in the results due to limiting with $\gamma_{\rm c}$. In panel c), we see our model $\rho_{\rm c}$ in red is well below that of the \citetalias{Harding2015} model result in blue, but it starts oscillating around their $\rho_{\rm c}$ close to $2.0R_{\rm LC}$. Our $\gamma$ value in panel d) is also above that of their model, but starts to converge to their result at $2.0R_{\rm LC}$.

\begin{figure*}
\centering
\includegraphics[width=.9\textwidth]{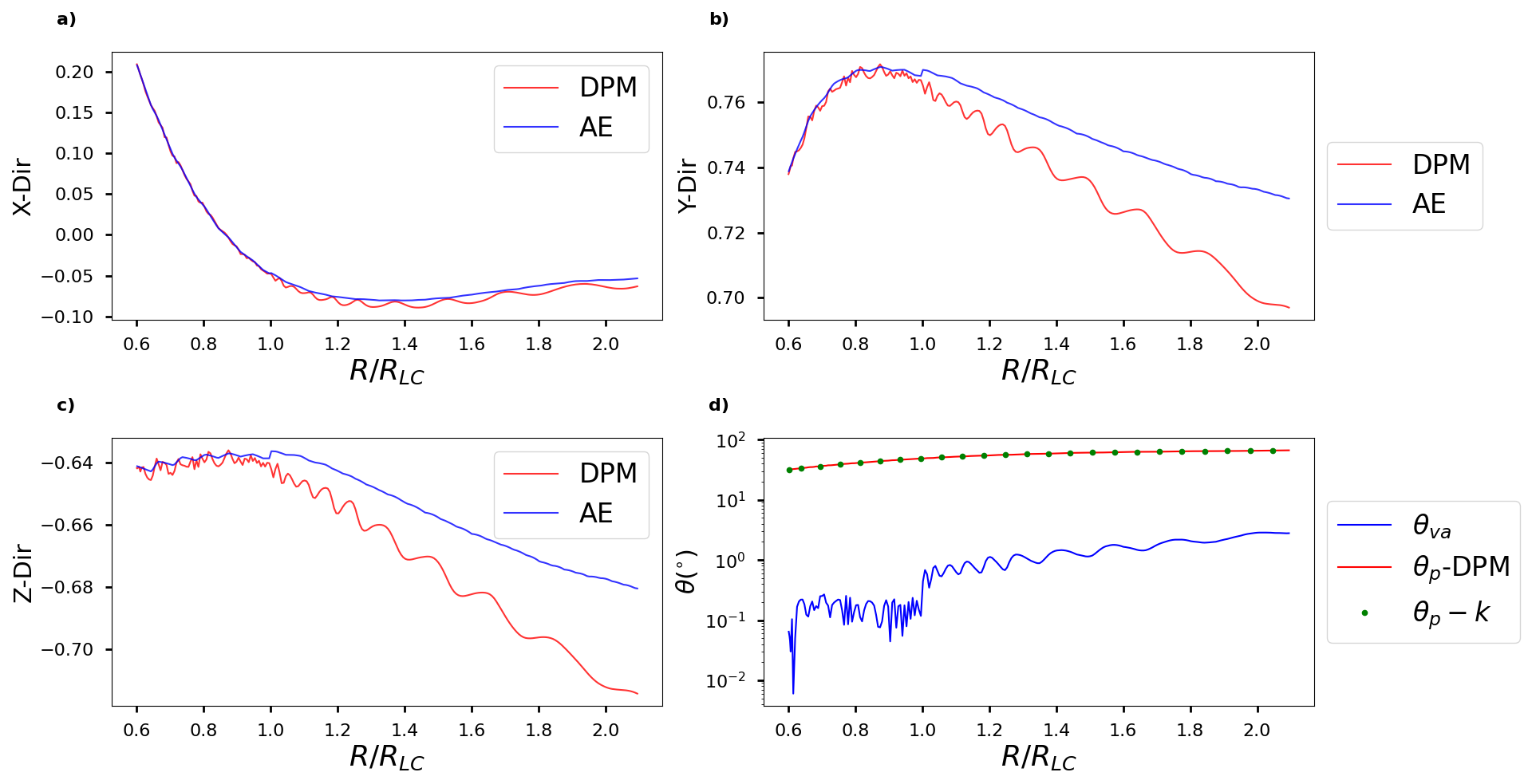}
\caption{The AE convergence results for the Vela-like case using $B_{\rm S} =8\times 10^{11} \, \rm{G}$, when limiting with $\gamma$ by $\gamma_{\rm c}$ and for a larger $E_{\parallel}$ namely $R_{\rm acc}^{\rm min} = 4.0\times 10^{-2} \rm{cm}^{-1}$ and $R_{\rm acc}^{\rm max} = 2.5\times 10^{-1} \rm{cm}^{-1}$. In this plot, DPM labels our model results, and AE those of \citet{Gruzinov2012} where these curves overlap. Panel a) shows the particle $x$-direction, panel b) the $y$-direction and panel~c) the $z$-direction. In panel~d), we show the various angles discussed in Section~\ref{sec:2}. Here $\theta_{\rm p}$-DPM overlaps with $\theta_{\rm p}$-k.}
\label{FF_AE_vel_8e11_CRR}
\end{figure*}

\begin{figure*}
\centering
\includegraphics[width=.9\textwidth]{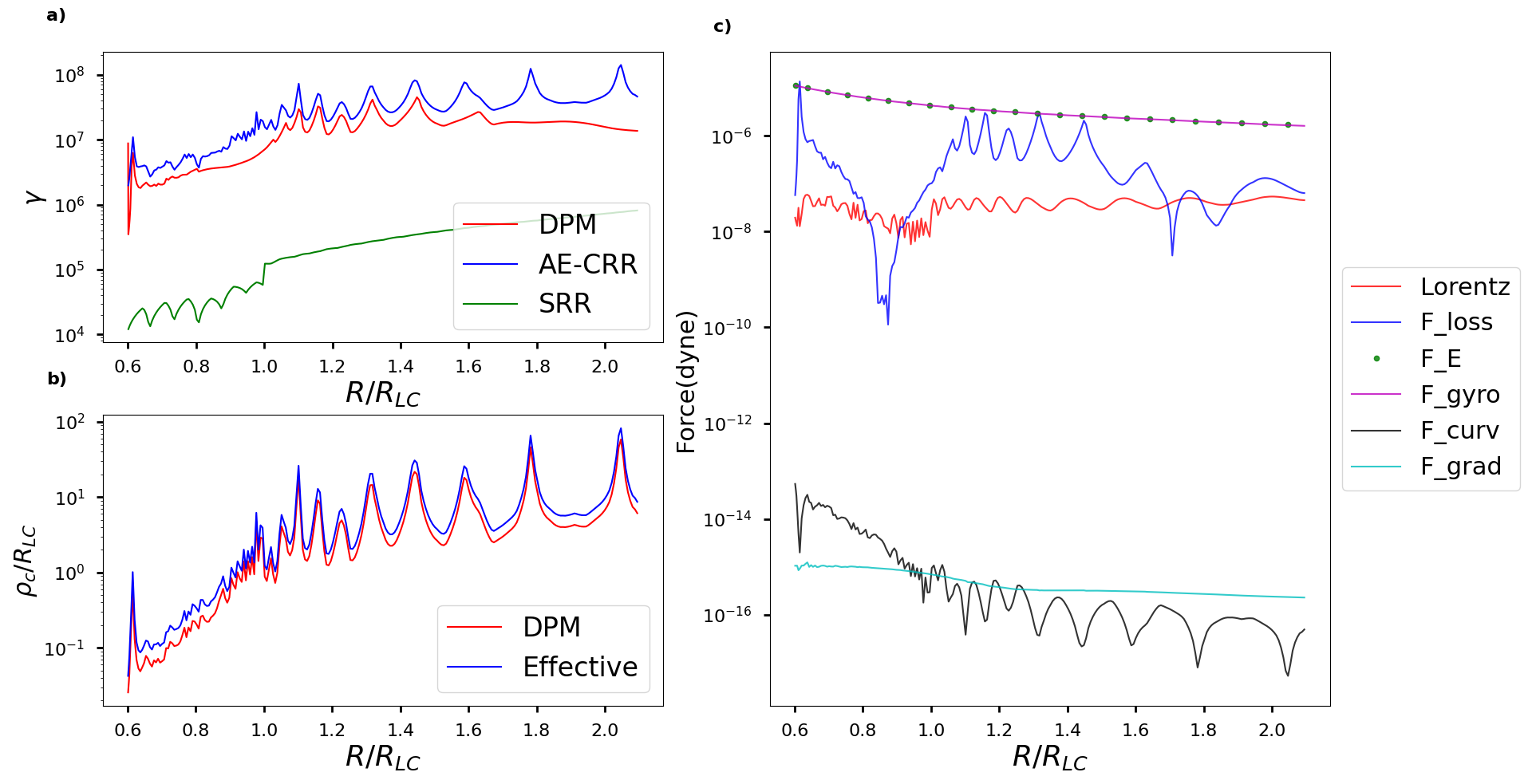}
\caption{Results for the Vela-like case using $B_{\rm S} =8\times 10^{11} \, \rm{G}$, with panel a) showing our $\gamma$ results in red $\gamma_{\rm c}$ in blue, and $\gamma_{\rm SRR}$ in green. Panel b) shows our particle $\rho_{\rm c}$ in red and the effective $\rho_{\rm c}$ from \citetalias{Kelner2015} in blue. Panel c) shows the different force components, namely the Lorentz force in red, the RRF in blue, the gyro-component of the Lorentz force in magenta, the $E$-field component of the Lorentz force in green dots, the curvature drift in black, and the gradient drift in cyan.}
\label{FF_AE_rho_c_8e11_CRR}
\end{figure*}

For the AE convergence results in Figure~\ref{FF_AE_vel_8e11_CRR}, we use the same case as for Figure~\ref{FF_Emission_phase_8e11_CRR}. We see that our model results in panels a), b), and c) agree well with the AE results until $1.0R_{\rm LC}$ where the $E_{\parallel}$ rapidly increases and the results start to diverge. This is also seen in the spike in $\theta_{\rm  VA}$ at $1.0R_{\rm LC}$ in panel d). For Figure~\ref{FF_AE_rho_c_8e11_CRR}, we use the same case and show our model $\gamma$ in red, $\gamma_{\rm c}$ in blue, and $\gamma_{\rm SRR}$ in green. Here we see our model is initially limited by $\gamma_{\rm c}$ and stays quite close to $\gamma_{\rm c}$ for the rest of its trajectory. In panel b), we see that our model is once again quite close to $\rho_{\rm eff}$ but deviates a little more than in the other cases. Looking at the forces in panel c), we see that the Lorentz force and RRF are not in equilibrium, but start to get close to equilibrium at $2.0R_{\rm LC}$.   

\subsection*{RD AE Results}
Here we show the AE convergence for an RD field using $B_{\rm S} =8\times 10^{8} \, \rm{G}$ and $R_{\rm acc} = 4.0\times 10^{-6} \,\rm{cm}^{-1}$ while injecting particles at the stellar surface, following the PC phase $0^{\circ}$ field line with a $\gamma_{0} = 10^{4}$ and $\theta_{\rm p} = 0.1^{\circ}$. These results show how well our model converges to the AE results for smooth analytical RD fields, initialising at the surface, and allowing the particle to naturally settle into equilibrium. 
% In panels a), b), and c) of Figure~\ref{RD_AE_vel}, we see our model results converge very well to the AE trajectory. 
In panel~d), we see that the $\theta_{\rm VA}$ is very small, indicating that our results are very close to the AE results. This shows that using these analytic RD fields we can use one set of equations of motion and accurately model the particle dynamics, naturally entering the radiation reaction limit approximated by the AE equations. Additionally, this illustrates how well our adaptive numerical scheme can deal with these high fields and high RRF without scaling or limiting the parameters, giving us high confidence in our results.   

\begin{figure}
\centering
\includegraphics[width=.5\textwidth]{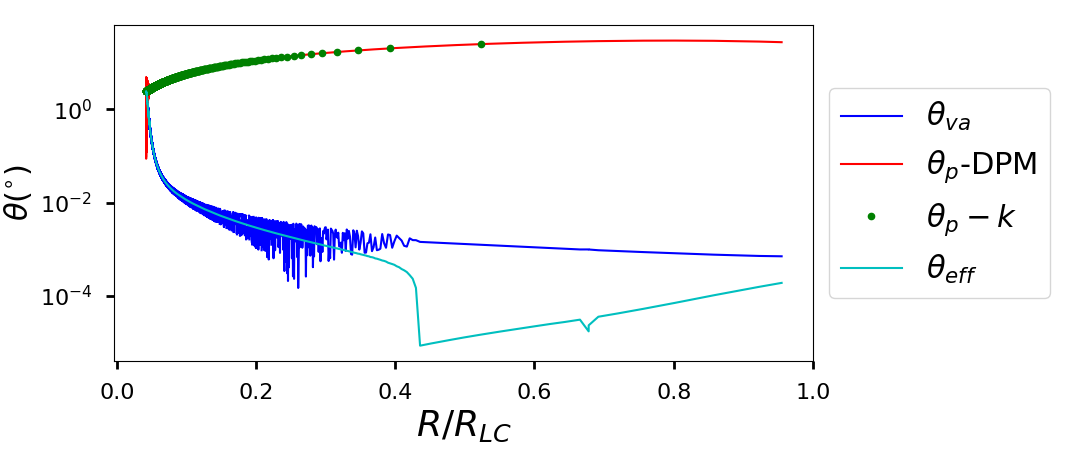}
\caption{The AE convergence results for the RD case using $B_{\rm S} =8\times 10^{8} \, \rm{G}$. In this plot DPM labels our model results, where $\theta_{VA}$ shows the angle between our particle velocity and the local AE velocity from \citet{Gruzinov2012}. Here $\theta_{p}-k$ is the theoretical pitch angle from \citetalias{Kelner2015} and $\theta_{\rm eff} = \arcsin (\tilde{B}_{\perp}/B_{\perp})$.}
\label{RD_AE_vel}
\end{figure}

\begin{figure*}
\centering
\includegraphics[width=.9\textwidth]{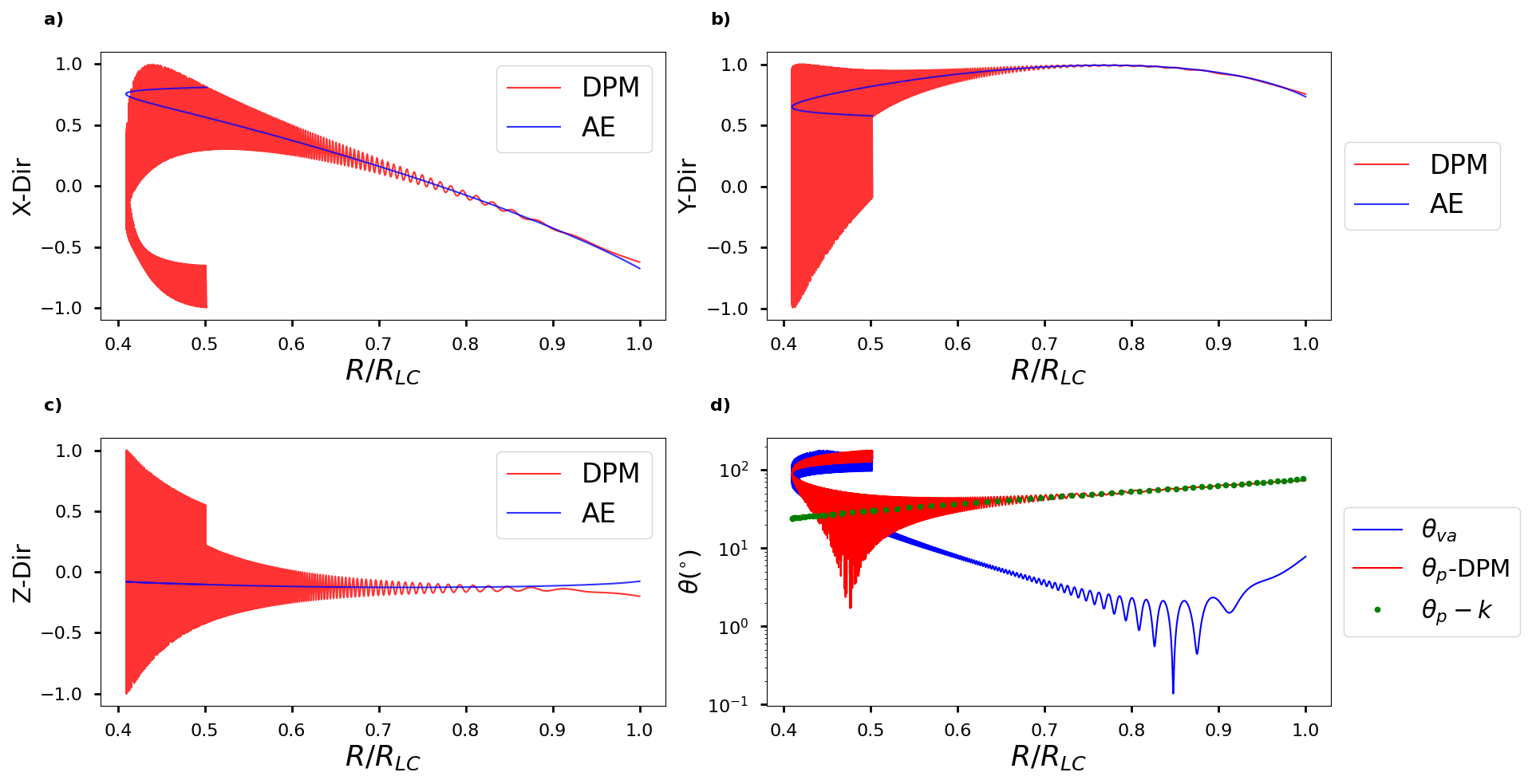}
\caption{In this plot for a mirror scenario, DPM labels our model results, and AE those of \citet{Gruzinov2012}. Panel a) shows the particle $x$-direction, panel b) the $y$-direction, and panel c) the $z$-direction. In panel d) we show the various angles discussed in Section~\ref{sec:2}.}
\label{AE_vel_mirror}
\end{figure*}

\begin{figure}
\centering
\includegraphics[width=.5\textwidth]{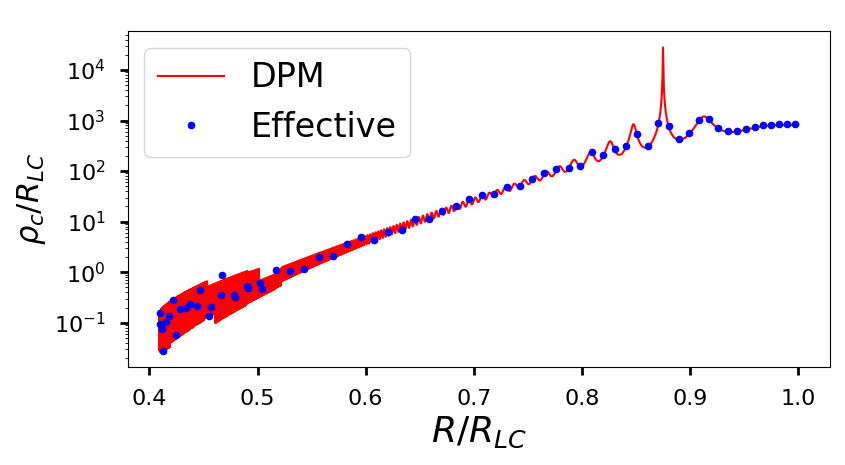}
\caption{Results for the case as shown in Figure~\ref{AE_vel_mirror}, showing our particle $\rho_{\rm c}$ in red and the effective $\rho_{\rm c}$ from \citetalias{Kelner2015} in blue.}
\label{AE_rho_c_mirror}
\end{figure}

For Figure~\ref{AE_vel_mirror}, we used a magnetic mirror scenario to investigate if we still converge to the AE results. These results show the inability of the AE equations to model the particle trajectories for a magnetic mirror scenario. For thoroughness we used the AE equations and changed the sign of the charge of the particle when the particle was mirrored, finding similarly inaccurate velocity component results. We use the same parameters as for Figure~\ref{RD_AE_vel}, but start the particle at $0.5R_{\rm LC}$ with an initial pitch angle of $160^{\circ}$, meaning the particle will move inwards. In Figure~\ref{AE_vel_mirror} panel a), b), and c), we indicate our model results in red, with the particle starting at $0.5R_{\rm LC}$ (initialised on the $x$-axis) and heading towards the stellar surface where it encounters a magnetic mirror around $0.4R_{\rm LC}$. It is then turned around and heads outward. When looking at the AE trajectories in blue and green in these panels, we notice that they are very different than our results as the particle moves inward, but agree with our result when the particle is moving outward. We believe this is because Equation~(\ref{AE}) are defined for a particle moving outward, thus these equations can not be used in a magnetic mirror scenario. This is illustrated by $\theta_{\rm VA}$ in panel d) as well, where one sees that $\theta_{\rm  VA}$ is initially very large and becomes small after the particle is mirrored and moves outward. In Figure~\ref{AE_rho_c_mirror} we see that our model's $\rho_{\rm c}$ still agrees very well with $\rho_{\rm eff}$ from \citetalias{Kelner2015}. This illustrates their model is a good approximation even in the magnetic mirror scenario we are interested in for future modelling.   
%Lastly, we see in panel c) that the Lorentz force and RRF are not in equilibrium, with the RRF being much lower than the Lorentz force. Looking at the $E$-field component and the gyro-component of the Lorentz force, one sees that the components are initially not close to one another, but as the particle travels outward these components balance one another as the field becomes FF in the outer segment of the magnetosphere.  

\subsection*{Emission maps} 
\begin{figure}
\centering
\includegraphics[width=.5\textwidth]{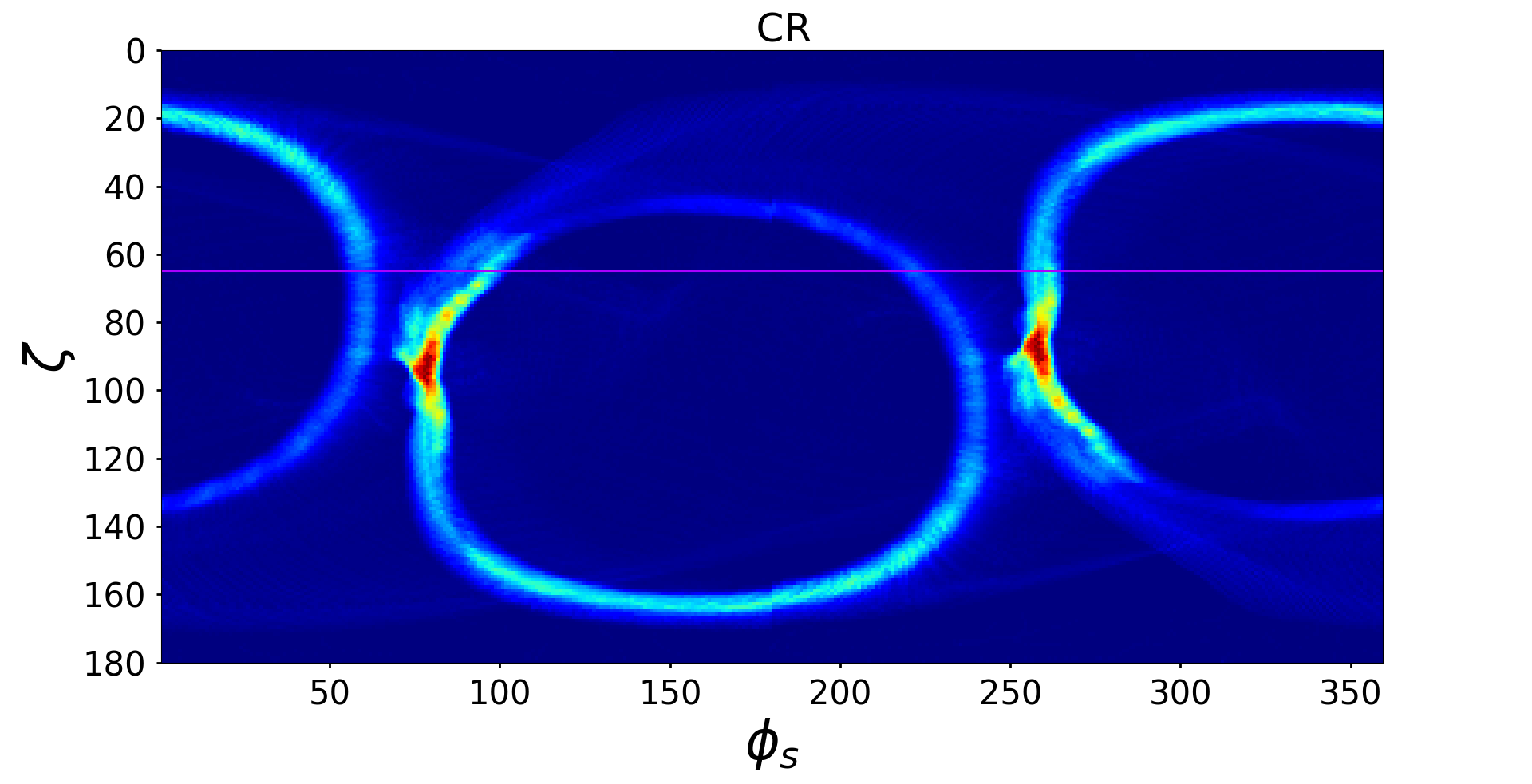}
\caption{CR emission map using our model's own $\rho_{\rm c}$ with a similar setup to Figure~\ref{CR_emission_maps} panel h), but using $R_{\rm acc}^{\rm min}=4.0\times 10^{-2} \,\rm{cm}^{-1}$ and $R_{\rm acc}^{\rm max}=2.5\times 10^{-1} \, \rm{cm}^{-1}$.}
\label{CR_map_own}
\end{figure}

To produce Figure~\ref{CR_map_own}, we use our own model's $\rho_{\rm c}$ and same setup as in Figure~\ref{CR_emission_maps} panel~d) but using $R_{\rm acc}^{\rm min}=4.0\times 10^{-2} \,\rm{cm}^{-1}$ and $R_{\rm acc}^{\rm max}=2.5\times 10^{-1} \, \rm{cm}^{-1}$. These are the same acceleration rates used in Figure~\ref{Al_B12_skymap}. These caustics appear much more similar to those in Figure~\ref{Al_B12_skymap} panel~b) than the caustics in Figure~\ref{CR_emission_maps} panel~d). The caustics have the correct position and notch emission region as Figure~\ref{Al_B12_skymap} panel~b), but there is still some extended emission in the upper arm of the caustic. See Section \ref{sec:3.5} for further discussion.

% Don't change these lines
\bsp	% typesetting comment
\label{lastpage}
\end{document}